\documentclass[a4paper,fleqn,usenatbib,useAMS]{mpazh}
\pdfoutput=1

\usepackage{hyperref}

\usepackage[utf8]{inputenc}
\usepackage{graphicx}
\usepackage{grffile}
\usepackage{natbib}

\usepackage{floatrow}
\usepackage{caption}
\usepackage{longtable}
\usepackage{amssymb}

\usepackage{multirow}

\usepackage[utf8]{inputenc}
\usepackage{graphicx}

\usepackage{floatrow}
\usepackage{longtable}
\usepackage{booktabs}

\usepackage[toc]{appendix}

\usepackage{multirow}

\def\doubleline{\vskip 3pt\hrule \vskip 1.5pt \hrule \vskip 5pt}

\def\ps1{\emph{Pan-STARRS1}}

\def\srg{\textit{SRG}}
\def\art{ART-XC}
\def\ero{eROSITA}
\def\srge{\textit{SRG/eROSITA}}
\def\rosat{\textit{ROSAT}}
\def\xmm{\textit{XMM-Newton}}
\def\xmmshort{\textit{XMM}}
\def\swift{\textit{Swift}}

\def\nh{N_{\rm H}}
\def\lx{L_{\rm X}}

\def\fwhm{FWHM}
\def\fwhmmes{FWHM_{\rm mes}}
\def\fwhmres{FWHM_{\rm res}}

\def\azt{AZT-33IK}
\def\rtt{RTT-150}

\def\o3hb{\rm $\lg({}[OIII]\lambda5007/H\beta)$}
\def\n2ha{\rm $\lg({}[NII]\lambda6584/H\alpha)$}

\title{New Active Galactic Nuclei Detected by the \art\ and \ero\ Telescopes during the First Five SRG All-Sky X-ray Surveys}
\author[Uskov et. al.]{
	G.S. Uskov,
			\thanks{\href{mailto:uskov@cosmos.ru}{\nolinkurl{uskov@cosmos.ru}}}
				$^{1}$
		S. Yu. Sazonov,
				$^{1}$
    	I. A. Zaznobin,
				$^{1}$
    	R. A. Burenin,
				$^{1}$
       \newauthor
        M. R. Gilfanov,
				$^{1,2}$
        P. S. Medvedev,
    				$^{1}$
        R. A. Sunyaev,
				$^{1,2}$
    	R. A. Krivonos,
				$^{1}$
        \newauthor
        E. V. Filippova,
				$^{1}$
        G. A. Khorunzhev,
				$^{1}$
		M. V. Eselevich,
				$^{3}$
	\\
			$^{1}$Space Research Institute, Russian Academy of Sciences, Moscow,
117997 Russia\\
			$^{2}$Max Planck Institut for Astrophysik, Karl-Schwarzschild-Str. 1,
85741 Garching, Germany\\
			$^{3}$Institute of Solar--Terrestrial Physics, Russian Academy of
Sciences, Siberian Branch, Irkutsk, 664033 Russia}

\date{Accepted XXX. Received YYY; in original form ZZZ}

\pubyear{2023}

\begin{document}
\label{firstpage}
\pagerange{\pageref{firstpage}--\pageref{lastpage}}
\maketitle

\begin{abstract}
We present the results of our identification of 14 X-ray sources detected 
in the eastern Galactic sky ($0<l<180 \circ$ ) in the 4--12 keV energy band
on the combined map of the first five all-sky surveys (from December 2019
to March 2022) with the Mikhail Pavlinsky \art\ telescope onboard the \srg\ observatory.
All 14 sources are reliably detected by the \srge\ telescope in the 0.2--8 keV energy band. 
Six of them have been detected in X-rays for the first time, while the remaining ones have already been known previously as X-ray sources, but their nature has remained unknown. 
We have taken optical spectra for 12 sources with the 1.6-m \azt\ telescope at the Sayan Observatory (the Institute of Solar–Terrestrial Physics, the Siberian Branch of the Russian Academy of Sciences). 
For two more objects we have analyzed the archival spectra taken during the 6dF survey. 
All objects have turned out to be Seyfert galaxies (one NLSy1, three Sy1, four Sy1.9, and six Sy2) at redshifts $z=0.015$--0.238. 
Based on data from the \ero\ and \art\ telescopes onboard the \srg\ observatory, we have obtained X-ray spectra for all objects in the energy range 0.2--12 keV. 
In four of them the intrinsic absorption exceeds $\nh>10^{22}$~cm$^{-2}$  at a 90\% confidence level, with one of them being probably heavily obscured ($\nh>5\times 10^{22}$~cm$^{-2}$ with 90\% confidence). 
This paper continues our series of publications on the identification of hard X-ray sources detected during the all-sky survey with the \srg\ orbital X-ray observatory.

{\it Keywords}:  active galactic nuclei, sky surveys, optical observations, redshifts, X-ray observations

\end{abstract}

\section{INTRODUCTION}

The Spectrum–RG (\srg) orbital observatory \citep{srg} has conducted an all-sky X-ray survey since December 2019.
There are two telescopes with grazing-incidence X-ray optics onboard the satellite: \ero\ \citep{pred21} and Mikhail Pavlinsky \art\ \citep{Pavlinsky_2021_art} operating in the 0.2--9 and 4--30~keV energy bands, respectively.
A total of eight full sky surveys, each with a duration of six months, are planned to be conducted.
The first two surveys were completed in December 2020, and the first catalog of X-ray sources (ARTSS12) detected with the \art\ telescope in the 4--12~keV energy band was produced from their results \citep{artsurvey}.
Among 867 sources it contains dozens of astrophysical objects (the exact number is unknown, since there are false X-ray sources in the catalog) whose nature was unknown when the catalog was released, with some of them having not been detected previously in X-rays.

The \srg/\art\ all-sky survey allows representative samples of such classes of objects as active galactic nuclei (AGNs) and cataclysmic variables (CVs) to be obtained.
Therefore, it is important to identify a maximally large number of new objects detected during the survey.
Such a work was begun when producing the ARTSS12 catalog and is continued at present.
The optical observations being carried out with the 1.6-m \azt\ telescope at the Sayan Observatory (the Institute of Solar-Terrestrial Physics, the Siberian Branch of the Russian Academy of Sciences) and the 1.5-m Russian–Turkish telescope (\rtt) at the T\"{U}BITAK National Observatory incorporated into the \srg\ ground support complex play a major role in this work.
The first results of this observational campaign were presented in \citet{zaznobin21, uskov22, zaznobin22cv}, where the identification of 25 AGNs (including eight objects from the archival data of the spectroscopic 6dF survey, \citealt{6dfsurvey}) and three CVs was reported.
In addition, during the \srg/\art\ sky survey several X-ray binaries with neutron stars and black holes were discovered and then identified \citep{lutovinov22,mereminsky22}.

The fourth \srg\ all-sky survey was completed in December 2021, and approximately a third of the sky had been scanned for the fifth time by March 7, 2022.
Then, the all-sky survey was suspended, and the \art\ telescope began to conduct a deep survey of the sky along the Galactic plane.
At present, the work on producing the catalog of sources detected by the \art\ telescope based on data from the first five surveys (below we will use precisely this wording, although the fifth survey has not been completed), ARTSS1-5, is being completed.
Many new objects appeared in the new catalog, and the work on their identification is underway.

In this paper we present the results of our identification and classification of 14 AGNs selected among the hard X-ray sources from the ARTSS1-5 catalog in the eastern Galactic half of the sky ($0<l<180\circ$). All these sources are reliably detected by the \srge\ telescope in the 0.2--8 keV energy band. We analyzed the broadband X-ray spectra obtained from the \ero\ and \art\ data and the optical spectra taken by us with the \azt\ telescope and previously during the 6dF survey. To calculate the luminosities of the objects, we use the model of a flat Universe with parameters $H_0=70$km s$^{-1}$ Mpc$^{-1}$, $\Omega_m = 0.3$.

\section{THE SAMPLE OF OBJECTS, OBSERVATIONS}

\begin{table*}
    \caption{The sample of objects.}
    \label{tab:list_src}
    \vskip 2mm
    \renewcommand{\arraystretch}{1.5}
    \renewcommand{\tabcolsep}{0.15cm}
    \centering
    \footnotesize
    \begin{tabular}[t]{rlccccccccl}
        \noalign{\doubleline}
    & &\multispan3\hfil \ero\ source \hfil & \multispan2\hfil Optical counterpart \hfil &  \\
 № & \art\ source & $\alpha$    & $\delta$ & $r98$     & $\alpha$    & $\delta$  &$r_{\rm A}$&$r_{\rm e}$& $F_{\rm A}^{4-12}$ & Discovered \\
 \noalign{\vskip 3pt\hrule\vskip 5pt}
1  & SRGA\,J001439.6$+$183503 & 3.66955   & 18.58246 & 11.0\arcsec & 3.66712   & 18.58203   & 10.6\arcsec &  8.4\arcsec &   $4.7_{-1.5}^{+1.8}$ & \xmmshort, \swift \\
2  & SRGA\,J002240.8$+$804348 & 5.68178   & 80.72962 & 2.2\arcsec  & 5.68204   & 80.72947   & 7.2\arcsec  &  0.6\arcsec &   $2.8_{-0.8}^{+0.9}$ & \rosat \\
3  & SRGA\,J010742.9$+$574419 & 16.92936  & 57.73894 & 2.7\arcsec  & 16.92964  & 57.73825   & 2.2\arcsec  &  2.6\arcsec &   $3.0_{-1.0}^{+1.2}$ & \srg\ \\
4  & SRGA\,J021227.3$+$520953 & 33.11066  & 52.16487 & 2.5\arcsec  & 33.11032  & 52.16483   & 7.6\arcsec  &  0.8\arcsec &   $1.4_{-0.9}^{+1.1}$ & \rosat \\
5  & SRGA\,J025208.4$+$482955 & 43.04074  & 48.49992 & 2.7\arcsec  & 43.04017  & 48.49983   & 13.1\arcsec &  1.4\arcsec &   $2.4_{-1.1}^{+1.4}$ & \rosat \\
6  & SRGA\,J045432.1$+$524003 & 73.63236  & 52.66875 & 2.8\arcsec  & 73.63262  & 52.66847   & 4.3\arcsec  &  1.2\arcsec &  $4.1_{-1.5}^{+1.8}$ & \srg \\
7  & SRGA\,J051313.5$+$662747 & 78.31903  & 66.46429 & 3.2\arcsec  & 78.31846  & 66.46398   & 17.9\arcsec &  1.4\arcsec &   $3.2_{-1.3}^{+1.5}$ & \swift \\
8  & SRGA\,J110945.8$+$800815 & 167.43408 & 80.13535 & 5.5\arcsec  & 167.43237 & 80.13489   & 10.8\arcsec &  2.0\arcsec &   $2.2_{-1.1}^{+1.3}$ & \srg \\
9 & SRGA\,J161251.4$-$052100 & 243.21307 & -5.35506 & 3.0\arcsec  & 243.21342 & -5.35485   & 17.7\arcsec &  1.5\arcsec &   $2.4_{-1.2}^{+1.4}$ & \rosat \\
10 & SRGA\,J161943.7$-$132609 & 244.93418 &-13.43768 & 3.4\arcsec & 244.93354 & -13.43781   & 8.7\arcsec  &  2.3\arcsec &  $2.9_{-1.2}^{+1.5}$ & \srg \\
11 & SRGA\,J182109.8$+$765819 & 275.29902 & 76.97126 & 5.4\arcsec  & 275.29846 & 76.97139   & 6.5\arcsec  &  0.7\arcsec &  $1.3_{-0.5}^{+0.6}$ & \srg \\
12 & SRGA\,J193707.6$+$660816 & 294.28375 & 66.13904 & 2.1\arcsec  & 294.28417 & 66.13925   & 6.4\arcsec  &  1.0\arcsec &  $0.8_{-0.4}^{+0.5}$ & \rosat \\
13 & SRGA\,J200331.2$+$701332 & 300.89093 & 70.22678 & 2.2\arcsec  & 300.89162 & 70.22692   & 15.0\arcsec &  1.0\arcsec &  $1.4_{-0.5}^{+0.5}$ & \rosat \\
14 & SRGA\,J211149.5$+$722815 & 317.96266 & 72.47104 & 3.0\arcsec  & 317.96575 & 72.47122   & 10.4\arcsec &  3.4\arcsec &   $0.9_{-0.6}^{+0.6}$ & \srg \\
\noalign{\vskip 3pt\hrule}
\end{tabular}
  \begin{flushleft}
  Column 1: the ordinal source number in the sample being studied.
  Column 2: the source name from the preliminary ARTSS1-5 catalog (the coordinates of the X-ray sources used in the names are given for epoch J2000.0).
  Column 3, 4: the source coordinates from the \ero\ data.
  Column 5: the radius of the 98\% eROSITA position error circle.
  Column 6, 7: the coordinates of the suspected optical counterpart.
  Column 8: the angular separation between the \art\ source and the optical counterpart.
  Column 9: the angular separation between the \ero\ source and the optical counterpart.
  Column 10: the average 4--12~keV X-ray flux from the sum of five ART-XC sky surveys, in units of $10^{-12}$~erg~s$^{-1}$~cm$^{-2}$. 
  Column 11: the orbital observatory that detected the X-ray source for the first time.
  \end{flushleft}
\end{table*}

Objects from the catalog of sources detected in the 4--12~keV energy band on the combined map of the first five \srg/\art\ sky surveys (from December 12, 2019, to March 7, 2022) (the ARTSS1-5 catalog, being prepared for publication) constitutes the sample. We considered only point sources from this catalog detected at a confidence level no less than 4.5 standard deviations in the half of the sky $0<l<180\circ$ (for which we also have \srge\ data). The sample includes a total of 14 objects.

Table~\ref{tab:list_src} for all objects gives the coordinates of the X-ray source from the \art\ and \ero\ data, the radius of the \ero\ position error circle (at 98\% confidence), the coordinates of the suspected optical counterpart (see the Section Results), the angular separation between the X-ray source (from the \art\ and \ero\ data) and the optical counterpart, the flux in the 4--12~keV energy band (from the \art\ data), and the name of the observatory that detected the source in X-rays for the first time. Six sources were detected for the first time with the \art\ and \ero\ telescopes onboard the \srg\ observatory.

Figure~\ref{fig:guid_images} shows optical images of the objects being studied and the corresponding \art\ and \ero\ position error circles of the X-ray sources. A specific extended optical object can be unambiguously associated with each X-ray source.

\subsection{X-ray Observations}

Depending on their positions in the sky, the sample sources were scanned during four or five \srg\ all-sky surveys. The combined data of these surveys were used to construct the spectra of the sources in the energy range 0.2--12~keV.
 
The \art\ X-ray spectra were extracted from the all-sky survey data with the standard software that was used to process the survey data \citep{Pavlinsky_2021_art,artsurvey}. The data from all seven \art\ modules were combined. We used the data in two broad energy bands, 4--7 and 7--12~keV, that were extracted in a circle of radius 120$\arcsec$. We calibrated the count rate-to-flux conversion factors using Crab Nebula observations (see \citealt{artsurvey}) and constructed a diagonal response matrix based on them. The background level was estimated using the data in the hard 30--70~keV energy band and the wavelet decomposition survey images (see \citealt{artsurvey}).

The \ero\ data were processed with the calibration and data processing system created and maintained at the Space Research Institute of the Russian Academy of Sciences, which uses the elements of the eSASS (eROSITA Science Analysis Software System) package and the software developed by the science group on the X-ray catalog of the Russian eROSITA consortium. We extracted the source spectra in a circle of radius 60$\arcsec$ and the background spectra in a ring with an inner radius of 120$\arcsec$ and an outer radius of 300$\arcsec$ around the source. If other sources fell into the background region, then the photons in a region of radius 40$\arcsec$ around them were excluded. The spectra were extracted from the data of all seven \art\ modules in the energy range 0.2--9.0~keV. When fitting the spectra, the data were binned in such a way that there were at least three counts in each energy channel.

\subsection{Optical Observations}

\begin{table*}
    \caption{Log of optical observations.}
    \label{tab:list_obs}
    \vskip 2mm
    \renewcommand{\arraystretch}{1.1}
    \renewcommand{\tabcolsep}{0.35cm}
    \centering
    \footnotesize
\begin{tabular}[t]{lccccc}
\noalign{\doubleline}
\art\ source & Date & Telescope & Grism & Slit, \arcsec & Exposure time, s \\
\noalign{\vskip 3pt\hrule\vskip 5pt}
    SRGA\,J001439.6+183503 & 2022-10-31 & \azt & VPHG600G & 3 & $5\times300$   \\
    SRGA\,J002240.8+804348 & 2022-10-31 & \azt & VPHG600G & 2 & $3\times600$   \\
                           & 2022-11-01 & \azt & VPHG600R & 2 & $3\times600$   \\
    SRGA\,J010742.9+574419 & 2022-03-04 & \azt & VPHG600G & 3 & $6\times600$   \\
    SRGA\,J021227.3+520953 & 2022-11-18 & \azt & VPHG600G & 2 & $4\times600$   \\
                           & 2022-11-21 & \azt & VPHG600R & 2 & $4\times600$   \\
    SRGA\,J025208.4+482955 & 2022-11-01 & \azt & VPHG600G & 2 & $3\times600$   \\
    SRGA\,J045432.1+524003 & 2022-11-01 & \azt & VPHG600G & 2 & $3\times600$   \\
    SRGA\,J051313.5+662747 & 2022-11-01 & \azt & VPHG600G & 2 & $3\times200$   \\
    SRGA\,J110945.8+800815 & 2022-11-03 & \azt & VPHG600G & 2 & $5\times300$   \\
                         & 2022-11-03 & \azt & VPHG600R & 2 & $2\times300$   \\
    SRGA\,J161251.4$-$052100 & 2003-05-30 & UKST & VPH580V & & $3\times1200$\\
                             & 2003-05-30 & UKST & VPH425R & & $3\times600$\\
    SRGA\,J161943.7$-$132609 & 2004-04-16 & UKST & VPH580V & & $5\times1200$\\
                             & 2004-04-16 & UKST & VPH425R & & $4\times600$\\
    SRGA\,J182109.8+765819 & 2022-11-17 & \azt & VPHG600G & 2 & $5\times600$   \\
    SRGA\,J193707.6+660816 & 2022-11-01 & \azt & VPHG600G & 1.5 & $6\times300$ \\
    SRGA\,J200331.2+701332 & 2022-11-18 & \azt & VPHG600G & 3 & $4\times600$   \\
                           & 2022-11-18 & \azt & VPHG600R & 3 & $4\times600$   \\
    SRGA\,J211149.5+722815 & 2022-11-21 & \azt & VPHG600G & 2 & $4\times600$   \\
                           & 2022-11-21 & \azt & VPHG600R & 2 & $5\times600$   \\
\noalign{\vskip 3pt\hrule\vskip 5pt}
\end{tabular}
\end{table*}

Our spectroscopy was carried out at the \azt\ telescope using the low- and medium-resolution ADAM spectrograph \citep{adam16,adam16a} (see the log of observations in Table~\ref{tab:list_obs}). We used long slits of width 1.5\arcsec, 2\arcsec, and 3\arcsec at the ADAM spectrograph. The slit center was brought into coincidence with the central region of the observed galaxy. The observations were performed at a seeing (FWHM) better than 2.5\arcsec.

We used volume phase holographic gratings (VPHGs, grisms), 600 lines per millimeter, to take the spectra at the ADAM spectrograph. As a dispersive element we used VPHG600G for the spectral
range 3650--7250~\AA\ with a resolution of 8.6~\AA\ for a 2\arcsec slit and VPHG600R for the spectral range 6460--10050~\AA\ with a resolution of 12.2~\AA\ for a 2\arcsec slit. When using VPHG600R, we set the OS11 filter, which removes the second interference order from the image. A thick e2v CCD30-11 array produced by the deep depletion technology is installed at the spectrograph. This allows the spectroscopic images to be obtained at a wavelength of 10 000~\AA\ without interference on the thin CCD substrate. All our observations were performed with zero slit position angle. After each series of spectroscopic images for each object, we obtained the calibration images of a lamp with a continuum spectrum and the line spectrum of a He–Ne–Ar lamp.

On each observing night we took the spectra of spectrophotometric standards from the ESO\footnote{https://www.eso.org/sci/observing/tools/standards} list for all of the sets of diffraction gratings and slits being used. The spectrophotometric standards were chosen so that they were approximately at the same elevation with the optical source observed by us. The data reduction was performed using the IRAF\footnote{http://iraf.noao.edu/} software and our own software. The flux calibration was performed by standard IRAF procedures from the onedspec package.

The spectra of the objects were corrected for interstellar extinction with the \emph{deredden} procedure from the onedspec IRAF package in a standard way \citep{deredden}. The color excess $E(B-V)$ was determined with the help of the \emph{GALExtin}\footnote{www.galextin.org} service using the model of \cite{Schlegel}. We took $R_V = 2.742$ from \cite{Schlafly11}.

For two objects from the sample we analyzed the archival spectroscopic data from the 6dF survey \citep{6dftotal}. This survey was conducted at the UKST 1.2-m Schmidt telescope using a multifiber spectrograph with a $5.7\circ$ field of view equipped with two low-resolution ($R\approx 1000$) gratings with overlapping spectral ranges. The range 4000--7500~\AA\ was completely covered. The spectra taken during the survey were not flux-calibrated and are presented in counts, which does not allow the absolute fluxes in emission lines to be measured. However, these data can be used to estimate the line equivalent widths and the ratios of the fluxes in pairs of closely spaced lines, which is quite enough for the classification of AGNs.

\begin{figure*}
  \centering

  \includegraphics[width=0.25\columnwidth]{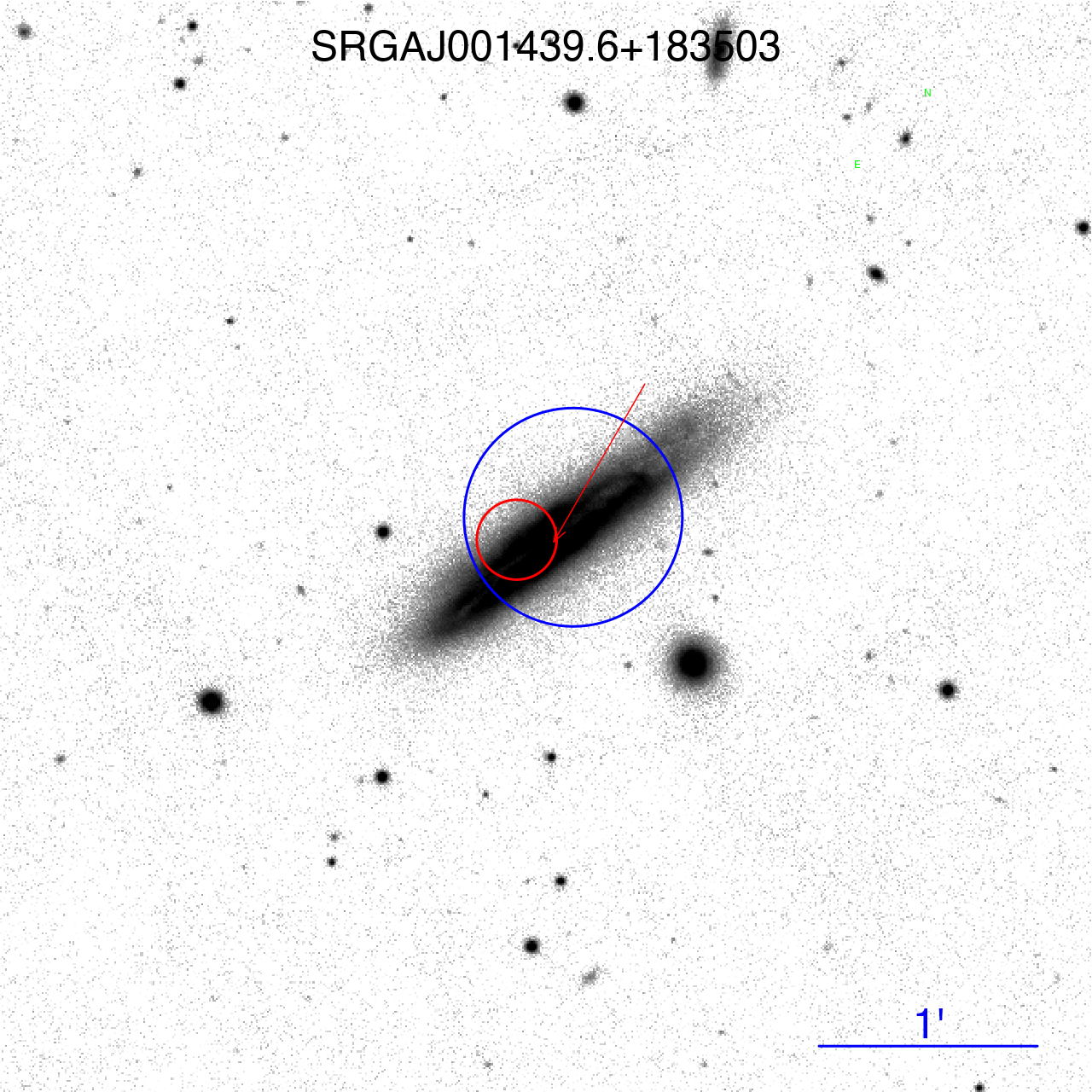}
  \includegraphics[width=0.25\columnwidth]{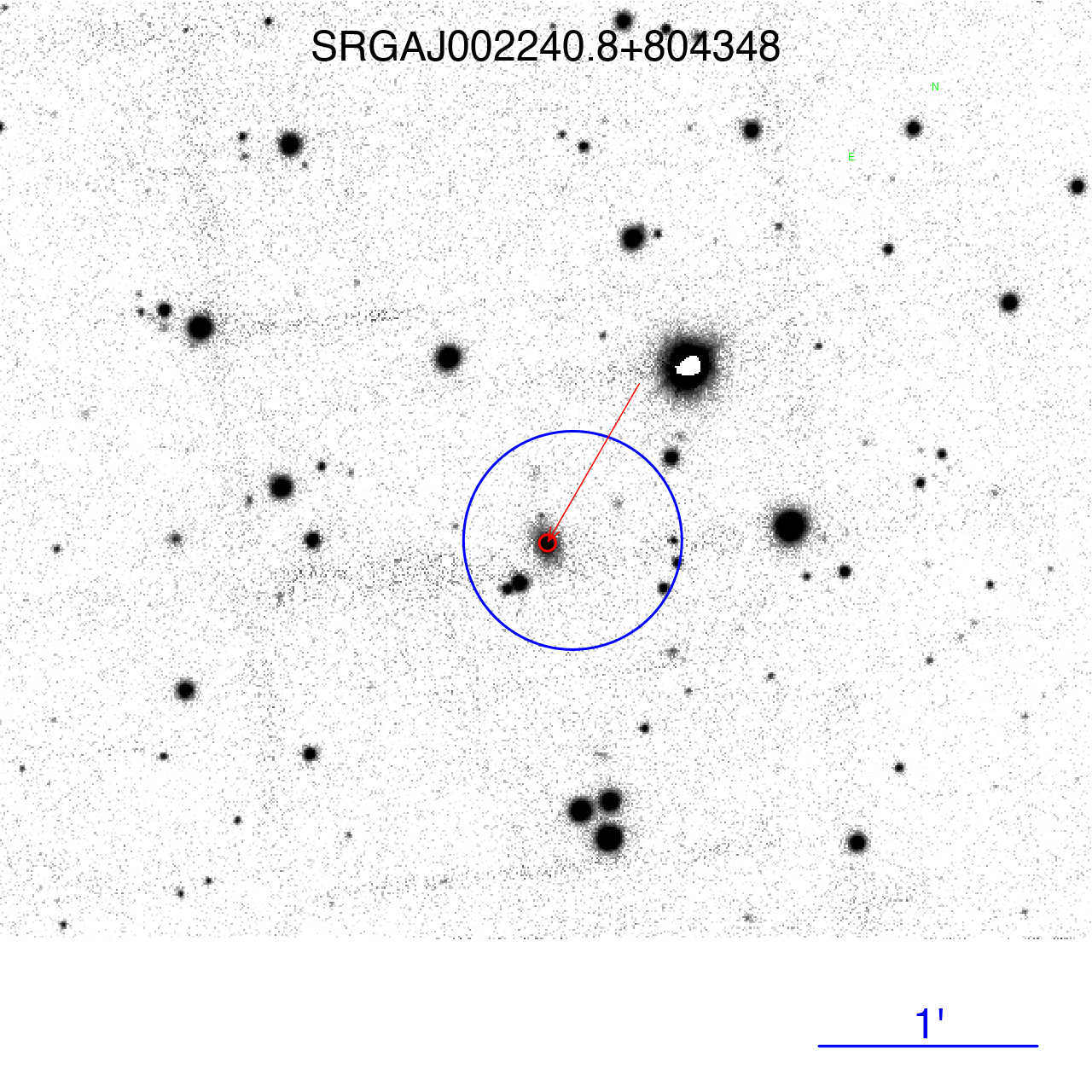}
  \includegraphics[width=0.25\columnwidth]{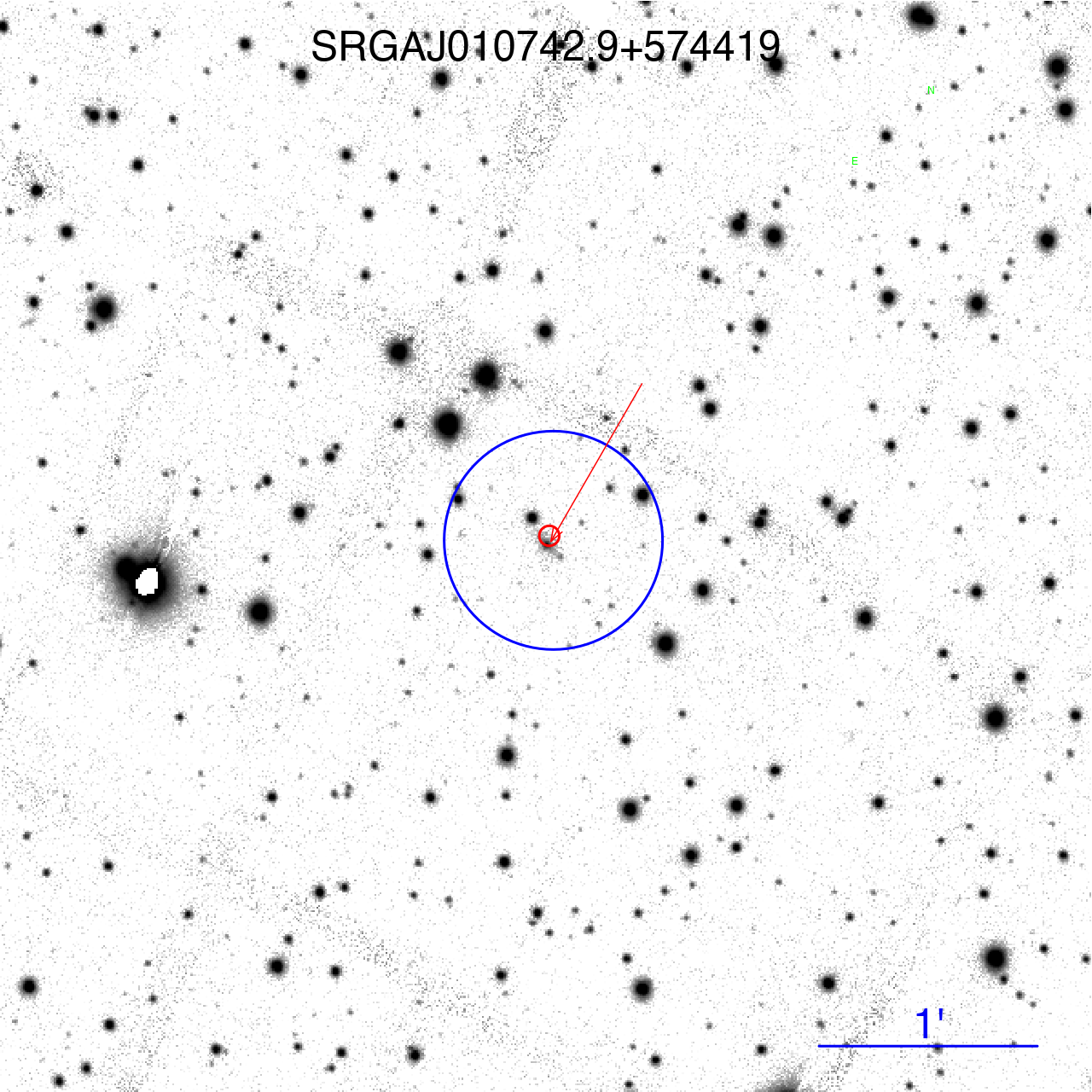}
  \includegraphics[width=0.25\columnwidth]{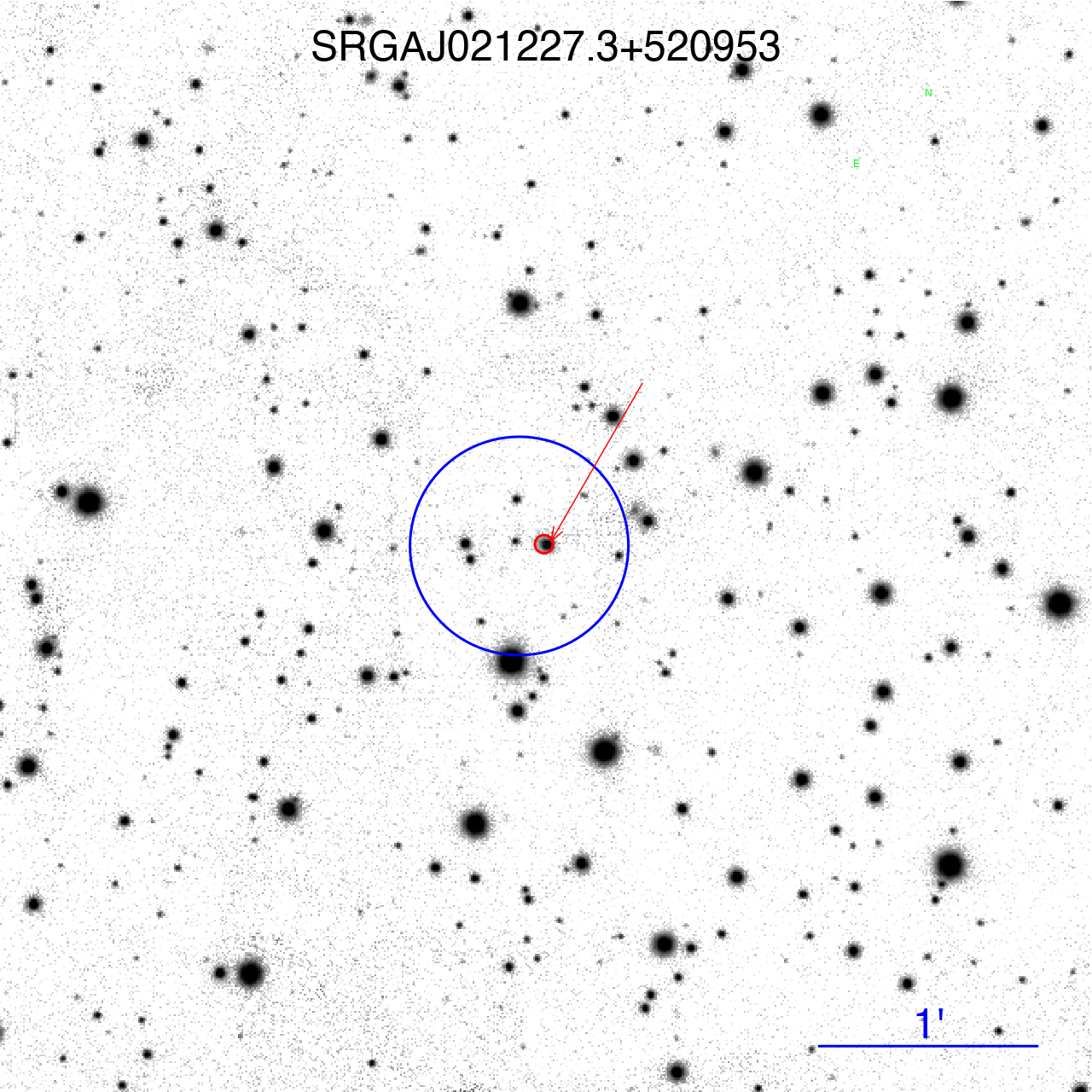}
  \includegraphics[width=0.25\columnwidth]{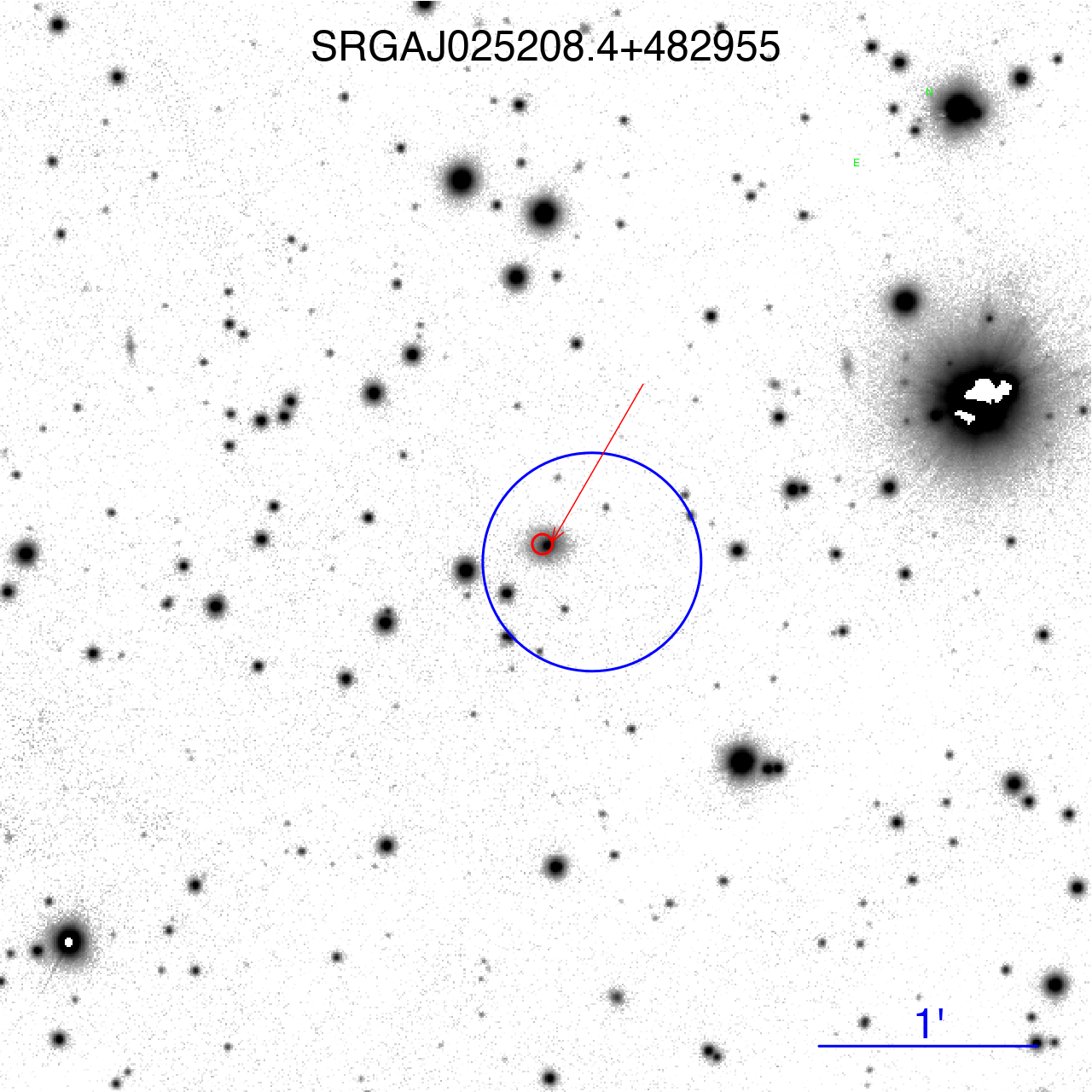}
  \includegraphics[width=0.25\columnwidth]{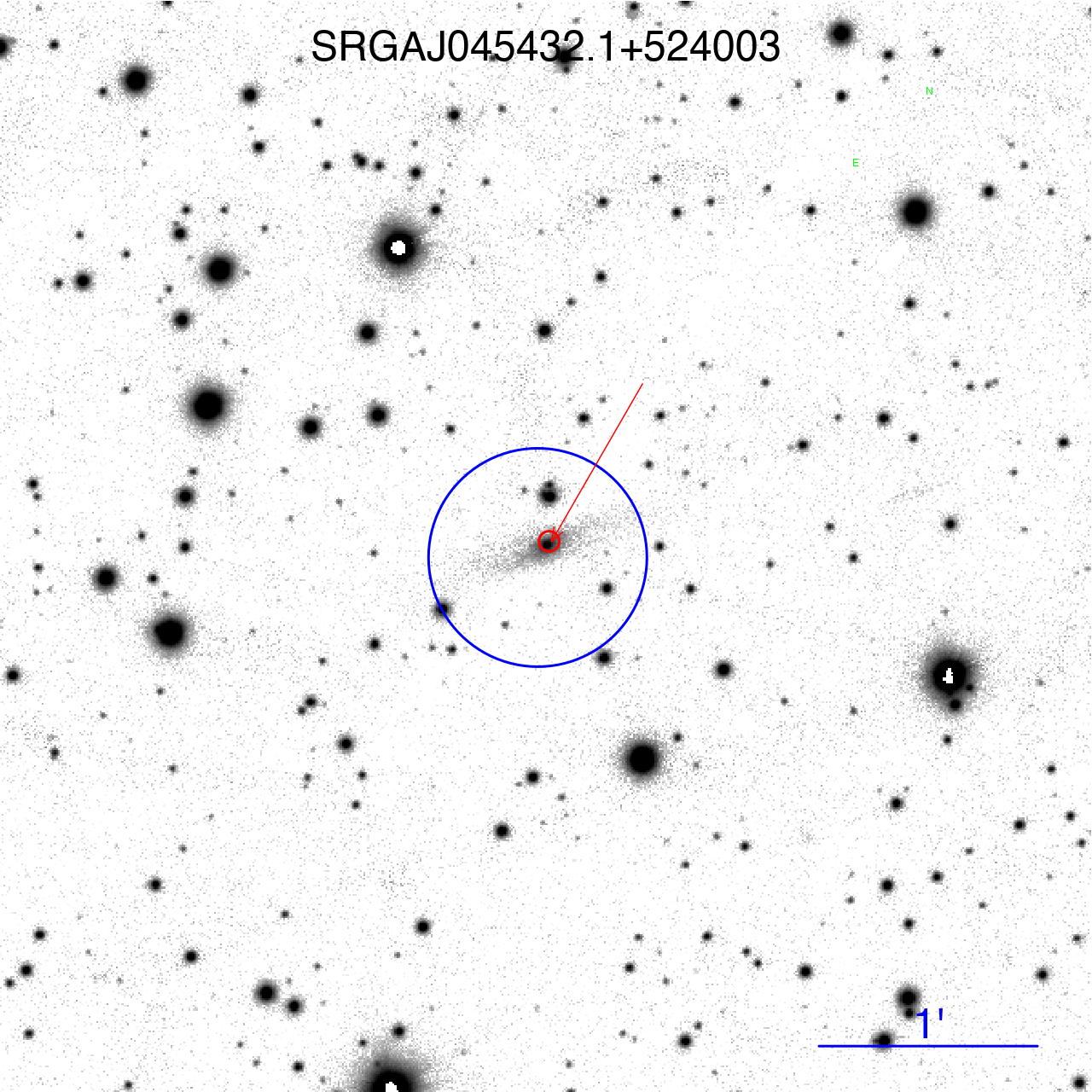}
  \includegraphics[width=0.25\columnwidth]{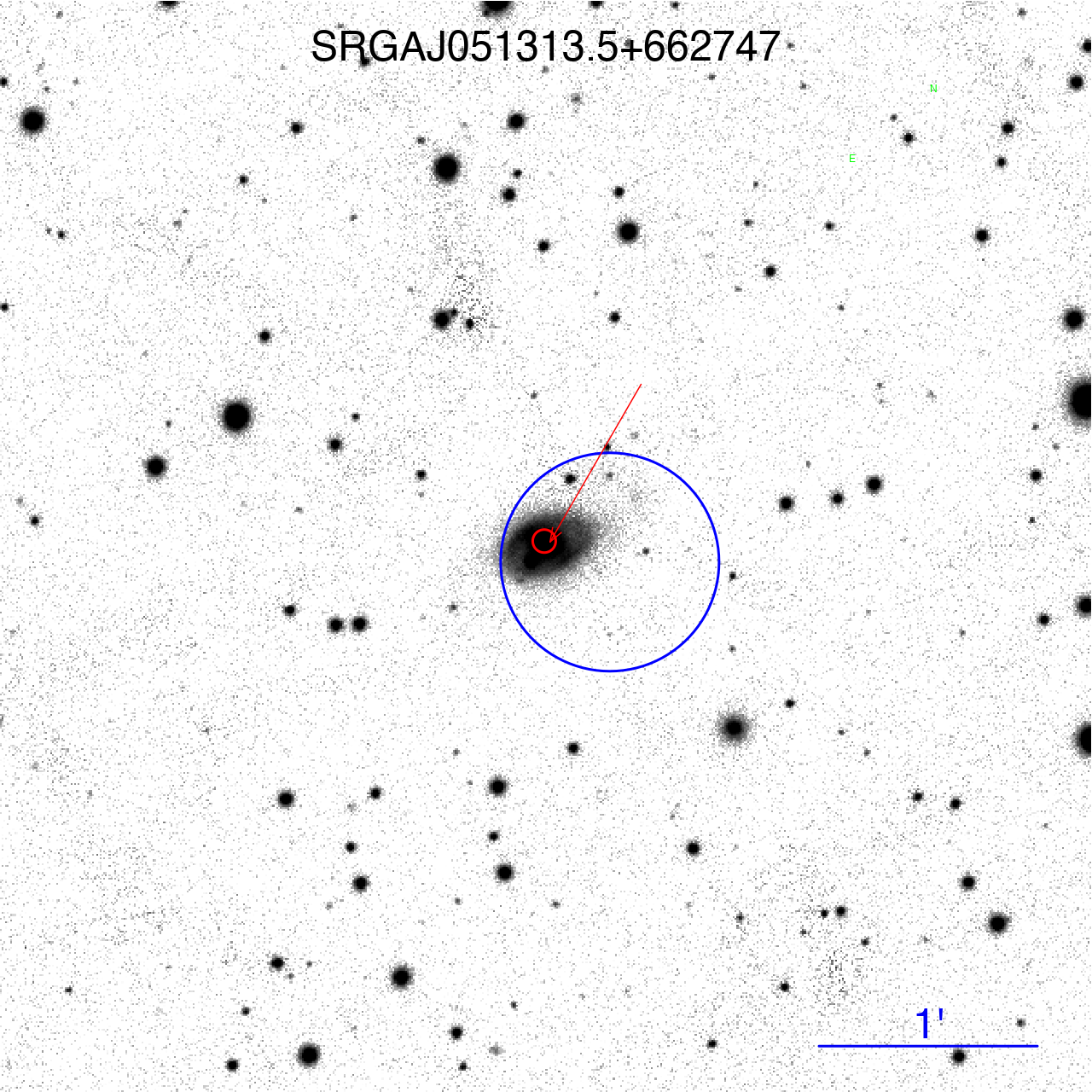}
  \includegraphics[width=0.25\columnwidth]{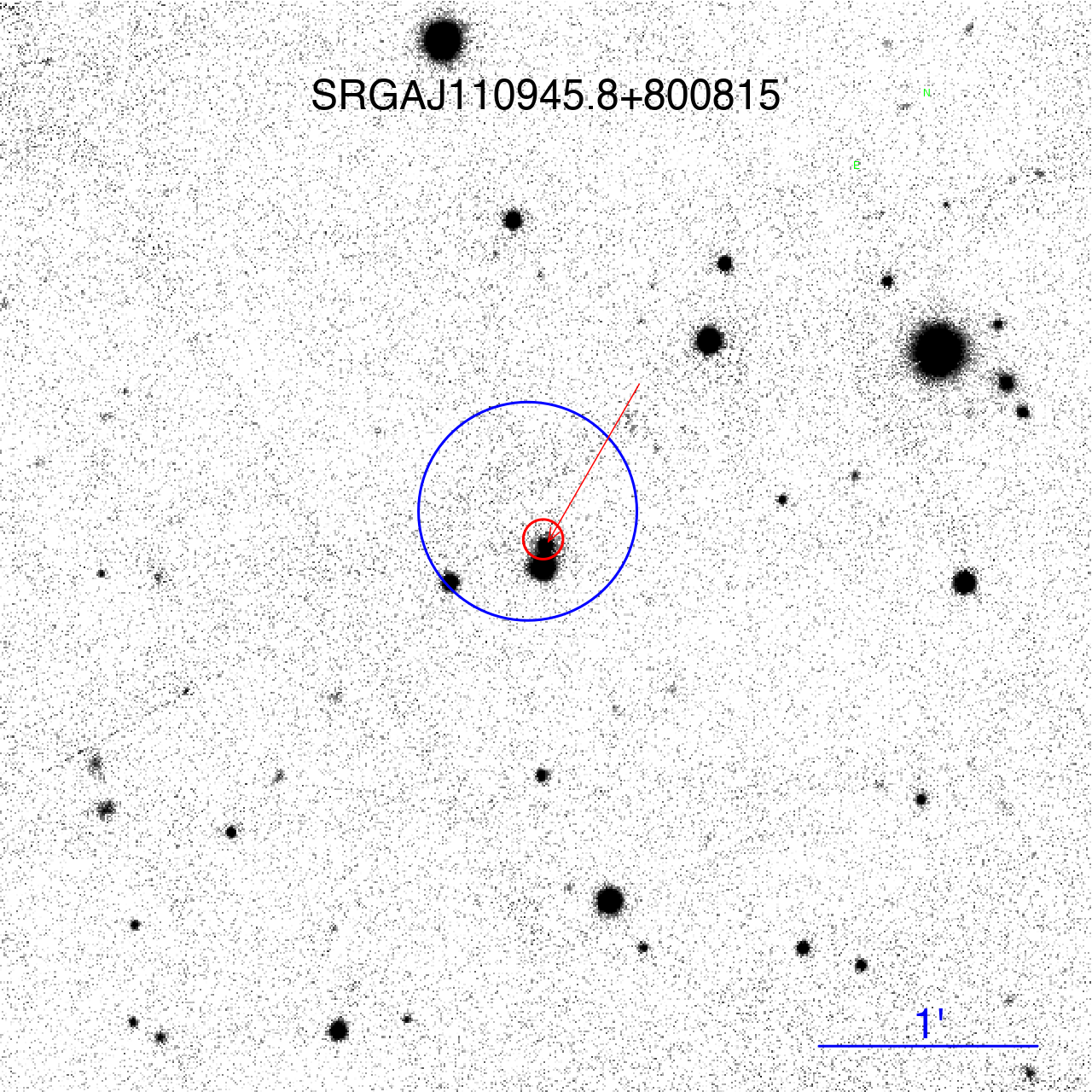}
  \includegraphics[width=0.25\columnwidth]{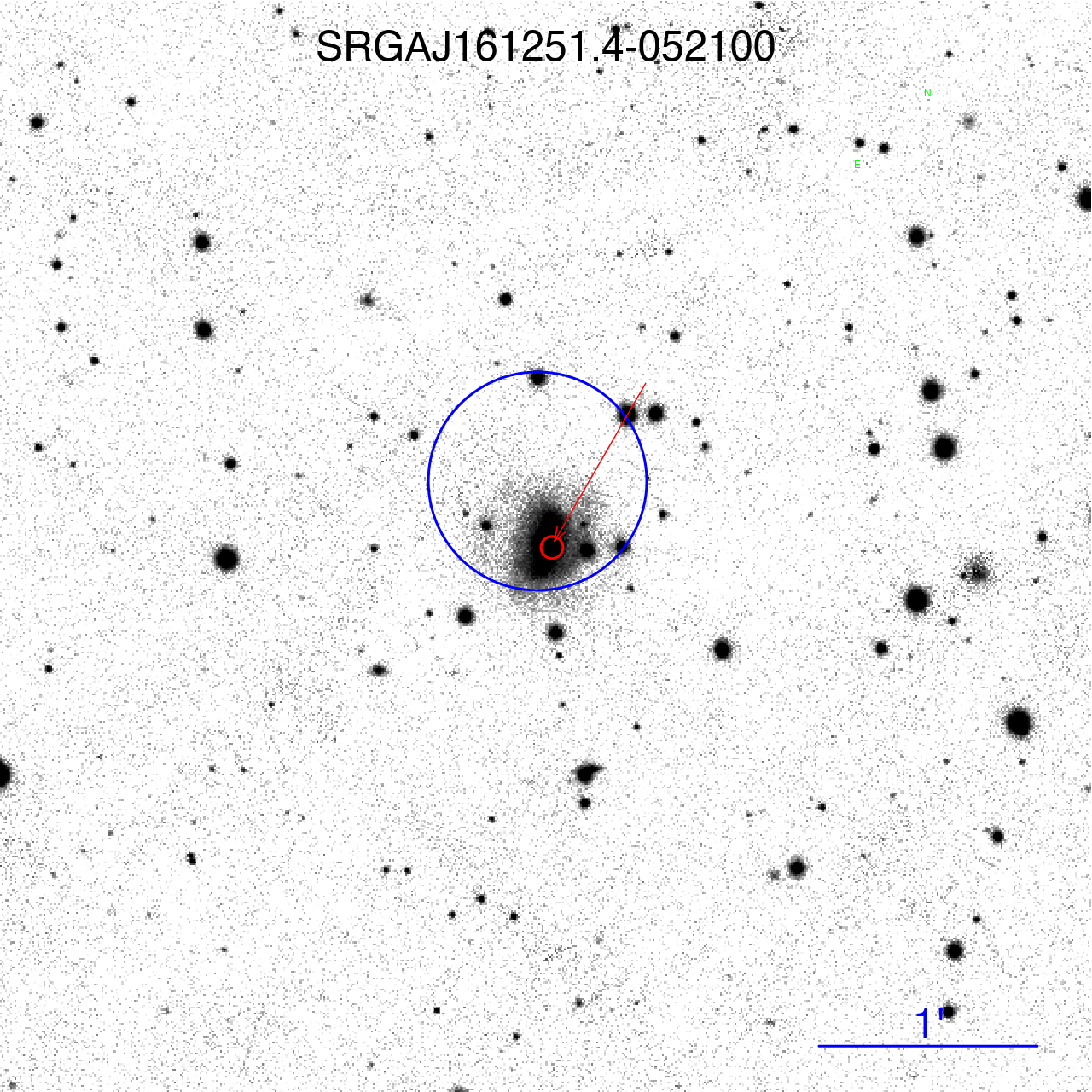}
  \includegraphics[width=0.25\columnwidth]{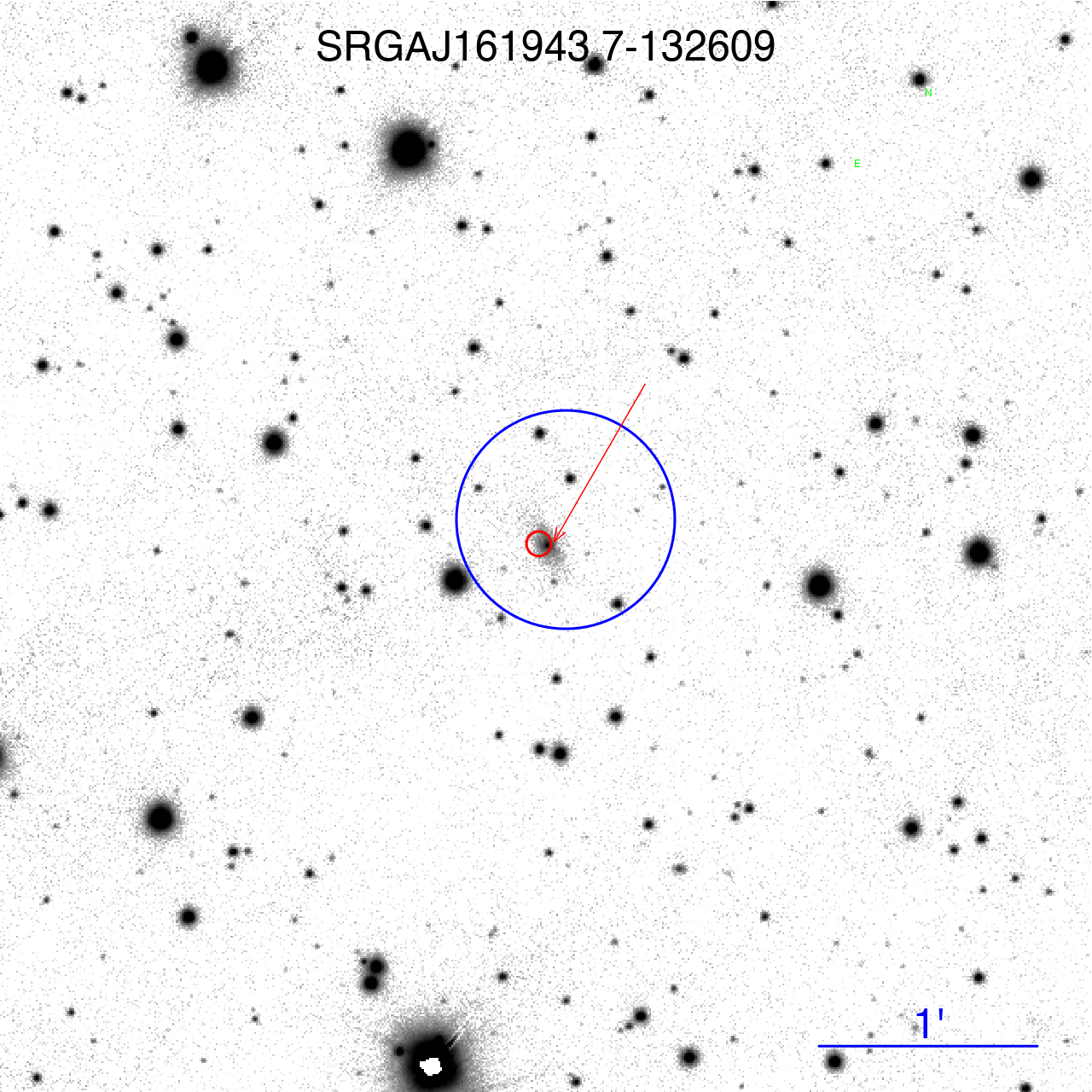}
  \includegraphics[width=0.25\columnwidth]{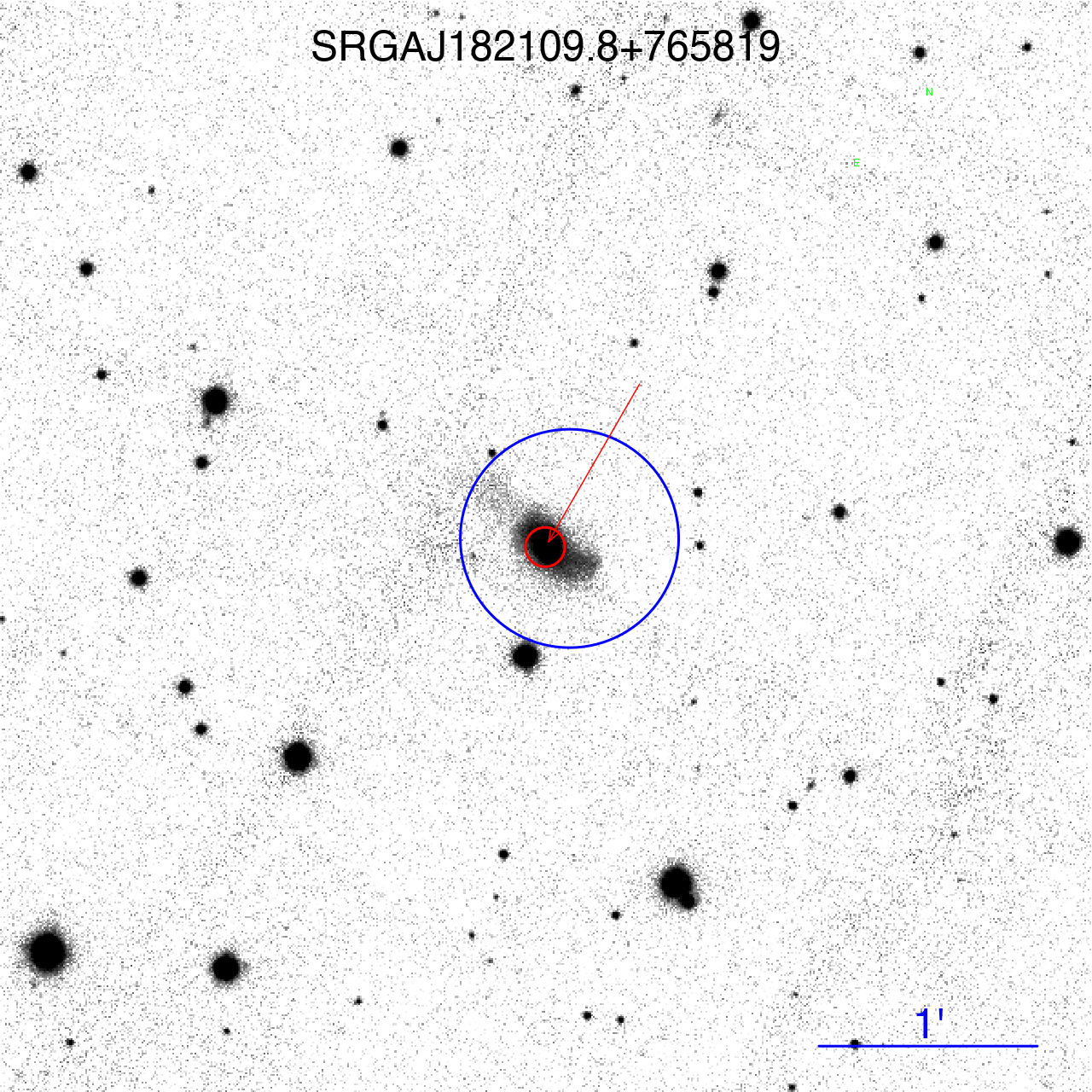}
  \includegraphics[width=0.25\columnwidth]{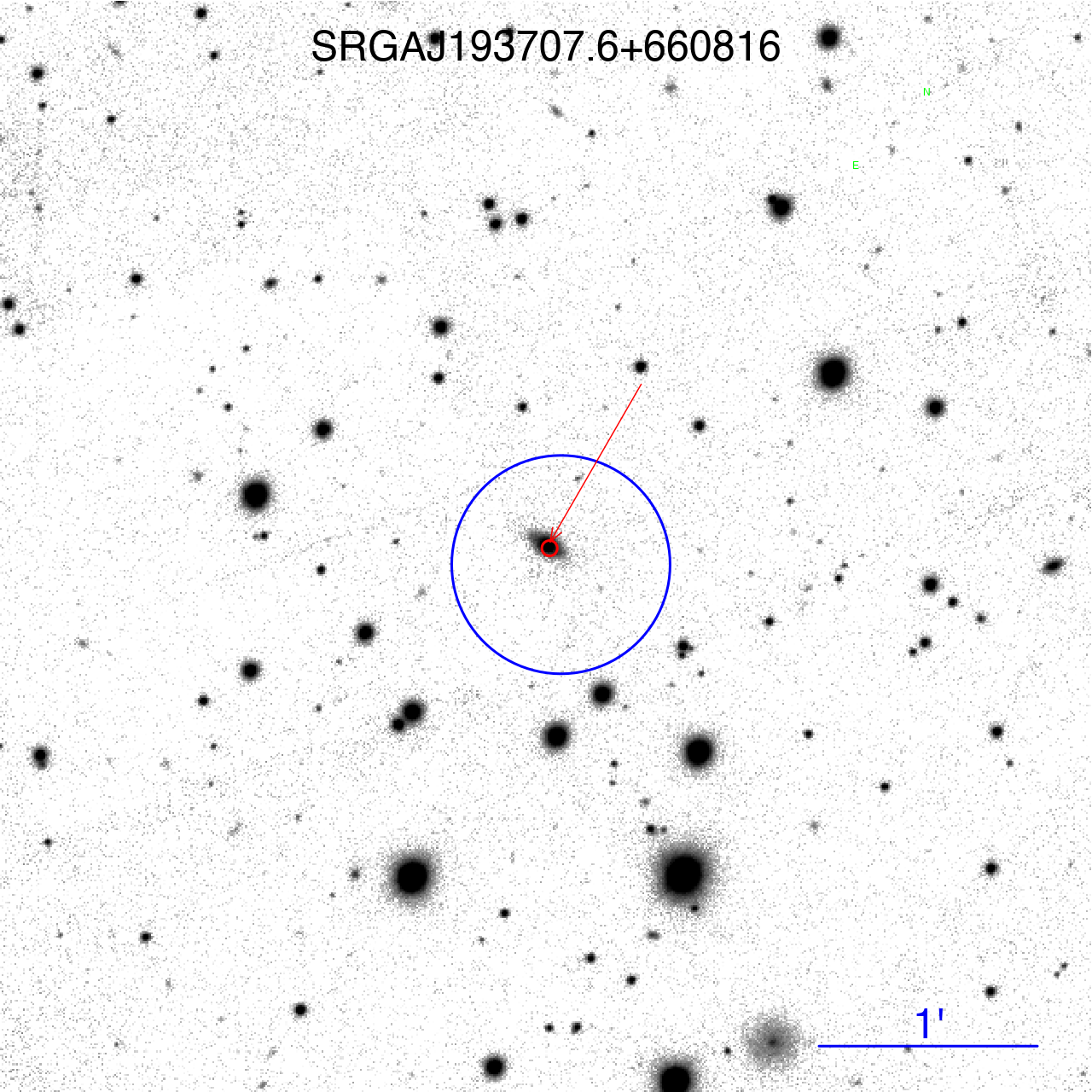}
  \includegraphics[width=0.25\columnwidth]{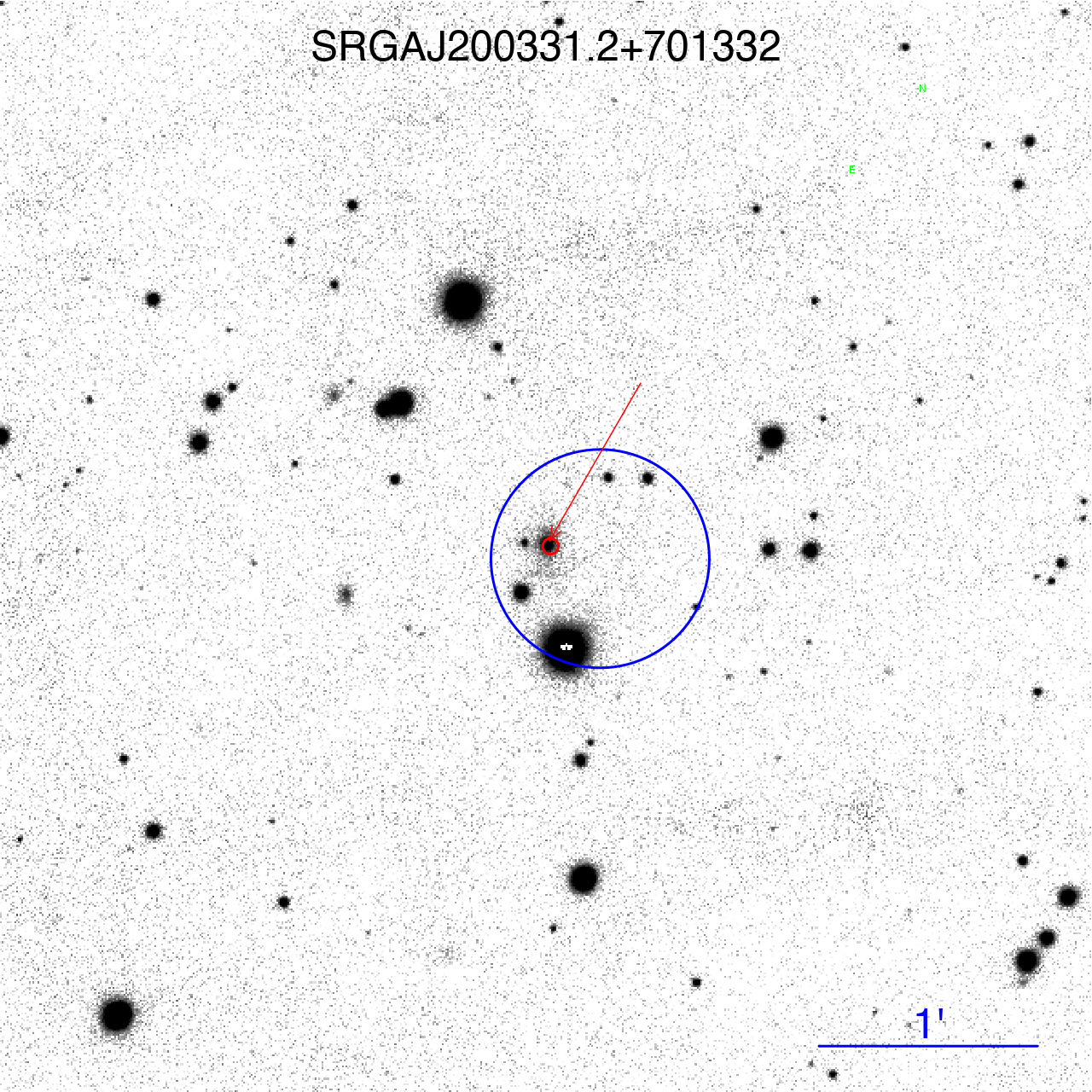}
  \includegraphics[width=0.25\columnwidth]{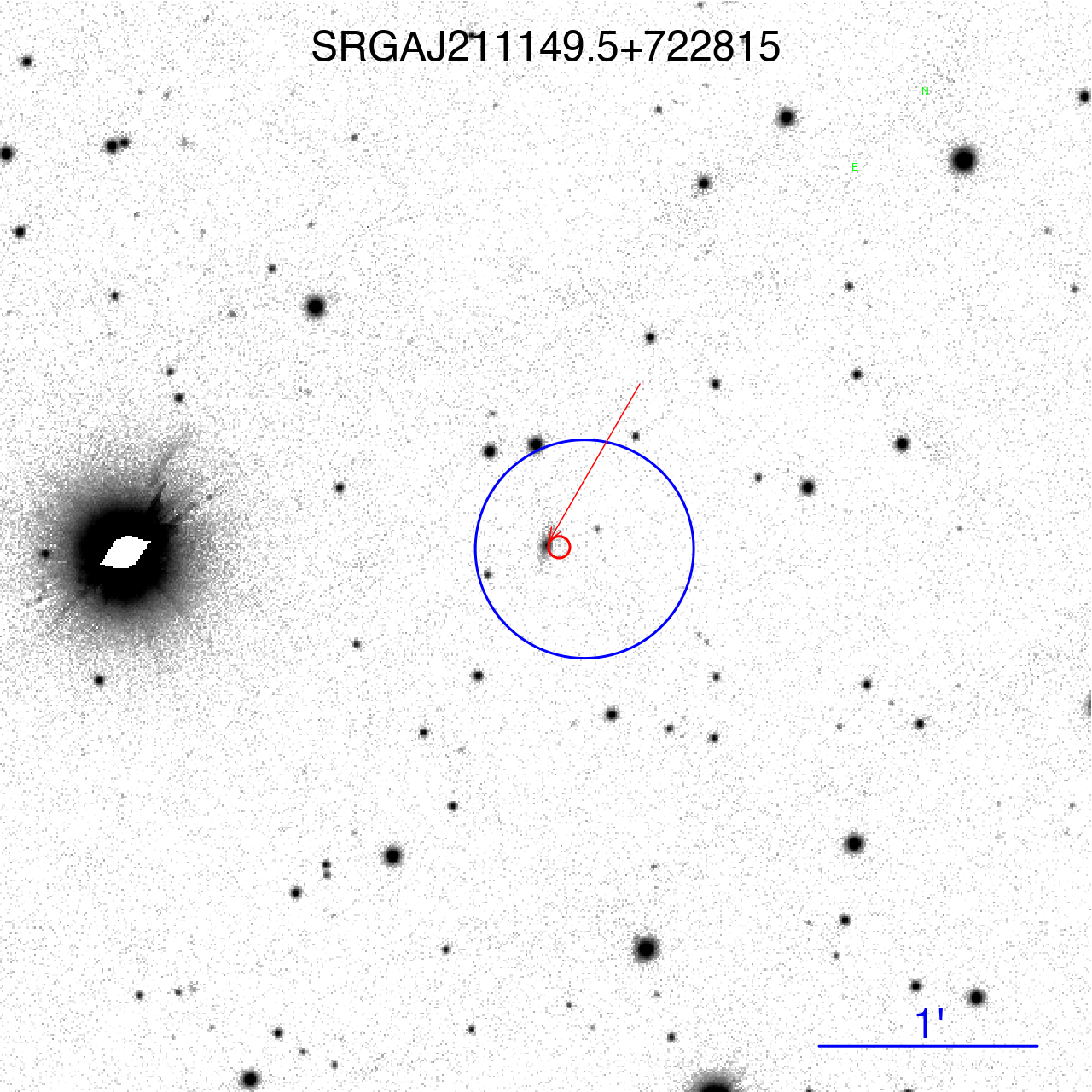}

  \caption{
  Optical images in the \emph{r} filter from the PanSTARRS PS1 survey \citep{chambers2016}. The large and small circles indicate the \art\ (radius 30\arcsec) and \ero\ (see $r98$ in Table~\ref{tab:list_src}) position error circles of the X-ray sources, respectively. The arrow indicates the optical objects whose spectra are analyzed in this paper.
  }
  \label{fig:guid_images}
\end{figure*}

\section{RESULTS}
\label{s:results}

\subsection{X-ray Spectra}

Our spectral analysis was performed jointly using the \ero\ and \art\ data. The spectra were fitted in the energy range 0.2--12~keV with the XSPEC v12.12.0\footnote{https://heasarc.gsfc.nasa.gov/xanadu/xspec} software \citep{arnaud96}. The $W$-statistic that takes into account the X-ray background was used for our model fitting.

To fit the spectra, we used the model of a power-law continuum with a low-energy cutoff due to photoabsorption in the Galaxy and the object itself. In the spectra of several sources with a great intrinsic absorption we detected an excess of the observed counts compared to the prediction of the power-law model at energies below $\sim 1$ kev. This excess can be caused by a slight inaccuracy in the current version of the \ero\ response matrix. On the other hand, in the X-ray spectra of type 2 AGNs at energies below 2~keV additional emission is often observed (see, e.g., \citealt{2005A&A...444..119G}) against the background of an absorbed power-law continuum. The nature of this emission can be varied, and it is by no means always possible to establish it. The following is discussed in the literature as possible mechanisms (see, e.g., \citealt{2007MNRAS.374.1290G}): (1) the emission from the central source scattered in the rarefied gas outside the dusty torus around the supermassive black hole (SMBH), (2) the emission from the gas in the galactic nucleus photoionized by the emission and/or shocks associated with the SMBH activity, and (3) the emission from the galaxy itself associated with active star formation.

Determining the nature of the observed excess emission at low energies requires refining the \ero\ response matrix, which is beyond the scope of this paper. Therefore, when fitting the spectra of the sources with such an excess, we included an additional soft component that was described by the thermal hot optically thin plasma emission spectrum (APEC, \citealt{apec}) in the spectral model. Thus, we used the following two models in XSPEC:
\[
TBabs(zTBabs(cflux~zpowerlaw))
\]
\[
TBabs(zTBabs(cflux~zpowerlaw)+apec)
\]

where TBabs is the absorption in the Galaxy from HI4PI data \citep{bekhti16}, zTBabs is the intrinsic absorption in the AGN frame, and cflux is the absorption-corrected flux of the power-law component in the 2--10~keV energy band.

When making a decision about the necessity of adding the soft component to the model, we used a likelihood ratio test: if $Cstat$ decreased by more than $6$ (corresponding to a statistical significance of more than 95\% for two degrees of freedom) when adding the soft component, then preference was given to the two-component model.

The X-ray spectrum fitting results are presented in Table~\ref{tab:xray_params}. The 90\% confidence intervals of the parameters are given. The spectra themselves are presented in Fig. \ref{fig:xray_plots}, with the \ero\ spectra having been rebinned for clarity. We will reiterate that when interpreting the spectral parameters given in Table~\ref{tab:xray_params}, it should be kept in mind that the final conclusion about the nature of the excess at low energies requires a further study and a refinement of the \ero\ response matrix, which is planned to be done in our succeeding papers. It is important to note that the power-law parameters (the slope and the hydrogen column density) do not change greatly when adding the soft component to the model.

\begin{table*}
 \caption{X-ray spectral parameters}
 \label{tab:xray_params}
 \renewcommand{\tabcolsep}{0.05cm}
 \renewcommand{\arraystretch}{1.7}
 \begin{tabular}[t]{lcccccccc}
 \toprule
 \multispan7\hfil TBabs zTBabs(ZPL),  TBabs (zTBabs ZPL + APEC) models \hfil \\
 \addlinespace[0.3em]
\art\ source & $N_{\rm H,MW}$ & $N_{\rm H}$ & $\Gamma$ & $F_{\rm PL}$ & $kT$  & $A_{\rm APEC}$ & dof & Cstat\\
\hline
SRGA\,J001439.6$+$183503   & $0.4$ & $109_{-57}^{+85}$      & $1.6_{-1.4}^{+1.8}$    & $5.15_{-1.81}^{+4.39}$ & $-$ & $-$ &  $19$ & $33.6$\\
 &  & $115_{-57}^{+88}$ & $1.7_{-1.4}^{+1.8}$ & $5.33_{-1.87}^{+2.95}$ & $0.77_{-0.22}^{+0.29}$ & $0.5_{-0.3}^{+0.4}~\times10^{-5}$ & $17$ & $21.8$\\
SRGA\,J002240.8$+$804348   & $1.4$ & $0.4_{-0.3}^{+0.3}$    & $1.90_{-0.15}^{+0.16}$ & $3.11_{-0.45}^{+0.51}$ & $-$ & $-$ &  $279$ & $265.2$\\
SRGA\,J010742.9$+$574419   & $3.2$ & $<3.1$                 & $1.9_{-0.4}^{+0.4}$    & $1.24_{-0.37}^{+0.49}$ & $-$ & $-$ &  $118$ & $123.1$\\
SRGA\,J021227.3$+$520953   & $1.5$ & $< 0.8$                & $2.04_{-0.14}^{+0.37}$ & $0.95\pm0.19$          & $-$ & $-$ &  $155$ & $146.5$\\
SRGA\,J025208.4$+$482955   & $1.8$ & $3.2_{-1.2}^{+1.4}$    & $1.7_{-0.3}^{+0.4}$    & $2.04_{-0.55}^{+0.69}$ & $-$ & $-$ &  $104$ & $110.5$\\
SRGA\,J045432.1$+$524003   & $3.4$ & $7.7_{-2.5}^{+2.9}$    & $1.5_{-0.3}^{+0.3}$    & $6.73_{-1.24}^{+1.44}$ & $-$ & $-$ &  $102$ & $102.1$\\
SRGA\,J051313.5$+$662747   & $0.9$ & $11_{-4}^{+5}$         & $1.5_{-0.5}^{+0.6}$    & $3.32_{-0.86}^{+1.05}$ & $-$ & $-$ & $60$  & $71.0$\\
& & $15_{-5}^{+7}$ & $1.9_{-0.6}^{+0.7}$ & $3.18_{-0.81}^{+1.01}$ & $0.24_{-0.19}^{+0.19}$ & $0.19_{-0.11}^{+78}~\times10^{-4}$ &  $58$ & $58.9$\\
SRGA\,J110945.8$+$800815   & $0.4$ & $1.8_{-1.7}^{+2.5}$    & $0.7_{-0.4}^{+0.5}$    & $1.48_{-0.55}^{+0.72}$ & $-$ & $-$ & $47$  & $50.3$\\
SRGA\,J161251.4$-$052100 & $1.0$ & $12_{-3}^{+4}$           & $1.9_{-0.5}^{+0.5}$    & $2.61_{-0.65}^{+0.82}$ & $-$ & $-$ & $80$  & $73.0$\\
SRGA\,J161943.7$-$132609 & $1.5$ & $<3.5$                   & $0.9_{-0.3}^{+0.5}$    & $2.49_{-0.79}^{+0.97}$ & $-$ & $-$ & $82$  & $87.0$\\
SRGA\,J182109.8$+$765819 & $0.5$ & $34_{-13}^{+16}$         & $1.1_{-0.6}^{+0.8}$    & $1.55_{-0.40}^{+0.46}$ & $-$ & $-$ & $58$  & $62.8$\\
SRGA\,J193707.6$+$660816 & $0.8$ & $0.32_{-0.13}^{+0.14}$   & $2.33_{-0.09}^{+0.10}$ & $1.24_{-0.12}^{+0.13}$ & $-$ & $-$ & $350$ & $379.8$\\
SRGA\,J200331.2$+$701332 & $1.0$ & $2.2_{-0.4}^{+0.4}$      & $2.00_{-0.14}^{+0.15}$ & $1.56_{-0.20}^{+0.23}$ & $-$ & $-$ & $314$ & $352.8$\\
SRGA\,J211149.5$+$722815 & $1.5$ & $8_{-4}^{+5}$            & $1.2_{-0.4}^{+0.5}$    & $1.34_{-0.33}^{+0.40}$ & $-$ & $-$ & $119$ & $112.1$\\
& & $14_{-4}^{+5}$ & $1.6_{-0.4}^{+0.5}$ & $1.23_{-0.30}^{+0.37}$ & $0.46_{-0.17}^{+0.25}$ & $0.7_{-0.4}^{+0.6}~\times10^{-5}$  & $117$ & $97.6$\\
\bottomrule
\end{tabular}
  \begin{flushleft}
 $N_{\rm H,MW}$ and $\nh$ are the gas column densities in the Galaxy and the object, respectively, in units of $10^{21}$~cm$^{-2}$; 
  $F_{\rm PL}$ is the absorption-corrected flux in the main power-law component in the observed 2--10~keV energy band, in units of  $10^{-12}$~erg~s$^{-1}$~cm$^{-2}$; $kT$ is the optically thin plasma temperature, in keV; $A_{\rm APEC}$ is the normalization of the plasma emission, in units of $10^{-14} (4 \pi)^{-1} [D_A(1+z)]^{-2} \int n_e n_H dV$, where $D_A$ is the angular diameter distance to the source (cm), $dV$ is a volume element (cm$^3$), $n_e$ and $n_H$ arethe number densities (cm$^{-3}$) of electrons and hydrogen nuclei, respectively.
  \end{flushleft}
\end{table*}

\subsection{Optical Spectra}

Standard criteria based on the emission line flux ratios \citep{oster, veron} were used to classify the Seyfert galaxies. The spectral continuum was fitted by a polynomial, while the emission lines were fitted by Gaussians. Thus, for each line we determined the central wavelength, the full width at half maximum $\fwhmmes$, the flux, and the equivalent width $EW$. The $\fwhm$ of the broad Balmer lines was corrected for the spectral resolution of the instrument: $\fwhm=\sqrt{\fwhmmes^2-\fwhmres^2}$, where $\fwhmres$ was determined for each dispersive element and each slit as the $FWHM$ of the lines in the calibration lamp spectrum.

The errors of the emission line parameters are given at 68\% confidence. The confidence level of the redshift was determined as the error of the mean narrow-line redshift. The measured $\fwhm$ of the narrow emission lines are consistent with the instrumental broadening and, therefore, the values of $\fwhm$ are not given for them. The confidence intervals for the line equivalent widths ($EW$) were obtained by the Monte Carlo method. Assuming that the flux errors obeyed a normal distribution, we selected $1000$ spectrum realizations. Then, for each of the realizations we estimated $EW$. The confidence intervals were estimated from the derived $EW$ distribution. To obtain an upper limit on the line flux, we fixed the center of the Gaussian and took its width to be equal to the instrumental broadening. The estimated line parameters for each of the sources are given below in Table \ref{tab:spec_lines}.

The redshifts of the objects were determined from the narrow emission lines and are given in the observatory frame. For the sources from the spectroscopic 6dF survey we used the redshifts from the same catalog. The results of the classification of sources and their redshift measurements are presented in Table ~\ref{tab:results_obs}.


\subsection{Results on Individual Objects}

\subsubsection{SRGA\,J001439.6+183503}

This X-ray source is present in the catalog of the \xmm\ slew survey \citep{xmmsl2, saxton2008} and the catalog of point sources detected by XRT onboard the \swift\ observatory \citep{evans2020}: the sources XMMSL2\,J001439.6+183450 and 2SXPS\,J001440.0+183455, respectively. There is the edge-on galaxy NGC52 in the \art\ and \ero\ position error circles (Fig.~\ref{fig:guid_images}). It is located at redshift $z=0.01817$ (according to the SIMBAD database) and has an infrared color $W1-W2=0.26$. The radio source NVSS\,J001440+183455 can also be associated with this object.

Weak narrow H$\alpha$ and {}[NII]$\lambda$6583 emission lines are seen in the optical spectrum (Fig. \ref{fig:spec0014_0107}, Table \ref{tab:spec_lines}). The Fraunhofer MgI and NaD absorption lines are also seen. The redshift was measured from the emission lines: $z = 0.01800\pm0.00007$.

The narrow-line flux ratio \n2ha=$0.43\pm0.10$ points to the presence of an active nucleus in the galaxy, according to the BPT diagram (see Fig.~\ref{chart:bpt}), while the absence of a broad component in the H$\alpha$ line suggests that this is a Seyfert 2 (Sy2) galaxy. In principle, the absence of the {}[OIII]$\lambda$5007 and H$\beta$ emission lines supposes that this is a LINER object, but this is highly unlikely, taking into account the object’s high X-ray luminosity ($\sim 3\times 10^{42}$~erg~s$^{-1}$ in the 4--12~keV energy band). The weakness of the emission lines in the optical spectrum probably stems from the fact that the active nucleus is observed through a thick layer of interstellar matter in the galaxy.

Our X-ray spectrum modeling (Fig.~\ref{fig:xray_plots}) shows the presence of substantial absorption in the source, $\nh> 5\times 10^{22}$ cm$^{-2}$, at 90\% confidence (Fig.~\ref{chart:xray}, Table~\ref{tab:xray_params}). This is consistent with the weakness of the emission lines and may also be related mainly to the thick layer of interstellar matter in the edge-on galaxy and not with the dusty torus around the supermassive black hole.

\subsubsection{SRGA\,J002240.8+804348} 

This X-ray source was discovered during the \rosat\ all-sky survey (RASS, \cite{2rxs}): 2RXS\,J002247.6+804418. There is the extended optical and infrared object WISEA\,J002243.69+804346.1 (Fig.~\ref{fig:guid_images}) with a color $W1-W2=0.61$ typical for AGNs in the \art\ and \ero\ position error circles.

Balmer emission lines, broad H$\alpha$ and H$\beta$, are observed in the galaxy’s spectrum (Fig. \ref{fig:spec0014_0107}, Table \ref{tab:spec_lines}). The forbidden {}[OIII]$\lambda$4959, {}[OIII]$\lambda$5007 lines are also present. The redshift of the source is $z=0.1147\pm0.0013$.

Against the background of the broad H$\alpha$ line it is impossible to distinguish its narrow component and the narrow {}[NII]$\lambda$6583 line. In the case of H$\beta$ we can set only an upper limit on the flux in the narrow component and, accordingly, a 2$\sigma$-lower limit on the flux ratio, \o3hb $> 0.6$. However, the presence of broad H$\alpha$ and H$\beta$ components allows us to say with confidence that this is a Seyfert 1 (Sy1) galaxy. In the X-ray spectrum there is evidence only for slight intrinsic
absorption ($\nh\lesssim 10^{21}$ cm$^{-2}$).

\subsubsection{SRGA\,J010742.9+574419}

This is a new X-ray source discovered in the first year of the \srg/\art sky survey. There is the extended optical and infrared object WISEA\,J010743.11+574417.7 (Fig.~\ref{fig:guid_images}) with a color $W1-W2=0.78$ typical for AGNs in the \art\ and \ero\ position error circles.

A broad H$\alpha$ line and narrow $H\beta$, {}[OIII]$\lambda$4959, {}[OIII]$\lambda$5007, H$\alpha$, and {}[NII]$\lambda$6583 lines are observed in the optical spectrum (Fig.~\ref{fig:spec0014_0107}, Table \ref{tab:spec_lines}). The redshift of the source is $z = 0.06992\pm 0.00030$.

The line flux ratios \o3hb = $0.50\pm0.12$ and \n2ha = $-0.70\pm0.08$, according to the BPT diagram (Fig. \ref{chart:bpt}), and the presence of a broad $H\alpha$ component with $FWHM > 2000$ km s$^{-1}$ with the absence of a broad $H\beta$ component allow the object to be classified as Sy1.9. On the BPT diagram the source falls into the region of galaxies with a composite spectrum most likely because we cannot reliably distinguish the narrow H$\alpha$ and {}[NII]$\lambda$6583 lines for it. No significant absorption was revealed in the X-ray spectrum.

\subsubsection{SRGA\,J021227.3+520953}

This X-ray source was discovered in RASS: 2RXS\,J021225.5+521004. There is the extended optical and infrared object 2MASS\,J02122646+5209533 (Fig.~\ref{fig:guid_images}) with a color $W1-W2=0.89$ typical for AGNs in the \art\ and \ero\ position error circles.

Balmer emission lines with broad H$\alpha$ and H$\beta$ components and narrow forbidden {}[OIII]$\lambda$4959, {}[OIII]$\lambda$5007, {}[NII]$\lambda$6548, and {}[NII]$\lambda$6583 lines are seen in the optical spectrum (Fig. \ref{fig:spec0212_0252}, Table \ref{tab:spec_lines}). The measured redshift is $z = 0.23810 \pm 0.00011$. The narrow-line flux ratios \n2ha=$-0.39\pm0.05$ and \o3hb = $0.87\pm0.11$ (Fig. \ref{chart:bpt}) and the presence of broad H$\alpha$ and H$\beta$ components allow the object to be classified as Sy1. No significant absorption was revealed in the X-ray spectrum.

\subsubsection{SRGA\,J025208.4+482955} 

This X-ray source was discovered in RASS: 2RXS\,J025208.8+482956. There is the extended optical and infrared object WISEA\,J025209.64+482959.4 (Fig.~\ref{fig:guid_images}) with a color $W1-W2=0.71$ typical for AGNs in the \art\ and \ero\ position error circles.

Emission lines are seen in the optical spectrum (Fig. \ref{fig:spec0212_0252}, Table \ref{tab:spec_lines}): broad H$\alpha$, narrow H$\alpha$ and H$\beta$, and forbidden {}[OIII]$\lambda$4959, {}[OIII]$\lambda$5007, {}[OI]$\lambda$6300, {}[NII]$\lambda$6548, {}[NII]$\lambda$6583, {}[SII]$\lambda$6716, and {}[SII]$\lambda$6730. The measured redshift is $z = 0.03366\pm0.00008$. The narrow-line flux ratios \n2ha=$-0.06\pm0.03$, \o3hb = $1.25\pm0.12$ (Fig. \ref{chart:bpt}) and the presence of a broad H$\alpha$ component allow the object to be classified as Sy1.9. Moderate absorption ($\nh \approx 3\times 10^{21}$ cm$^{-2}$) was revealed in the X-ray spectrum.

\subsubsection{SRGA\,J045432.1+524003}

This is a new X-ray source discovered in the \srg/\art\ survey. In the \art\ and \ero\ position error circles there is the galaxy LEDA\,16297 (Fig.~\ref{fig:guid_images}) with redshift $z=0.03123$ (SIMBAD) and color $W1-W2=0.39$ with which the radio source NVSS\,J045432+524009 can also be associated.

Emission lines are seen in the optical spectrum (Fig. \ref{fig:spec0454_1109}, Table \ref{tab:spec_lines}): broad and narrow H$\alpha$, narrow H$\beta$, and forbidden {}[OIII]$\lambda$4959, {}[OIII]$\lambda$5007, {}[OI]$\lambda$6300, {}[NII]$\lambda$6548, {}[NII]$\lambda$6583, {}[SII]$\lambda$6716 and {}[SII]$\lambda$6730. The measured redshift is $z = 0.03117 \pm 0.00012$. Based on the narrow-line flux ratios \n2ha=$0.255\pm0.020$, \o3hb = $1.19\pm0.13$ (Fig. \ref{chart:bpt}) and the presence of a broad H$\alpha$ component, the object can be classified as Sy1.9. Significant absorption ($\nh \sim 10^{22}$ cm$^{-2}$) was revealed in the X-ray spectrum.

\subsubsection{SRGA\,J051313.5+662747}

This X-ray source is present in the 2SXPS catalog: 2SXPS\,J051316.0+ 662750. In the \art\ and \ero\ position error circles there is the galaxy 2MASX\,J05131637+6627498 (Fig.~\ref{fig:guid_images}) at redshift $z=0.01491$ (SIBMAD) with a color $W1-W2=0.50$ with which the radio source NVSS\,J051316+662801 can also be associated.

Narrow Balmer H$\beta$ and H$\alpha$ emission lines and narrow forbidden {}[OIII]$\lambda$4959, {}[OIII]$\lambda$5007, {}[OI]$\lambda$6300, {}[NII]$\lambda$6548, {}[NII]$\lambda$6583, {}[SII]$\lambda$6716, and {}[SII]$\lambda$6730 lines are seen in the optical spectrum (Fig. \ref{fig:spec0454_1109}, Table \ref{tab:spec_lines}). The measured redshift is $z = 0.01479\pm0.00008$. The narrow-line flux ratios \n2ha=$-0.053\pm0.011$, \o3hb = $0.90\pm0.04$ (Fig. \ref{chart:bpt}) and the absence of broad H$\alpha$ and H$\beta$ components allow the object to be classified as Sy2. This is consistent with significant absorption ($\nh \sim 10^{22}$ cm$^{-2}$) in the X-ray spectrum.

\subsubsection{SRGA\,J110945.8+800815}

This is a new X-ray source discovered in the \srg/\art\ sky survey. There is the infrared and radio source WISEA\,J110943.77+800805.6 = NVSS\,J110944+800807 (Fig.~\ref{fig:guid_images}) with a color $W1-W2=0.76$ typical for AGNs in the \art\ and \ero\ position error circles. It should be noted that there is a star $\sim 15$ mag only 6\arcsec\ away from this object (located at a distance $\sim 1.5$ kpc from the Sun, Gaia DR3, \citealt{gaia22}). It lies at the boundary of the 98\% \ero\ position error circle, and the possibility that it makes some contribution to the X-ray flux measured by \art\ and \ero\ if it has an active corona or, for example, is a cataclysmic variable must not be ruled out.

Narrow H$\beta$ and H$\alpha$ emission lines and narrow forbidden {}[OII]$\lambda$3727, {}[OIII]$\lambda$4959, {}[OIII]$\lambda$5007, {}[OI]$\lambda$6300, {}[NII]$\lambda$6548, {}[NII]$\lambda$6583, {}[SII]$\lambda$6716, and {}[SII]$\lambda$6730 lines are seen in the optical spectrum (Fig. \ref{fig:spec0454_1109}, Table \ref{tab:spec_lines}). The measured redshift is $z=0.18879\pm0.00031$. The narrow-line flux ratios \n2ha=$0.00\pm0.05$, \o3hb = $1.02\pm0.14$ (Fig. \ref{chart:bpt}) and the absence of broad H$\alpha$ and H$\beta$ components allow the object to be classified as Sy2. In spite of this, no statistically significant intrinsic absorption is revealed in the X-ray spectrum, while the upper limit on the absorption column density is $\nh<4\times 10^{21}$~cm$^{-2}$ at 90\% confidence. At the same time, the power-law continuum is unusually hard for AGNs, with a slope $\Gamma=0.7^{+0.5}_{-0.4}$. This may suggest that the spectrum of this source actually has a more complex shape, which is impossible to ascertain due to the insufficient number of photons in the spectrum being analyzed.

\subsubsection{SRGA\,J161251.4$-$052100}

This X-ray source was discovered in RASS: 2RXS\,J161250.6$-$052118. In the \art\ and \ero\ position error circles there is the galaxy LEDA\,3097794 (Fig.~\ref{fig:guid_images}) with a redshift $z=0.03054$ (SIMBAD, based on the 6dF survey) and an infrared color $W1-W2=0.78$ typical for AGNs.

Narrow Balmer H$\alpha$ and H$\beta$ emission lines and narrow forbidden {}[OIII]$\lambda$4959, {}[OIII]$\lambda$5007, {}[NII]$\lambda$6548, and {}[NII]$\lambda$6583 lines are seen in the optical spectrum (Fig. \ref{fig:spec1612_1821}, Table \ref{tab:spec_lines}). The narrow-line flux ratios \n2ha=$0.26\pm0.14$, \o3hb $>0.8$ and the absence of broad H$\alpha$ and H$\beta$ components allow the object to be classified as Sy2 (Fig. \ref{chart:bpt}). This is consistent with significant absorption ($\nh \sim 10^{22}$ cm$^{-2}$) in the X-ray spectrum.

\subsubsection{SRGA\,J161943.7$-$132609}

This is a new X-ray source discovered in the \srg/\art\ sky survey. In the \art\ and \ero\ position error circles there is the galaxy 2MASX\,J16194407$-$1326166 (Fig.~\ref{fig:guid_images}) with a redshift $z=0.07891$ (SIMBAD, based on the 6dF survey) and an infrared color $W1-W2=0.81$ typical for AGNs.

A broad H$\alpha$ emission line and narrow forbidden {}[OIII]$\lambda$4959, {}[OIII]$\lambda$5007, {}[NII]$\lambda$6548, and {}[NII]$\lambda$6583 lines are seen in the optical spectrum (Fig. \ref{fig:spec1612_1821}, Table \ref{tab:spec_lines}). The narrow-line flux ratios \n2ha$>-0.7$, \o3hb$>0.7$ (Fig. \ref{chart:bpt}) and the presence of a broad H$\alpha$ component allow the object to be classified as Sy1.9. No significant absorption was revealed in the X-ray spectrum.

\subsubsection{SRGA\,J182109.8+765819}

This is a new X-ray source discovered in the \srg/\art sky survey. In the \art\ and \ero\ position error circles there is the galaxy LEDA\,2772547 (Fig.~\ref{fig:guid_images}) with a color $W1-W2=0.91$ typical for AGNs. The radio source VLASS1QLCIR\,J182111.52+765816.6. can also be associated with it.
 
A narrow H$\alpha$ emission line and narrow forbidden {}[OIII]$\lambda$4959, {}[OIII]$\lambda$5007, {}[OI]$\lambda$6300, {}[NII]$\lambda$6548, {}[NII]$\lambda$6583, {}[SII]$\lambda$6716, and {}[SII]$\lambda$6730 lines are seen in the optical spectrum (Fig. \ref{fig:spec1612_1821}, Table \ref{tab:spec_lines}). The Fraunhofer MgI and NaD, F absorption lines are also seen.

The redshift was measured from the emission lines, $z = 0.0631\pm0.0004$. The narrow-line flux ratios \n2ha=$0.14\pm0.04$, \o3hb$>0.9$ (Fig. \ref{chart:bpt}) and the absence of broad H$\alpha$ and H$\beta$ components allow the object to be classified as Sy2. Significant X-ray absorption ($\nh \approx 3 \times 10^{22}$ cm$^{-2}$) is observed.

\subsubsection{SRGA\,J193707.6+660816}

This X-ray source was discovered in RASS (2RXS\,J193708.1+660821). In the \art\ and \ero\ position error circles there is the optical and radio source 2MASS\,J19370820+6608213 = NVSS\,J193710+660830 (Fig.~\ref{fig:guid_images}) with an infrared color $W1-W2=0.64$ typical for AGNs.

Balmer emission lines are seen in the optical spectrum (Fig. \ref{fig:spec1937_2111}, Table \ref{tab:spec_lines}): broadH$\delta$, broad H$\gamma$, broad and narrow H$\beta$, broad and narrow H$\alpha$. Narrow forbidden {}[OIII]$\lambda$4959, {}[OIII]$\lambda$5007, {}[NII]$\lambda$6548, {}[NII]$\lambda$6583, {}[SII]$\lambda$6716, and {}[SII]$\lambda$6730 lines are also observed.

The measured redshift is $z = 0.07136 \pm 0.00012$. The narrow-line flux ratios \n2ha=$-0.48\pm0.04$, \o3hb = $1.01\pm0.09$ (Fig. \ref{chart:bpt}) and the presence of broad H$\alpha$, H$\beta$, H$\gamma$, and H$\delta$ components with a typical line FWHM$\approx2000$ km s$^{-1}$ allow the object to be classified as a narrow-line Seyfert 1 (NLSy1) galaxy. There may be slight absorption ($\nh\approx 3 \times 10^{20}$ cm$^{-2}$) in the X-ray spectrum.

\subsubsection{SRGA\,J200331.2+701332}

This X-ray source was discovered in RASS: 2RXS\,J200332.1+701331. It is also known as a hard X-ray source, SWIFT\,J2003.4+7023 \citep{oh2018}. In \art\ and \ero\ position error circles there is the optical object 2MASS\,J20033397+7013369 (Fig.~\ref{fig:guid_images}) with a color $W1-W2=0.89$ typical for AGNs.

Broad H$\beta$ and H$\alpha$ emission lines and narrow forbidden {}[OIII]$\lambda$4959, {}[OIII]$\lambda$5007, {}[NII]$\lambda$6548 and {}[NII]$\lambda$6583 lines are observed in the optical spectrum (Fig. \ref{fig:spec1937_2111}, Table \ref{tab:spec_lines}). The measured redshift is $z = 0.09759\pm0.00002$. The narrow-line flux ratios \n2ha$>0.23 $, \o3hb $>0.8$ (Fig. \ref{chart:bpt}) and the presence of broad H$\alpha$ and H$\beta$ components allow the object to be classified as Sy1. Slight absorption ($\nh\approx 2 \times 10^{21}$ cm$^{-2}$) was revealed in the X-ray spectrum.

\subsubsection{SRGA\,J211149.5+722815}

This is a new X-ray source discovered in the \srg/\art\ sky survey. In the \art\ and \ero\ position error circles there is the optical and radio source WISEA\,J211151.78+722816.4 = NVSS\,J211152+722819 (Fig.~\ref{fig:guid_images}) with an infrared color $W1-W2=1.08$ typical for AGNs.

A narrow H$\alpha$ emission line and narrow forbidden {}[OIII]$\lambda$4959, {}[OIII]$\lambda$5007, {}[NII]$\lambda$6548, and {}[NII]$\lambda$6583 lines are observed in the optical spectrum (Fig. \ref{fig:spec1937_2111}, Table \ref{tab:spec_lines}). The measured redshift is $0.10611\pm0.00011$. The narrow-line flux ratios \n2ha=$0.20\pm0.07$, \o3hb $>0.9$ (Fig. \ref{chart:bpt}) and the absence of broad H$\alpha$ and H$\beta$ components allow the object to be classified as Sy2. Significant absorption ($\nh \sim 10^{22}$ cm$^{-2}$) is present in the X-ray spectrum.

\section{PROPERTIES OF THE AGN SAMPLE}

\begin{table*}
  \caption{
  Properties of the AGNs whose spectra were obtained as a result of the \azt\ observations and from the archival \emph{6dF} spectra
  } 
  \label{tab:results_obs}
  \vskip 2mm
  \renewcommand{\arraystretch}{1.7}
  \renewcommand{\tabcolsep}{0.35cm}
  \centering
  \footnotesize
  \begin{tabular}{rlccc}
    \noalign{\doubleline}     
   № & Object & Optical type & $z^1$ & $\log\lx^2$ \\
  \hline
 1& SRGA\,J001439.6$+$183503 &     Sy2       & $0.01800 \pm 0.00007$ & $42.58_{-0.19}^{+0.27}$ \\
 2& SRGA\,J002240.8$+$804348 &     Sy1       & $0.11470  \pm 0.00130$& $44.03_{-0.07}^{+0.07}$ \\
 3& SRGA\,J010742.9$+$574419 &     Sy1.9     & $0.06992 \pm 0.00030$ & $43.17_{-0.16}^{+0.14}$ \\
 4& SRGA\,J021227.3$+$520953 &     Sy1       & $0.23810 \pm 0.00011$ & $44.21_{-0.09}^{+0.08}$ \\
 5& SRGA\,J025208.4$+$482955 &     Sy1.9     & $0.03366 \pm 0.00008$ & $42.73_{-0.14}^{+0.13}$ \\
 6& SRGA\,J045432.1$+$524003 &     Sy1.9     & $0.03117 \pm 0.00012$ & $43.18_{-0.09}^{+0.08}$ \\
 7& SRGA\,J051313.5$+$662747 &     Sy2       & $0.01479 \pm 0.00008$ & $42.21_{-0.13}^{+0.12}$ \\
 8& SRGA\,J110945.8$+$800815 &     Sy2       & $0.18879 \pm 0.00031$ & $44.18_{-0.20}^{+0.17}$ \\  
 9& SRGA\,J161251.4$-$052100$^+$ & Sy2       & $0.03055$             & $42.75_{-0.13}^{+0.12}$ \\
10& SRGA\,J161943.7$-$132609$^+$ & Sy1.9     & $0.07891$             & $43.58_{-0.17}^{+0.14}$ \\
11& SRGA\,J182109.8$+$765819 &     Sy2       & $0.06310 \pm 0.00040$ & $43.17_{-0.13}^{+0.11}$ \\
12& SRGA\,J193707.6$+$660816 &     NLSy1     & $0.07136 \pm 0.00012$ & $43.19_{-0.05}^{+0.04}$ \\
13& SRGA\,J200331.2$+$701332 &     Sy1       & $0.09759 \pm 0.00002$ & $43.58_{-0.06}^{+0.06}$ \\
14& SRGA\,J211149.5$+$722815 &     Sy2       & $0.10611 \pm 0.00011$ & $43.59_{-0.12}^{+0.11}$ \\
    \noalign{\vskip 3pt\hrule\vskip 5pt}
  \end{tabular}
  \begin{flushleft}
  $^1$ The redshifts were measured from emission lines.
  
  $^2$ The absorption-corrected luminosity in the 2--10~keV energy band in erg s$^{-1}$.
  
  $^+$ The redshifts were taken from the \emph{6dF}catalog. 
  
The error corresponds to the 68\% confidence interval for the redshift and 90\% for the luminosity without including the error in $z$.
  \end{flushleft}
 \end{table*}

 \begin{table*}
 \caption{
 The masses, bolometric luminosities, and Eddington ratios for the central black holes in Sy1 and NLSy1 galaxies
 }
 \label{tab:mbh_edd}
\centering
\begin{tabular}{lccc}
\hline
\addlinespace[0.3em]
Object & BH mass, $10^8 M_\odot$ & $L_{\rm bol}$, $10^{44}$ erg s$^{-1}$ & $\lambda_{\rm Edd}$\\
\addlinespace[0.3em]
\hline
\addlinespace[0.3em]
SRGAJ002240.8+804348 &  $2.6 \pm 0.6$  & $12  \pm 2$   & $0.034 \pm 0.009 $\\
SRGAJ021227.3+520953 &  $1.4 \pm 0.3$  & $18   \pm 4 $  & $0.10  \pm 0.03  $ \\
SRGAJ193707.6+660816 & $0.12 \pm 0.02$ & $1.7 \pm 0.2$ & $0.11  \pm 0.02 $ \\
SRGAJ200331.2+701332 &  $2.2 \pm 0.5$  & $4.1 \pm 0.6$ & $0.014 \pm 0.004 $\\
\hline
\end{tabular}
\begin{flushleft}
$L_{\rm bol}$ is the bolometric luminosity derived for a fixed bolometric correction $L_{\rm bol}/\lx=11$; $\lambda_{\rm Edd}$ is the bolometric-to-Eddington luminosity ratio. The errors correspond to the 68\% confidence interval.
\end{flushleft}
\end{table*}

Table \ref{tab:results_obs} presents basic characteristics of the identified AGNs: the optical type, the redshift, and the X-ray luminosity $\lx$. The latter was calculated using the single-component X-ray spectrum model from Table \ref{tab:xray_params} in the 2--10~keV energy band\footnote{Taking into account the low redshifts of the objects, we do not make the $k$-correction.} (in the ob- served frame) and was corrected for Galactic and intrinsic absorption.

\begin{figure}
  \centering
    \includegraphics[width=1\columnwidth]{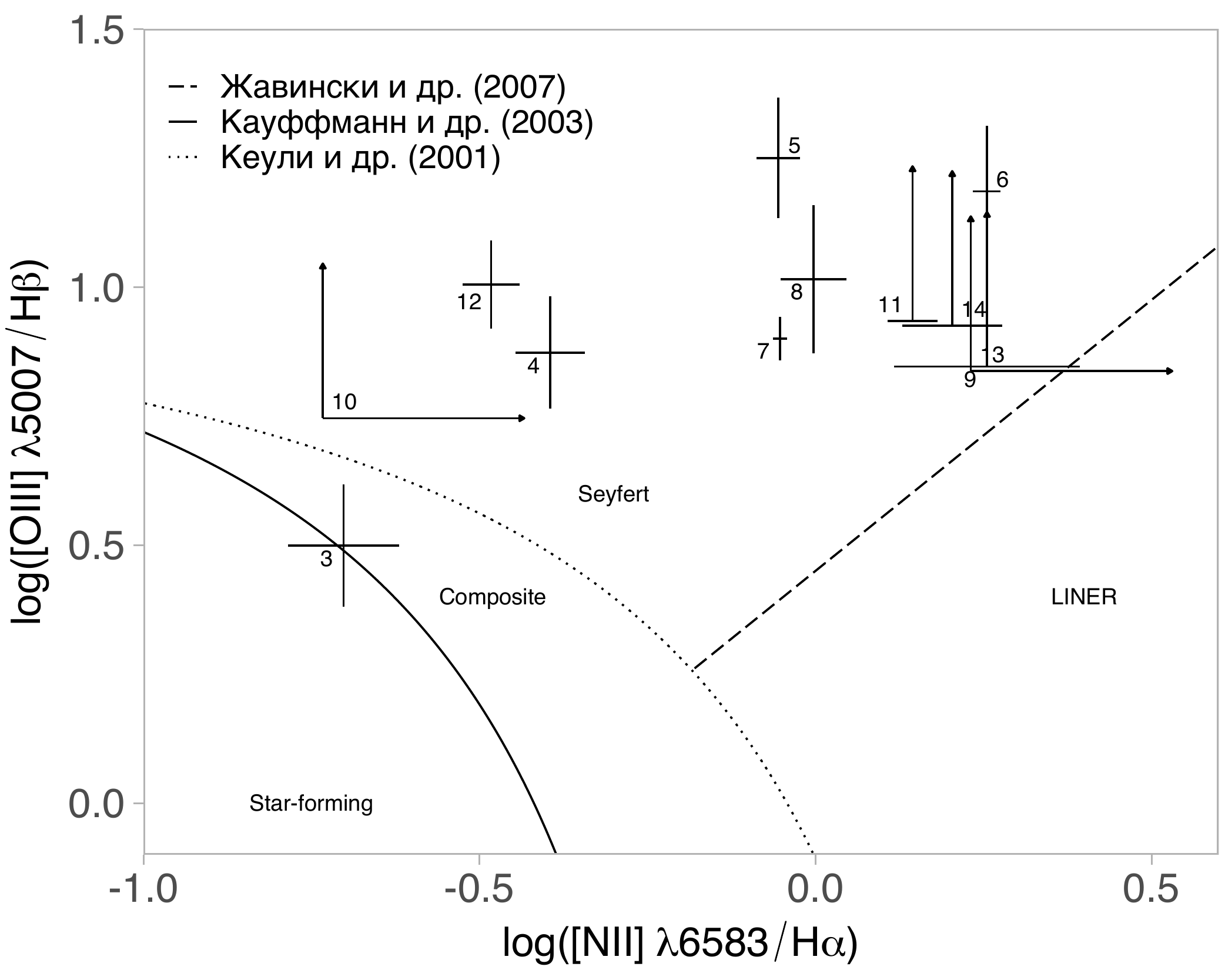}
    \caption{
     Positions of the AGNs being studied on the BPT diagram \citep{bpt}. The 1$\sigma$ confidence intervals for the flux ratios are presented on the graph. The arrows indicate the lower 2$\sigma$ limits. 
     The demarcation lines between different classes of galaxies were taken from \citep{kauff03} (solid line), \citep{kewly01} (dotted line), and \citep{scha07} (dashed line). 
     The sources are marked by their numbers specified in Table~\ref{tab:list_src}. 
    The following sources did not fell on the diagram: SRGA\,J001439.6+183503 (1) due to the absence of the {}[OIII]$\lambda$5007 and H$\beta$ lines and SRGA\,J002240.8+804348 (2) due to the impossibility to reliably 
    identify the narrow H$\alpha$ and {}[NII]$\lambda$6583 lines in the 
    spectrum.
    }
    \label{chart:bpt}
\end{figure}

The X-ray luminosities of the objects vary in the range from $\sim 10^{42}$ to $\sim 10^{44}$ erg s$^{-1}$, typical for AGNs at the present epoch. According to the narrow-line flux ratios, \n2ha and \o3hb, all sources fall into the region of Seyfert galaxies on the BPT diagram (Fig. \ref{chart:bpt}), except for SRGA\,J001439.6+183503, SRGA\,J002240.8+804348, and SRGA\,J010742.9+574419. However, the high X-ray luminosity, the presence of broad hydrogen line components in SRGA\,J002240.8+804348 and SRGA\,J010742.9+574419, and the ratio \n2ha$\approx 0.4$ in SRGA\,J001439.6+183503 point to the presence of active nuclei in these galaxies.

\begin{figure}
  \centering
    \includegraphics[width=1\columnwidth]{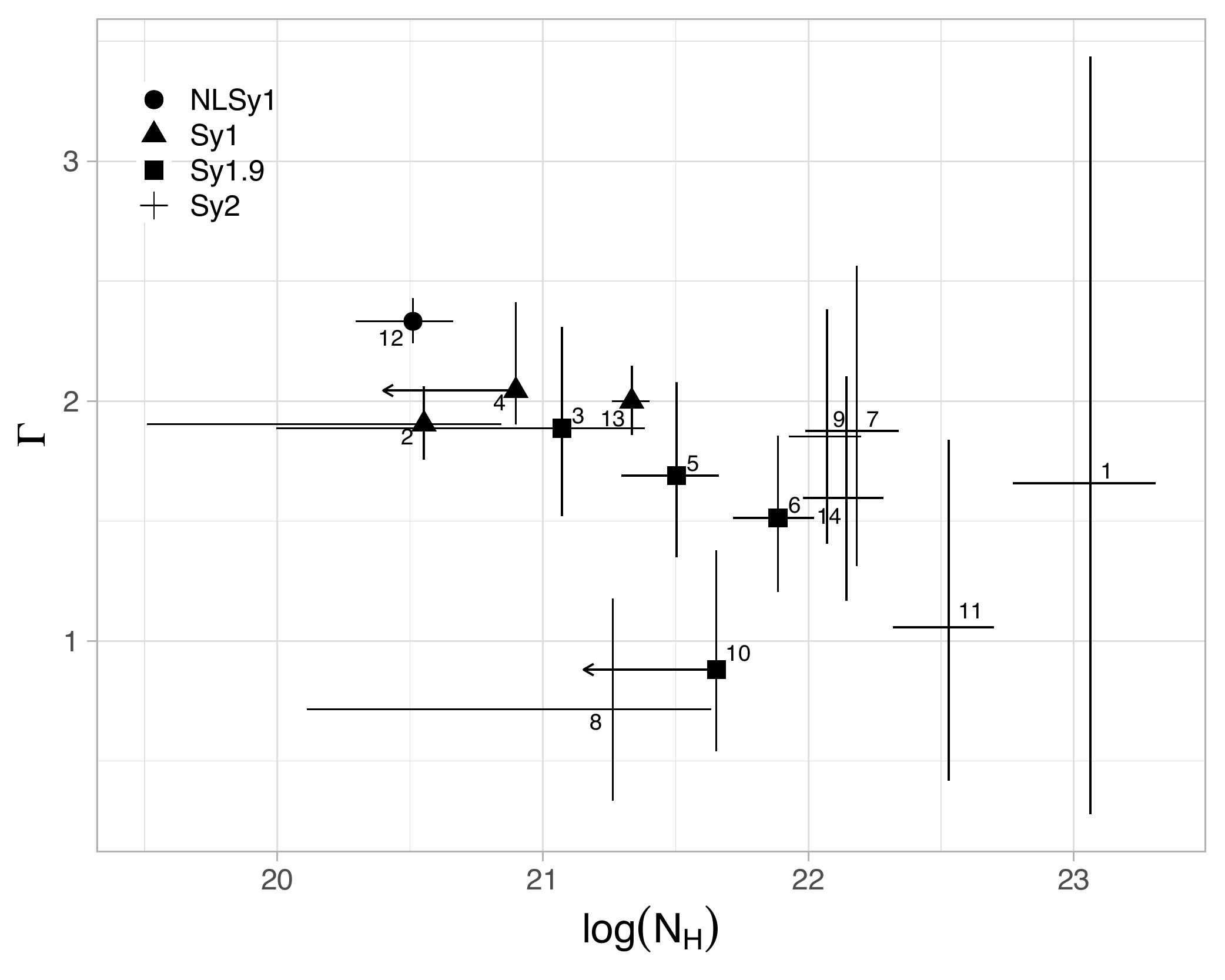}
    \caption{
    Slope of the X-ray power-law continuum versus intrinsic column density for the 14 AGNs (from the best-fit models) investigated using the \art\ and \ero\ data (see Table~\ref{tab:xray_params}). The dots of different shapes indicate the optical types of the sources. The errors and the upper limits correspond to the 90\% confidence intervals. The sources are indicated by their numbers from Table~\ref{tab:list_src}.
  }
  \label{chart:xray}
\end{figure}

In Fig. \ref{chart:xray} the slope of the power-law continuum $\Gamma$ is plotted against the intrinsic absorption column density $\nh$ for the objects being studied. Almost all of the slopes are close, within the error limits, to the canonical slope for AGNs, $\Gamma \approx 1.8$. The slope is considerably larger for only one narrow-line Seyfert 1 galaxy in the sample, SRGA\,J193707.6+660816: $\Gamma=2.33\pm 0.10$, typical for AGNs of this type (see, e.g., \citealt{1997MNRAS.285L..25B, 1999ApJS..125..317L}). Significant intrinsic absorption was revealed only in Seyfert 2 galaxies (Sy2 and Sy1.9).

For four Seyfert 1 galaxies, including NLSy1 SRGA\,J193707.6+660816, we can estimate the masses of the central black holes from the luminosity and the width of the broad H$\alpha$ emission line based on the well-known empirical relation (see Eq. (6) in \citealt{green2005}) using the flux and the width of this line from Table \ref{tab:spec_lines}\footnote{We do not make such estimates for Sy1.9 objects, since the H$\alpha$ emission for them can be subject to significant intrinsic absorption.}. In addition, we can estimate the bolometric luminosities of these objects. For this purpose, we took the bolometric correction for the 2--10~keV energy band $L_{\rm bol}/\lx=11$ from \cite{sazonov2012}, which was obtained for a representative sample of Seyfert galaxies in the nearby Universe. It should be kept in mind that this correction has an uncertainty $\sim 2$ that we ignore.

Table~\ref{tab:mbh_edd} gives the derived black hole masses, bolometric luminosities, and bolometric-to-Eddington luminosity ratios ($\lambda_{\rm Edd}$). The latter quantity characterizes the accretion regime. The derived $\lambda_{\rm Edd}$ vary from $\sim$1 to $\sim$10\%, which, on the whole, is typical for Seyfert galaxies (see, e.g., \citealt{khorunzhev2012}).

\section{CONCLUSIONS}

Using the observations carried out at the \azt\ telescope and the archival spectroscopic data from the 6dF survey, we managed to identify 14 new AGNs among the X-ray sources detected in the 4--12~keV energy band during the first five \srg/\art\ all-sky surveys. All sources are also detected with confidence by \ero\ in the 0.2--8.0~keV energy band. All objects turned out to be nearby ($z = 0.015-0.238$) Seyfert galaxies (one NLSy1, three Sy1, four Sy1.9, and six Sy2).

For all objects we constructed broadband (0.2--12~keV) X-ray spectra based on data from the \art\ and \ero\ telescopes onboard the \srg\ observatory. In four objects the intrinsic absorption exceeds $\nh> 10^{22}$~cm$^{-2}$ at 90\% confidence, and one of them (SRGA\,J001439.6+183503) is probably heavily obscured ($\nh> 5\times 10^{22}$~cm$^{-2}$ with 90\% confidence). Interestingly, in the latter case the absorption can be mainly associated not with the dusty torus around the central supermassive black hole, but with the great thickness of the interstellar medium of an edge-on galaxy.

This paper continues our series of publications on the optical identification of X-ray sources detected during the \srg/\art\ all sky survey. The result obtained will help to obtain a large ($\sim$2000 objects), statistically complete sample of AGNs selected by their emission in the hard 4--12~keV X-ray energy band on completion of the planned eight sky surveys.

\section*{Acknowledgements}

This work was supported by RSF grant no. 19-12-00396. The measurements with the \azt\ telescope were supported by the Ministry of Education and Science of Russia and were obtained using
the equipment of the Angara sharing center\footnote{http://ckp-rf.ru/ckp/3056/}. In this study we used observational data from the \art\ and \ero\ telescopes onboard the \srg\ observatory. The \srg\ observatory was built by Roskosmos in the interests of the Russian Academy of Sciences represented by its Space Research Institute (IKI) within the framework of the Russian Federal Space Program, with the participation of the Deutsches Zentrum fur Luft- und Raumfahrt (\textit{DLR}). The \srg\ spacecraft was designed, built, launched, and is operated by the Lavochkin Association and its subcontractors. The science data are downlinked via the Deep Space Network Antennae in Bear Lakes, Ussuriysk, and Baykonur, funded by Roskos- mos. The \ero\ X-ray telescope was built by a consortium of German Institutes led by MPE, and supported by DLR. The \ero\ data used in this work were processed using the \textit{eSASS} software developed by the German \ero\ consortium and the proprietary data reduction and analysis software developed by the Russian \ero\ Consortium.

\bibliographystyle{mnras}
\bibliography{main}

\section{APPENDIX}

\begin{figure*}
  \centering
  
  \includegraphics[width=0.44\columnwidth]{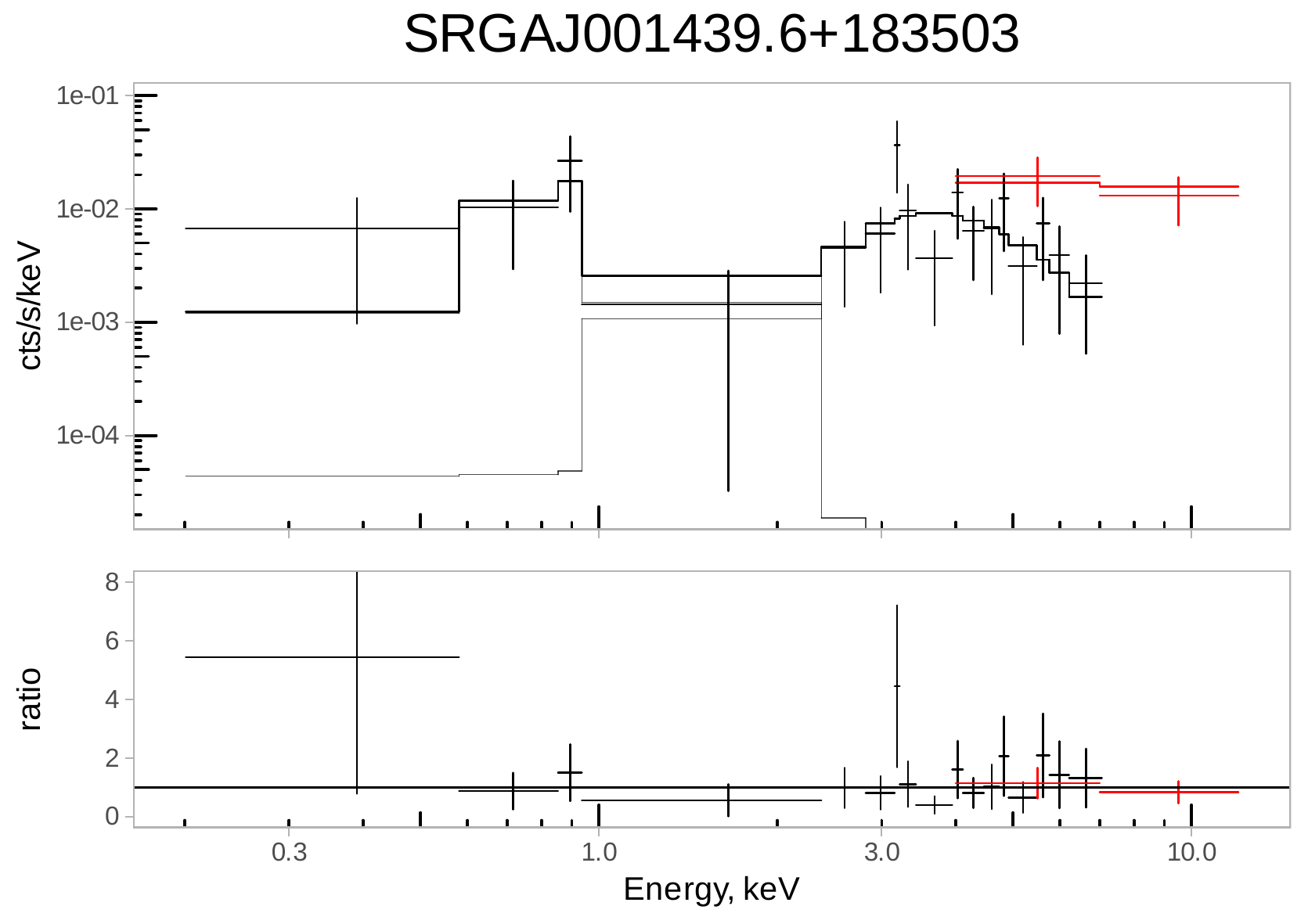}
  \includegraphics[width=0.44\columnwidth]{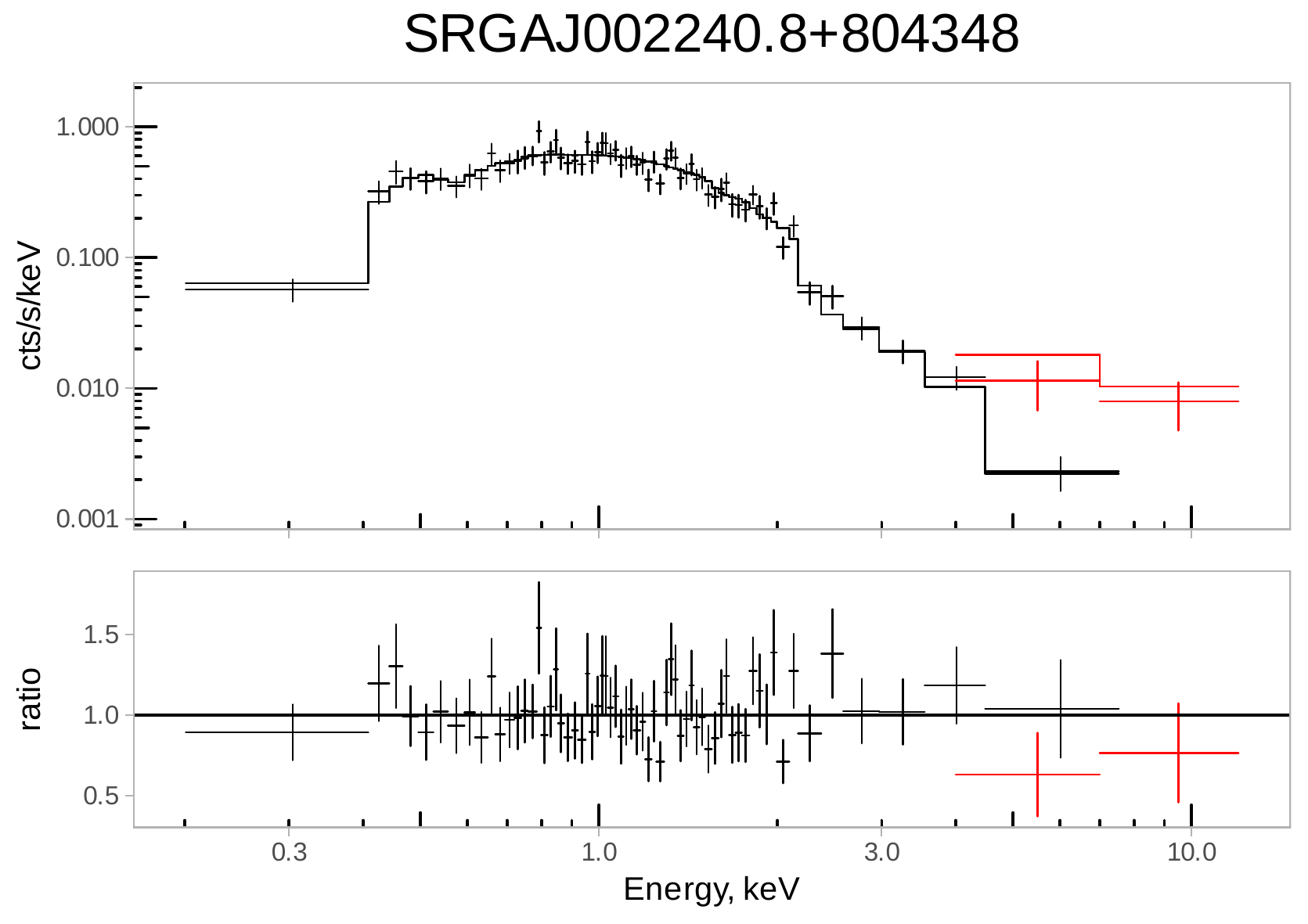}
  \includegraphics[width=0.44\columnwidth]{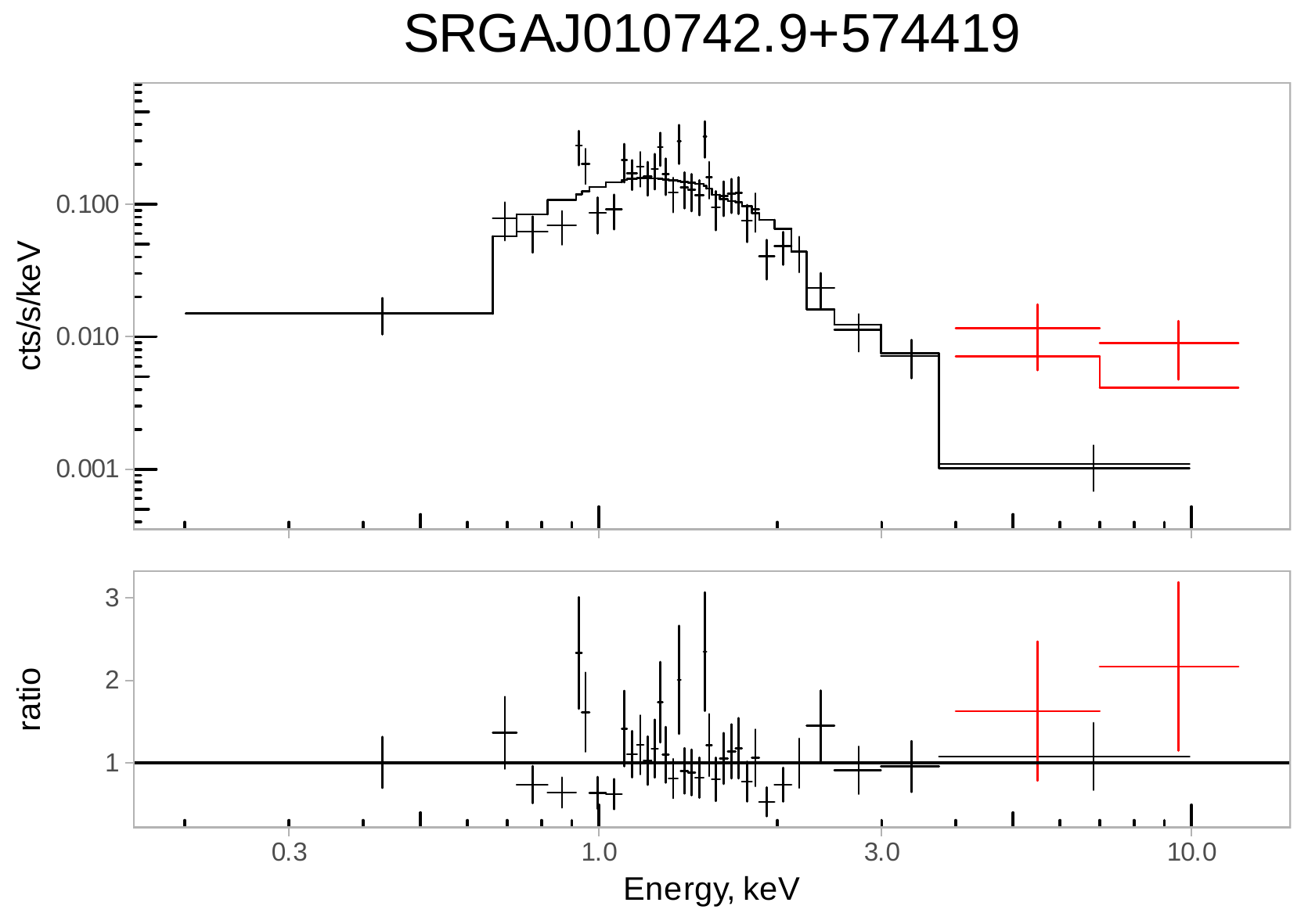}
  \includegraphics[width=0.44\columnwidth]{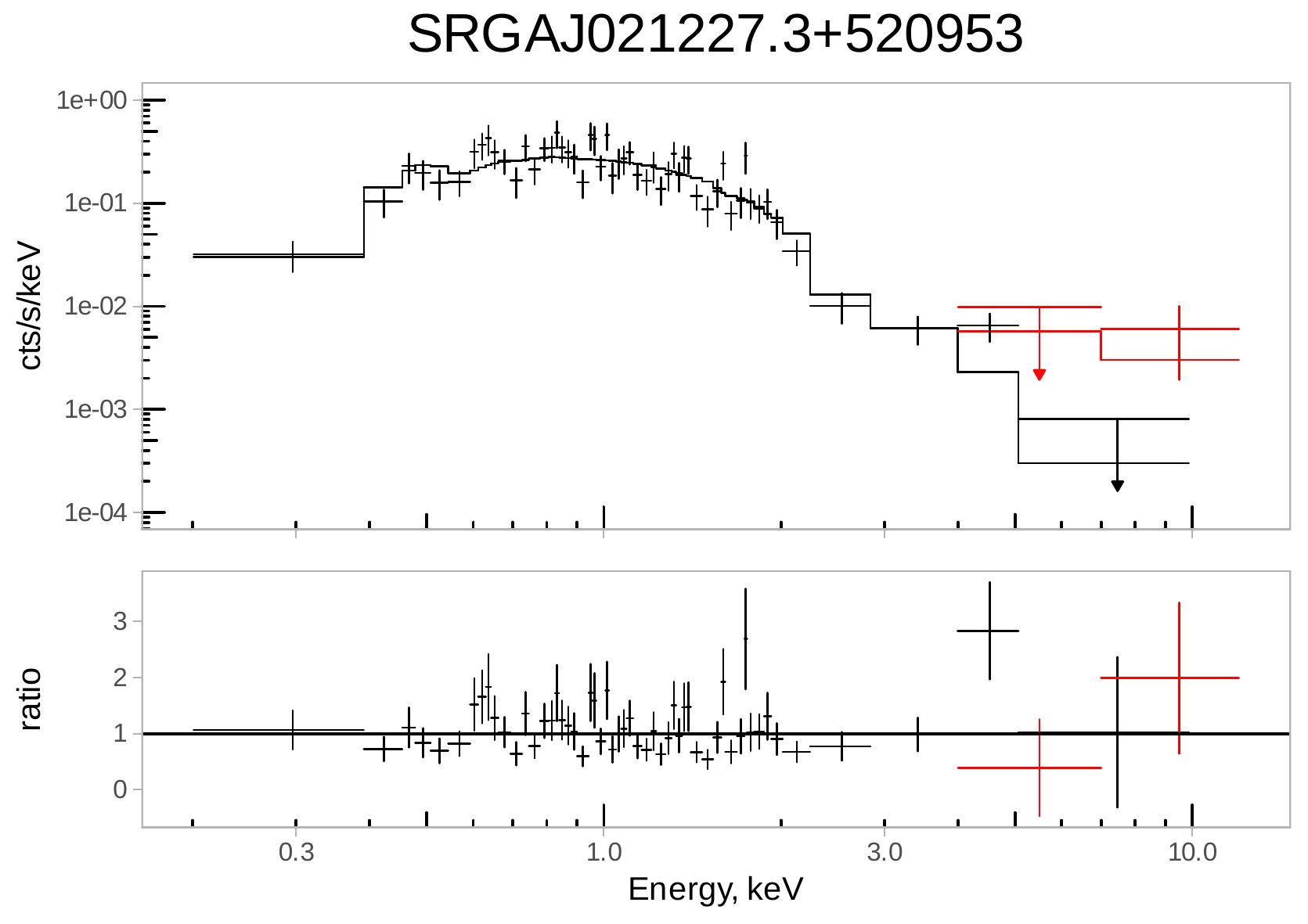}
  \includegraphics[width=0.44\columnwidth]{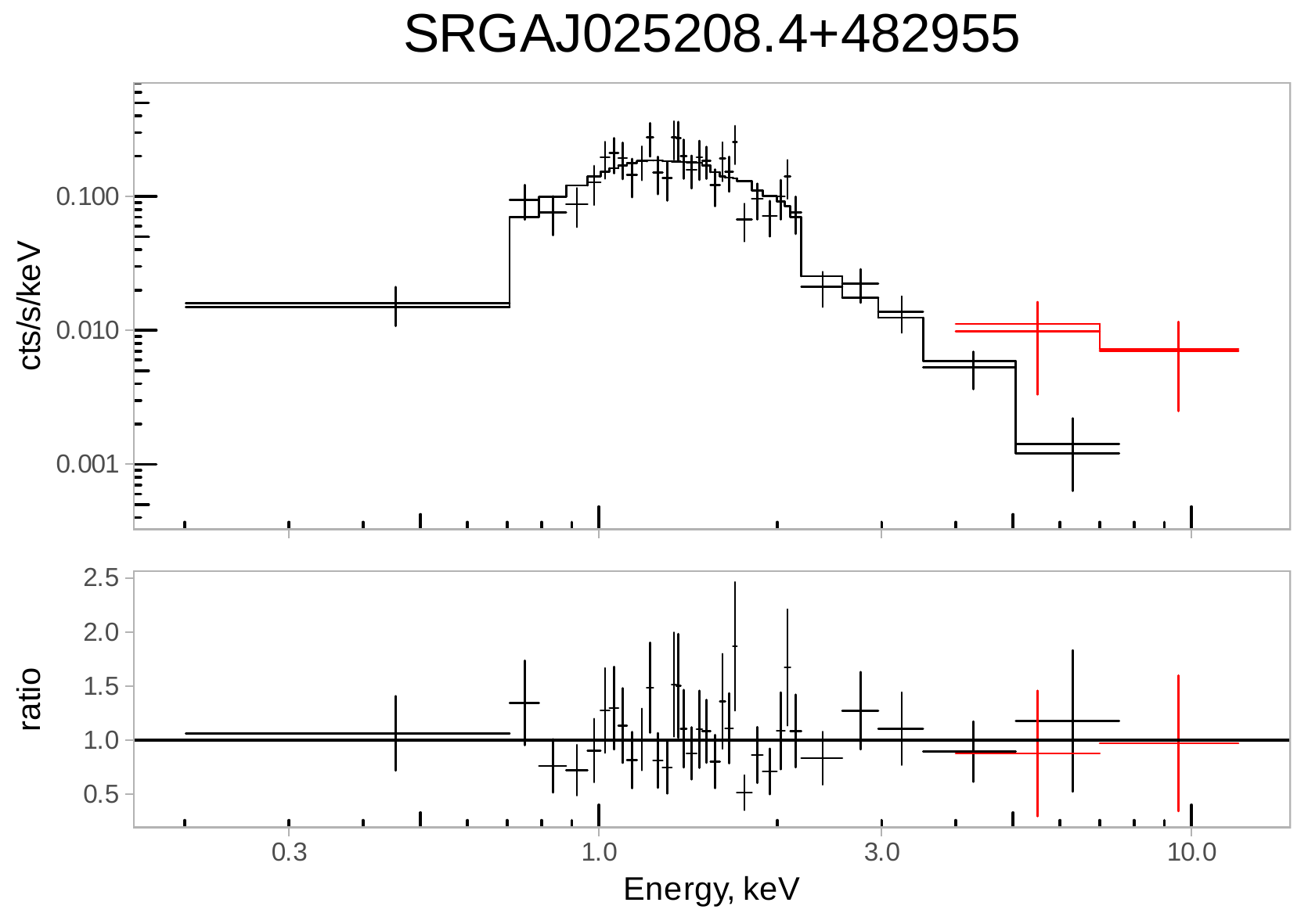}
  \includegraphics[width=0.44\columnwidth]{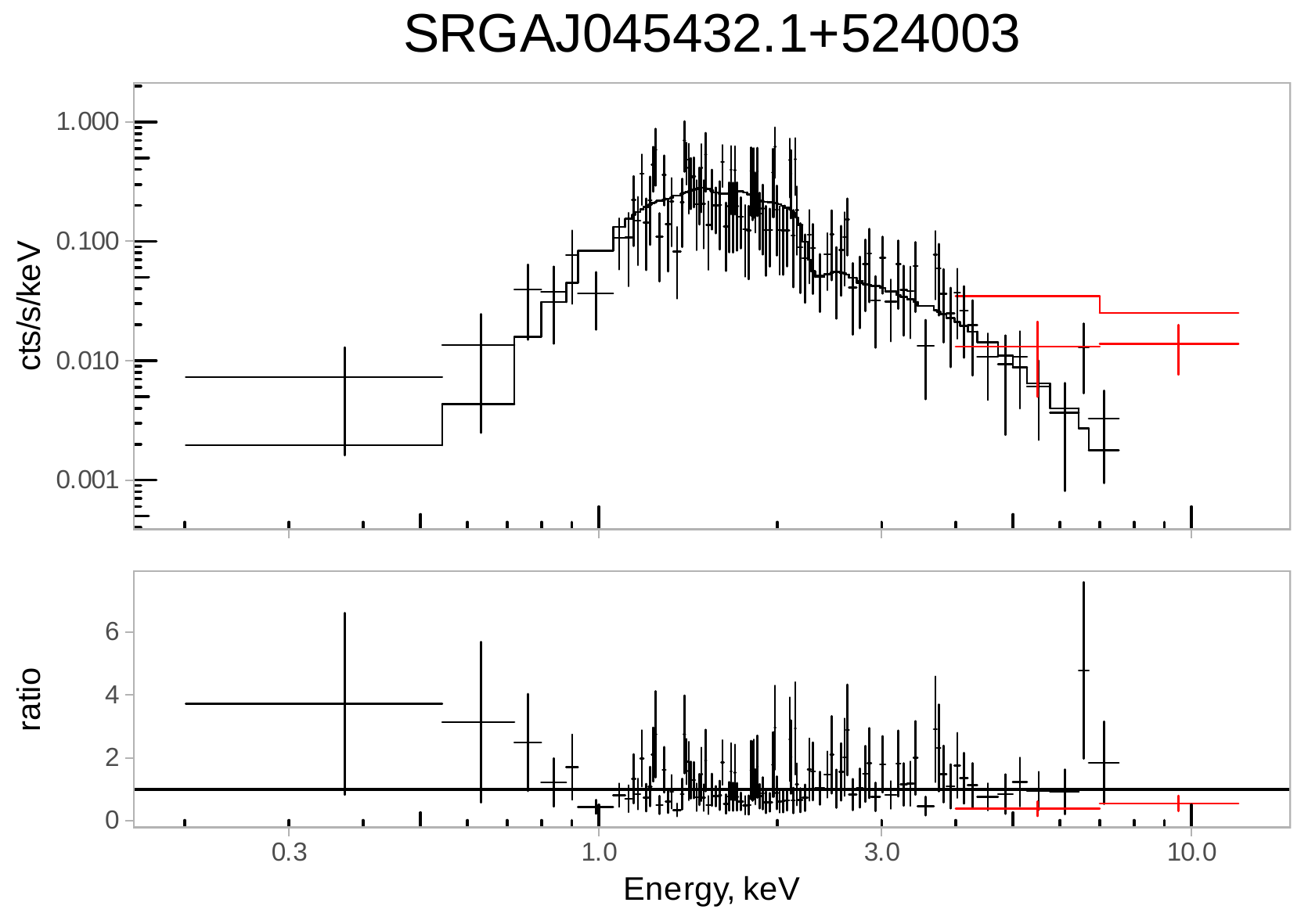}
  \includegraphics[width=0.44\columnwidth]{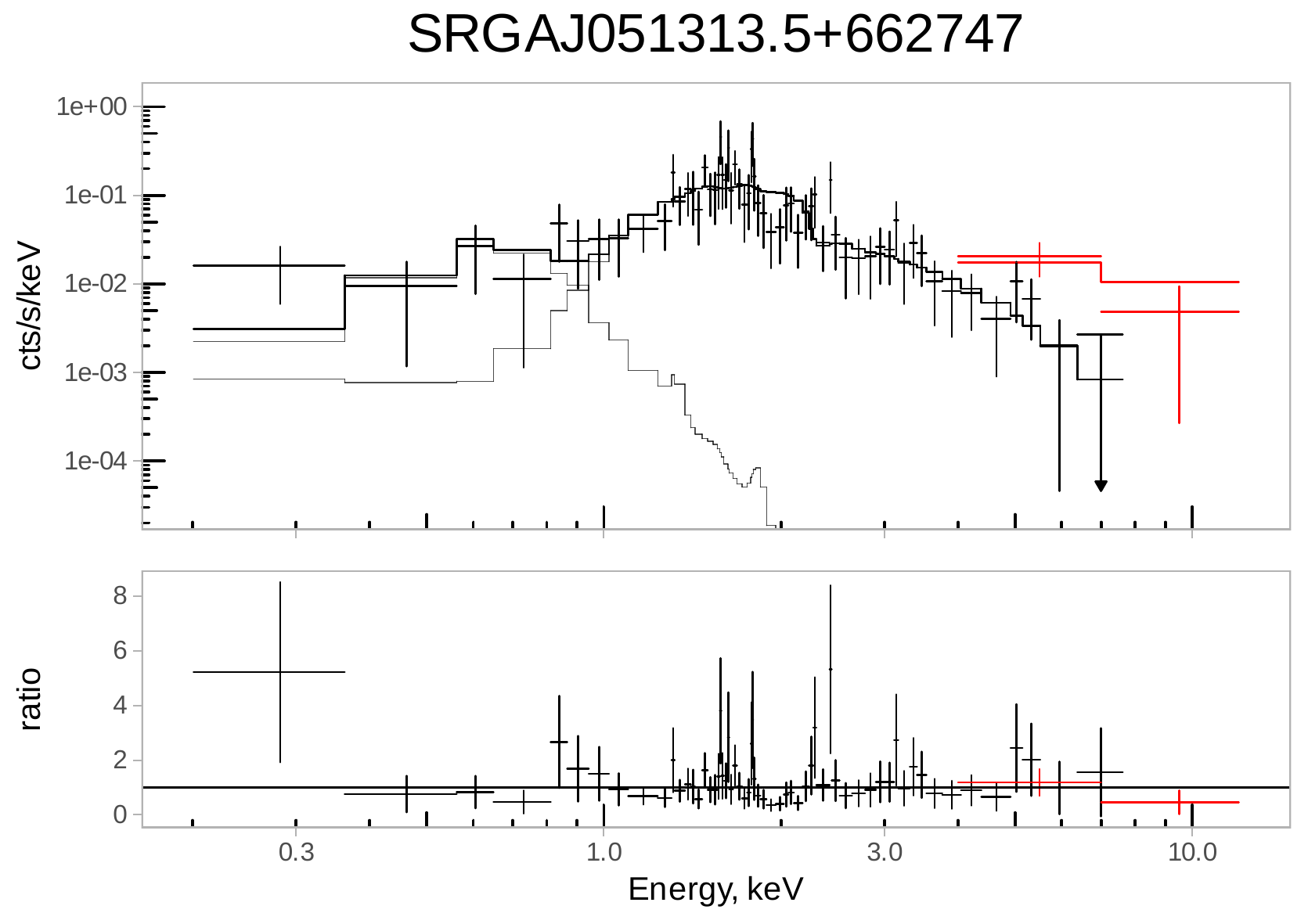}
  \includegraphics[width=0.44\columnwidth]{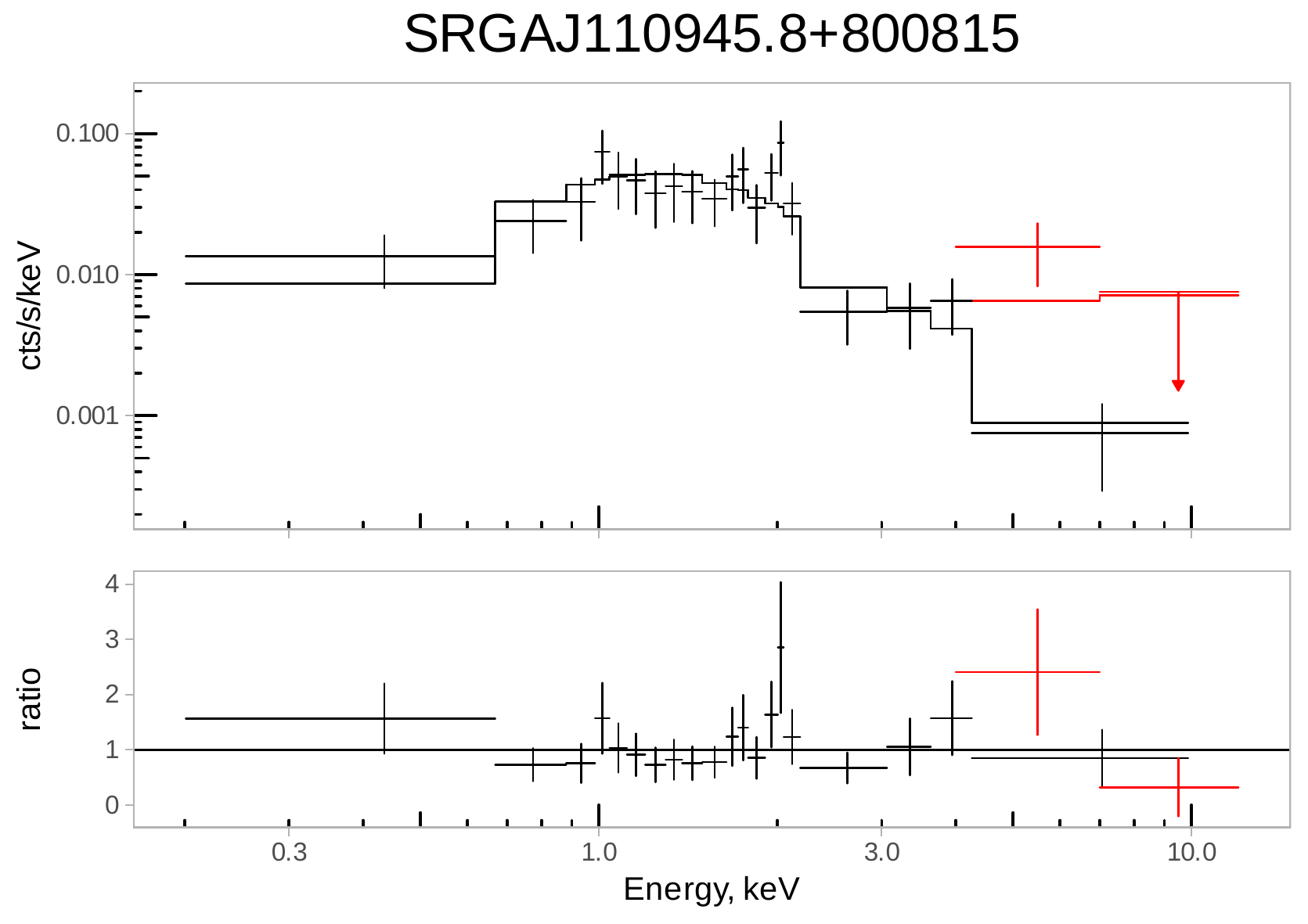}
  
  \caption{
  X-ray spectra from the \ero\ (black) and \art\ (red) data and the best-fit models (see Table \ref{tab:xray_params}). The soft component is additionally highlighted, if required. The arrows indicate the 2$\sigma$ upper limits. The ratio of the measurements to the model is shown on the lower panels.
  }
  \label{fig:xray_plots}
\end{figure*}

\addtocounter{figure}{-1}

\begin{figure*}
  \centering
  \includegraphics[width=0.45\columnwidth]{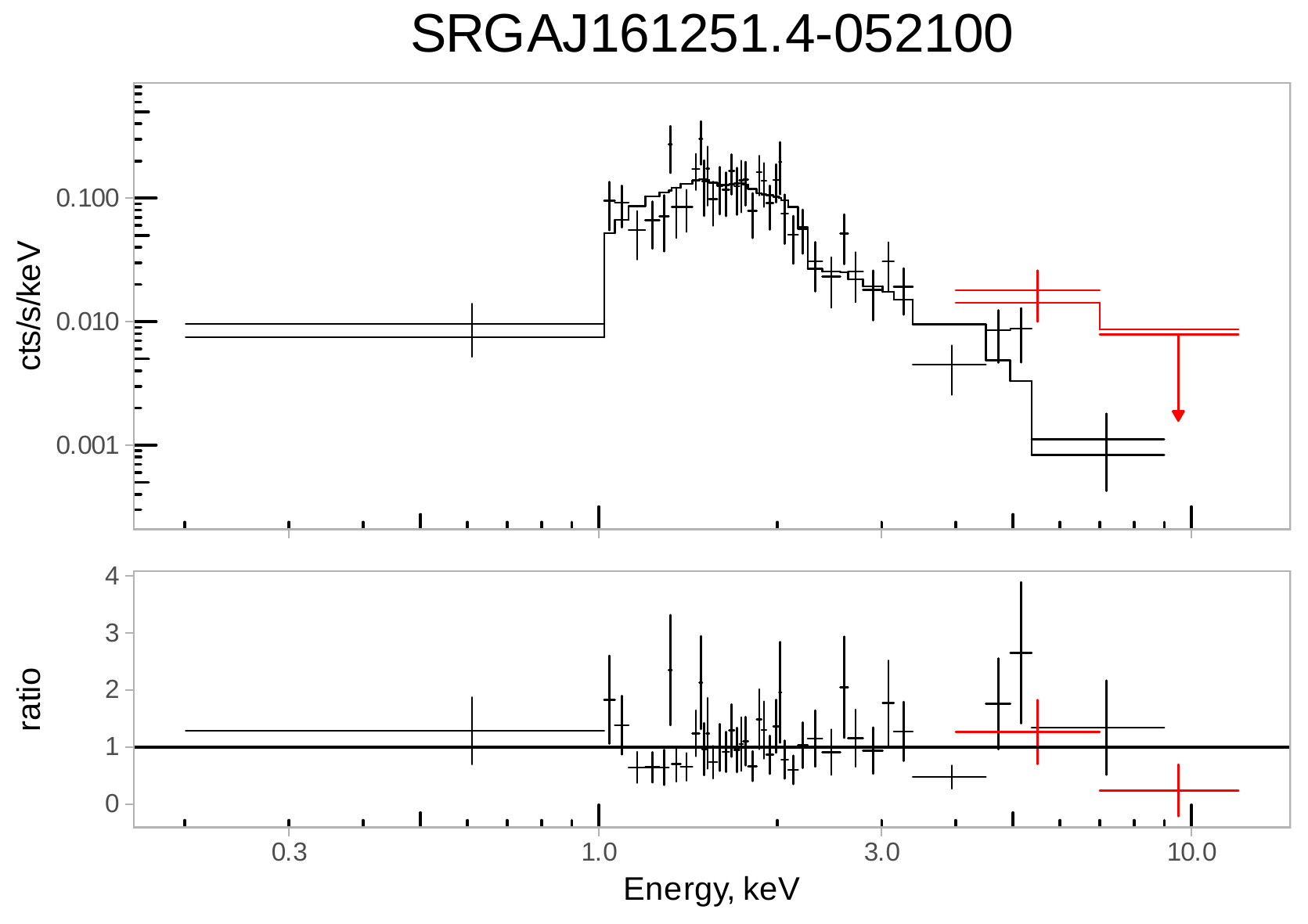}
  \includegraphics[width=0.45\columnwidth]{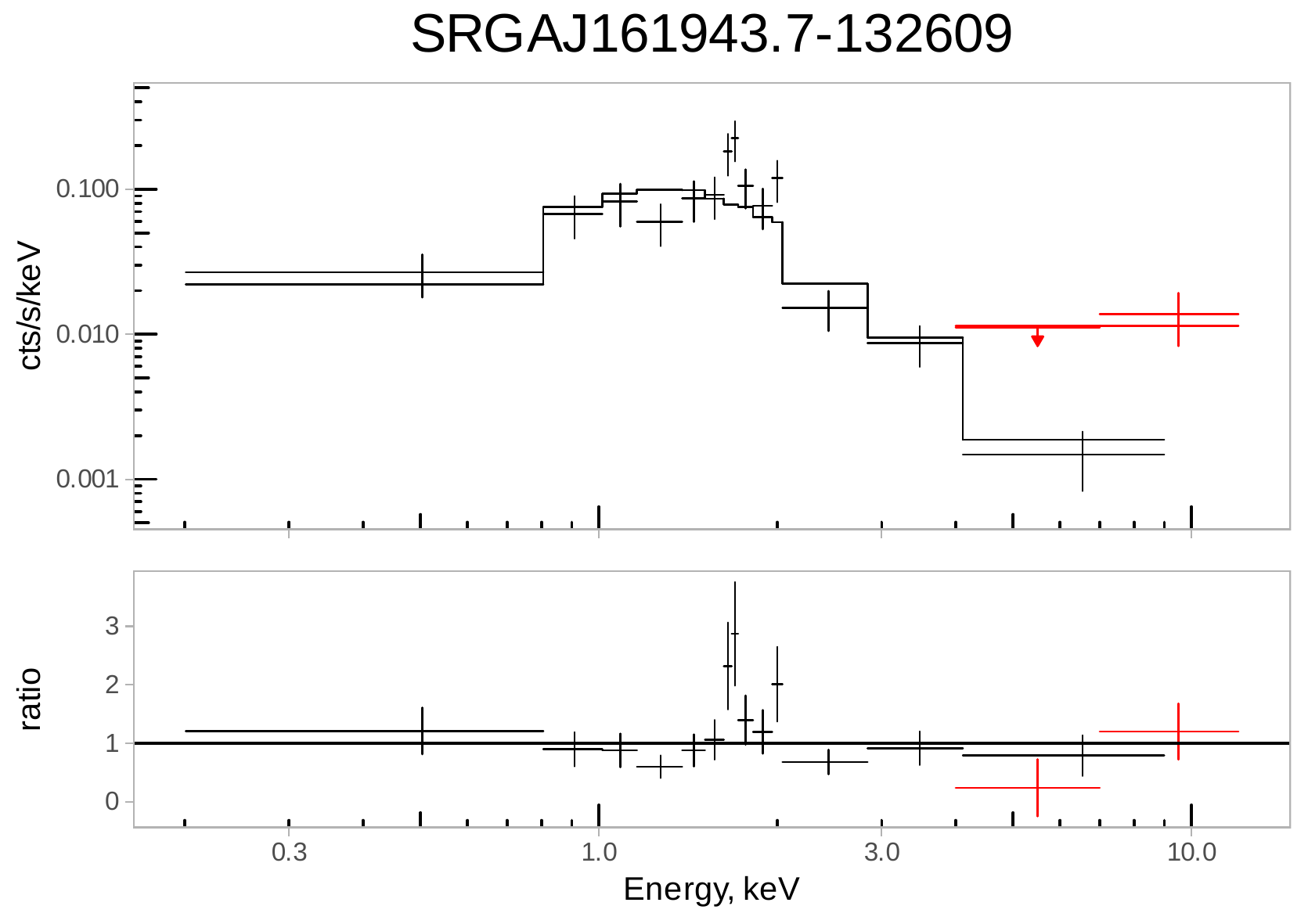}
  \includegraphics[width=0.45\columnwidth]{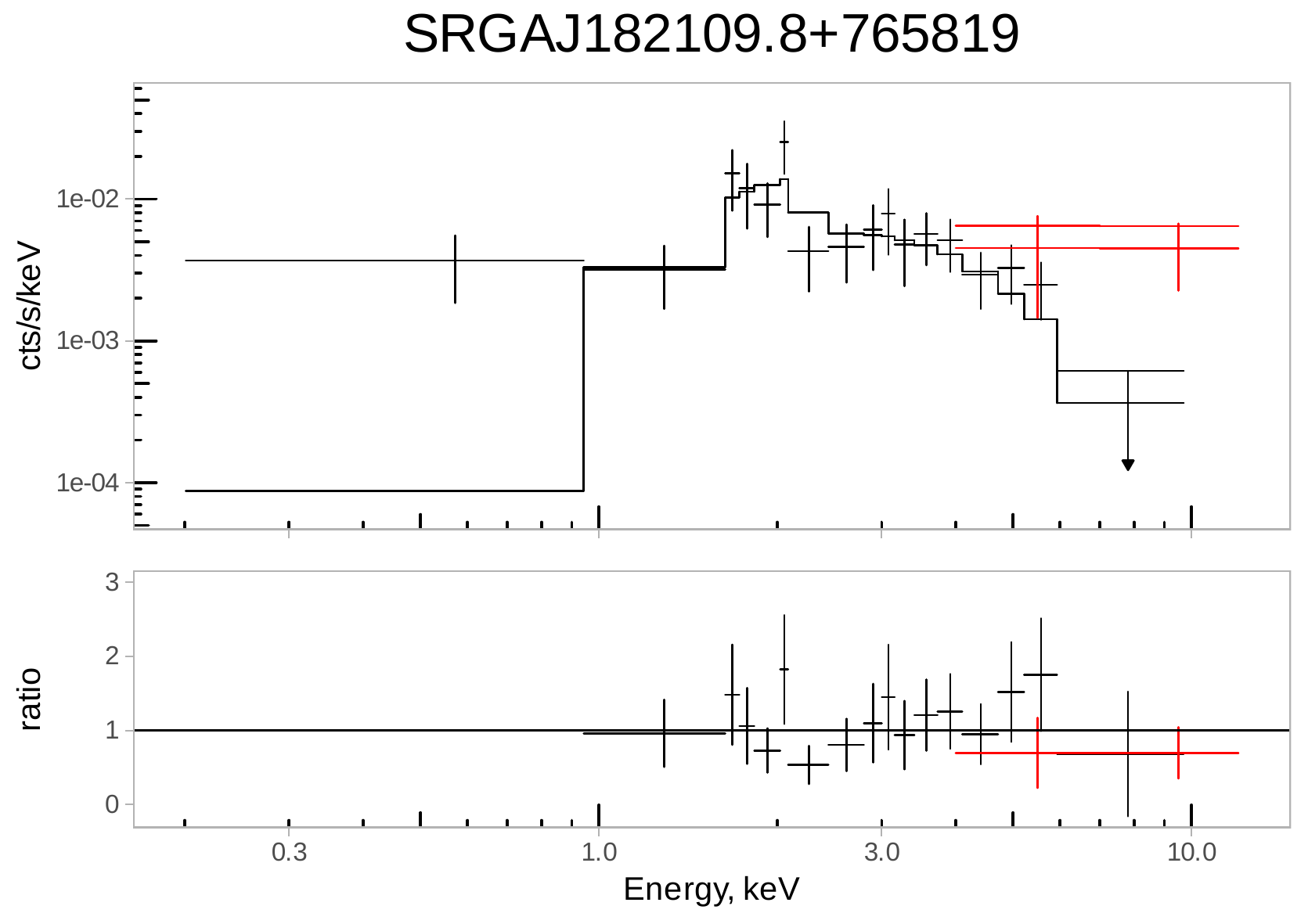}
  \includegraphics[width=0.45\columnwidth]{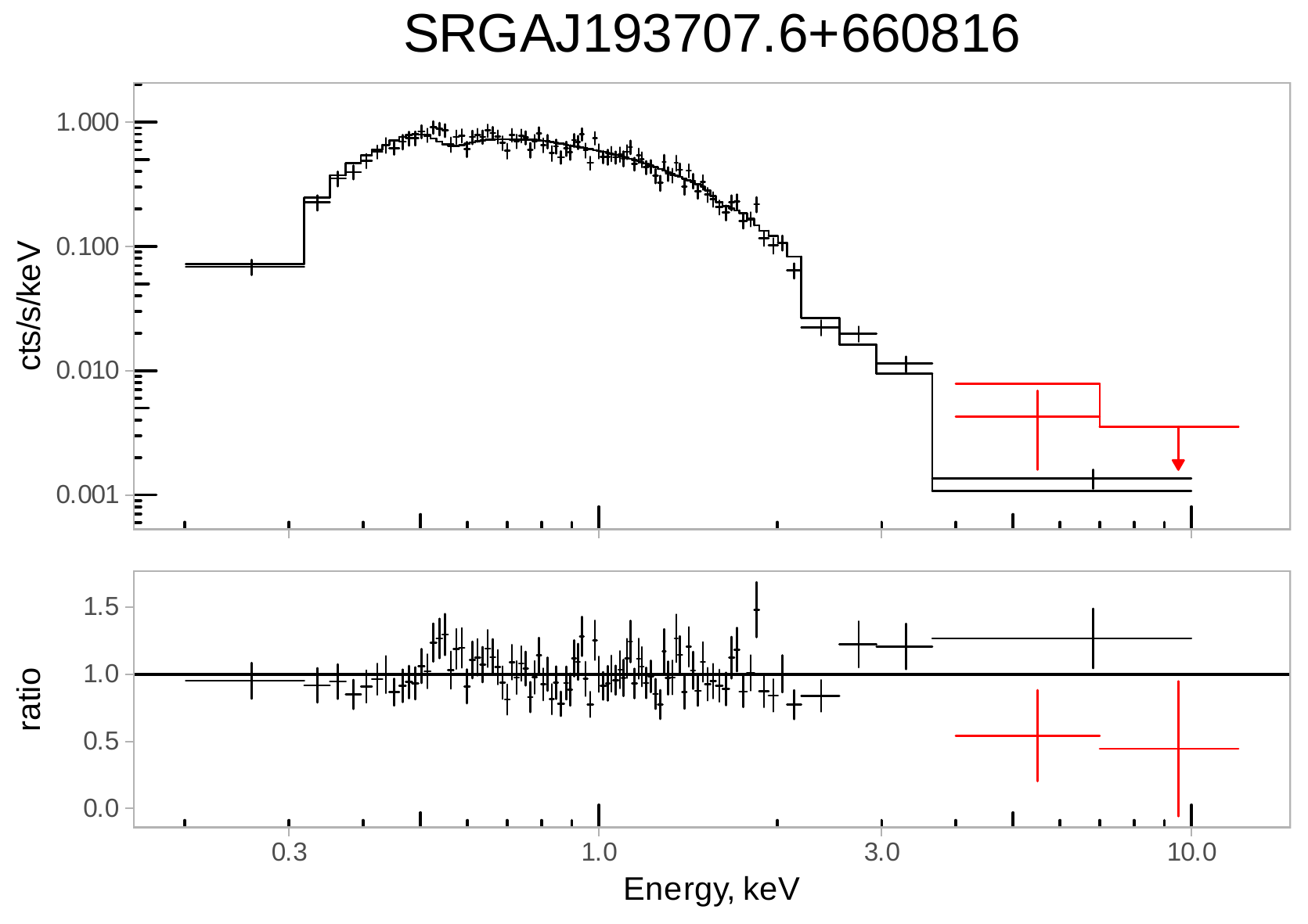}
  \includegraphics[width=0.45\columnwidth]{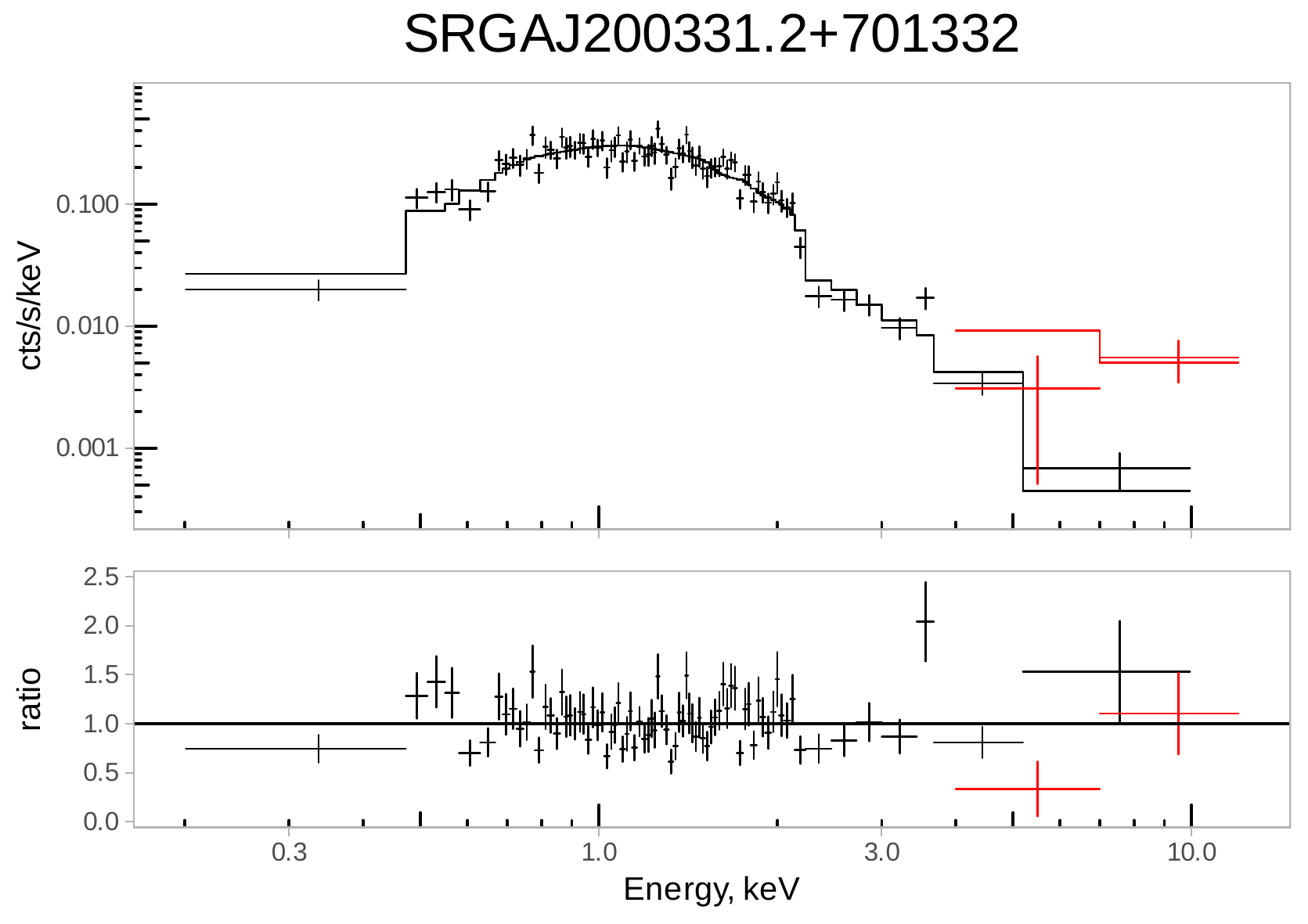}
  \includegraphics[width=0.45\columnwidth]{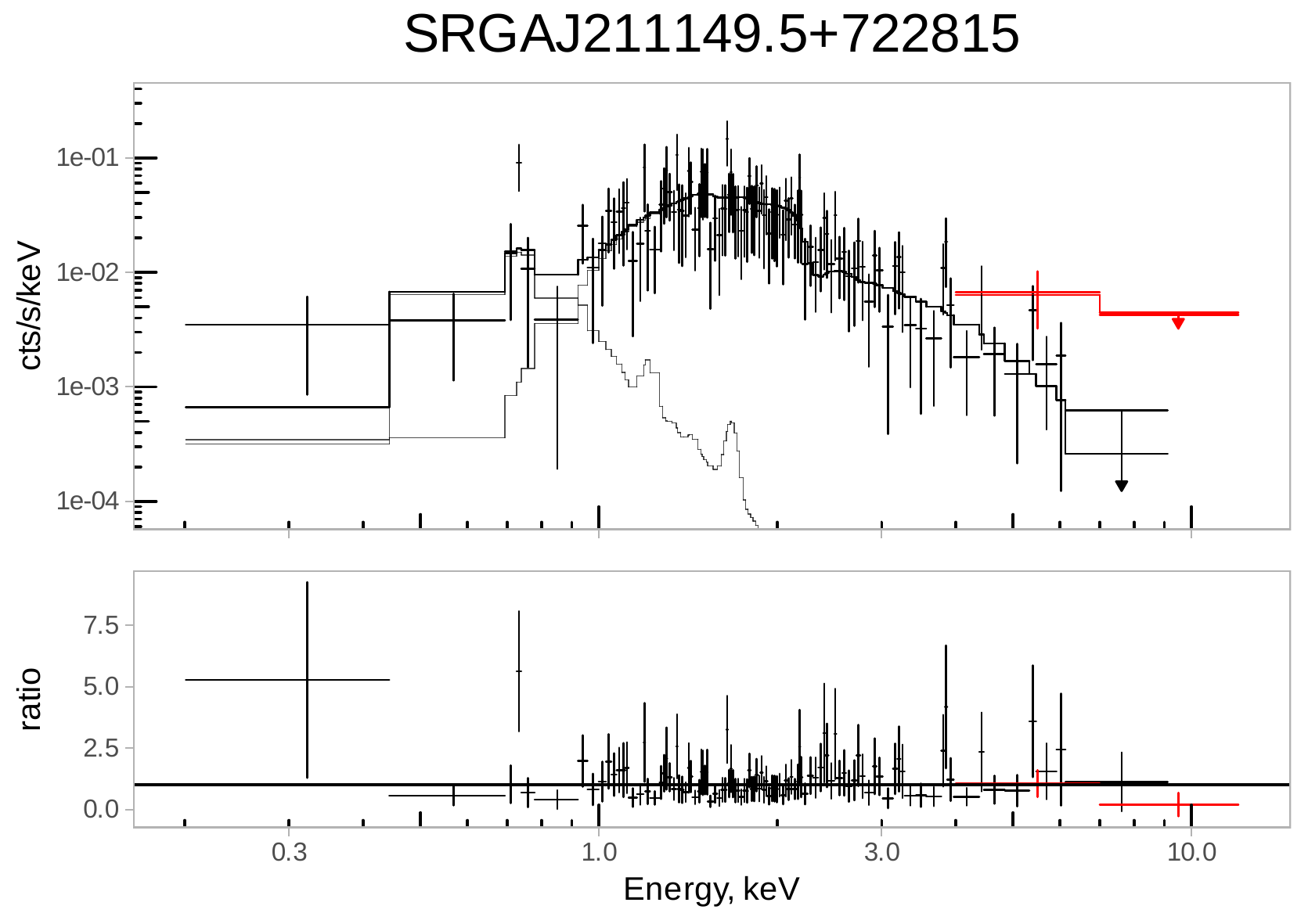}

  \caption{
  (Contd.)
  }
\end{figure*}

\begin{figure*}
  \centering
  SRGA\,J001439.6+183503
  \includegraphics[width=0.8\columnwidth]{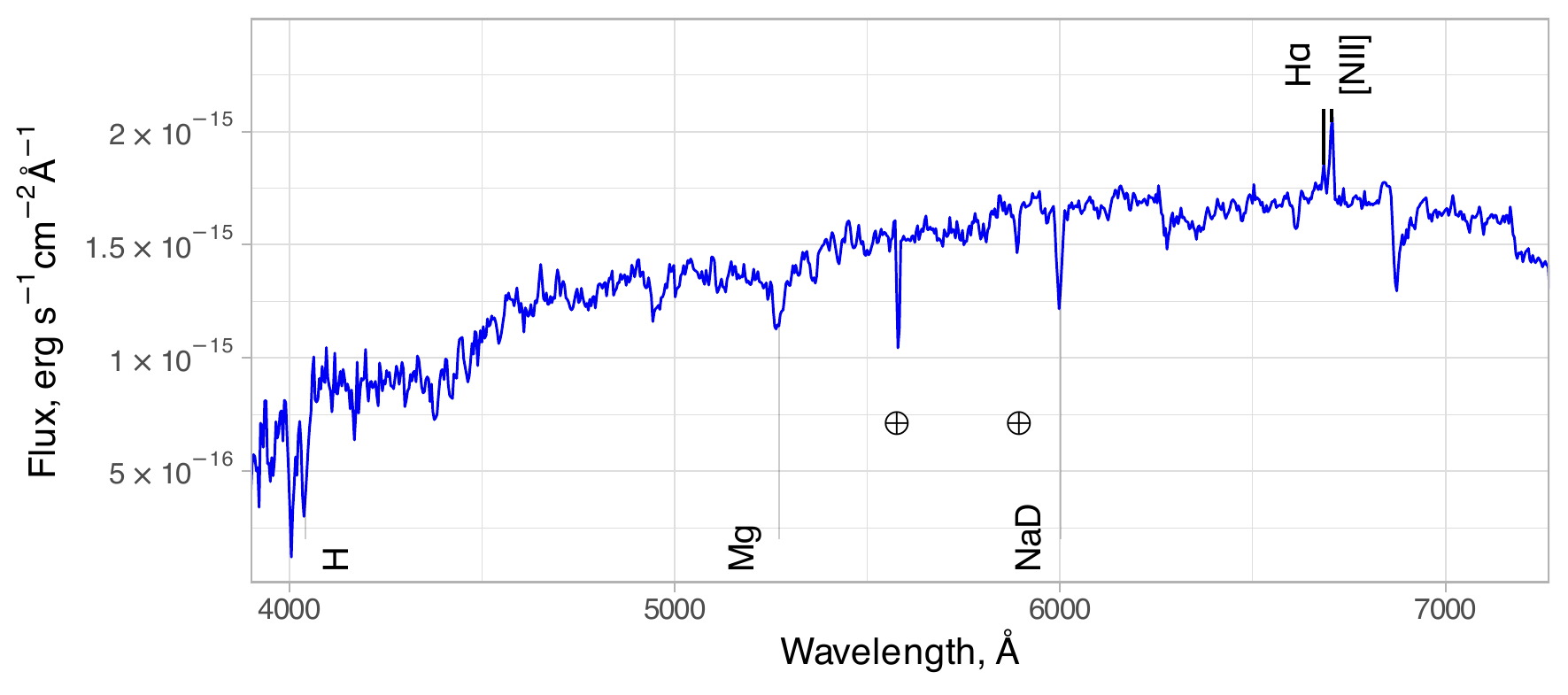}
  SRGA\,J002240.8+804348
  \includegraphics[width=0.8\columnwidth]{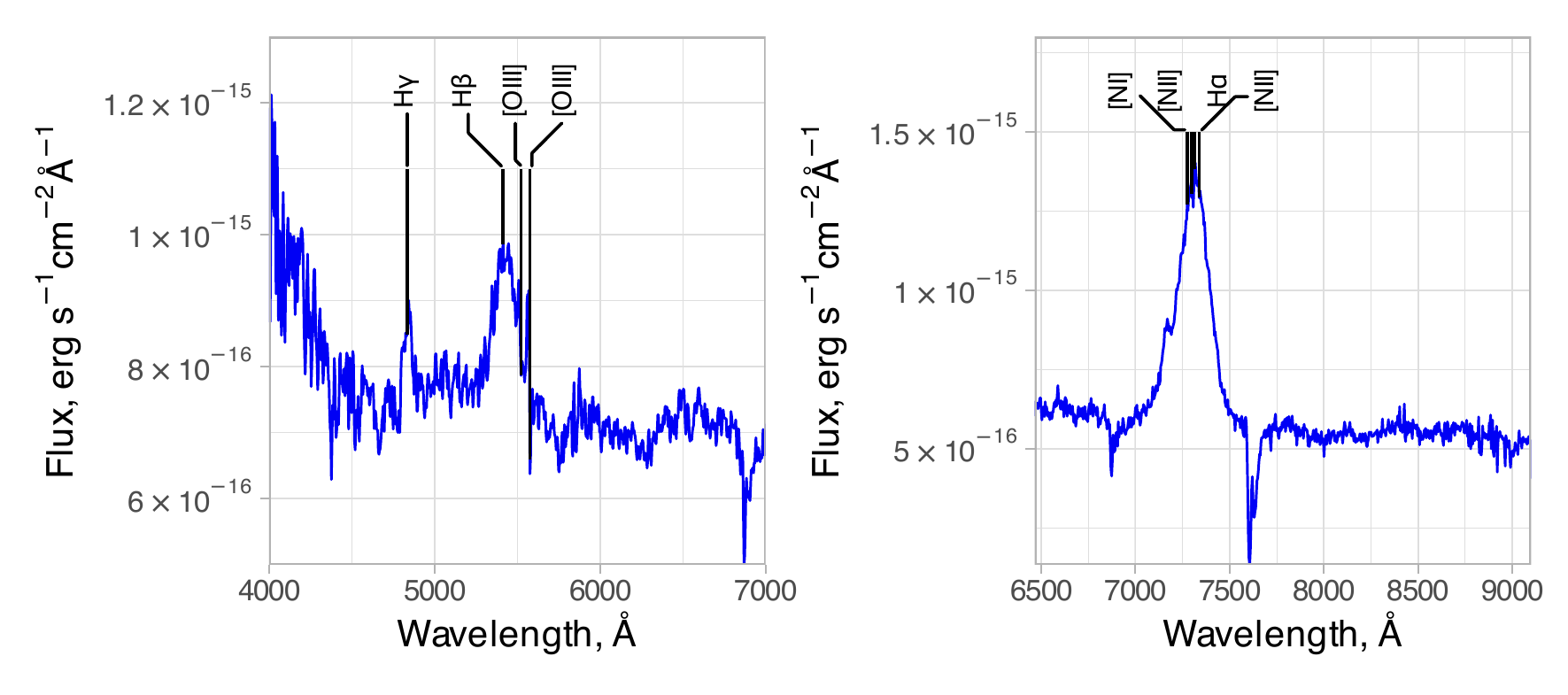}
  SRGA\,J010742.9+574419
  \includegraphics[width=0.8\columnwidth]{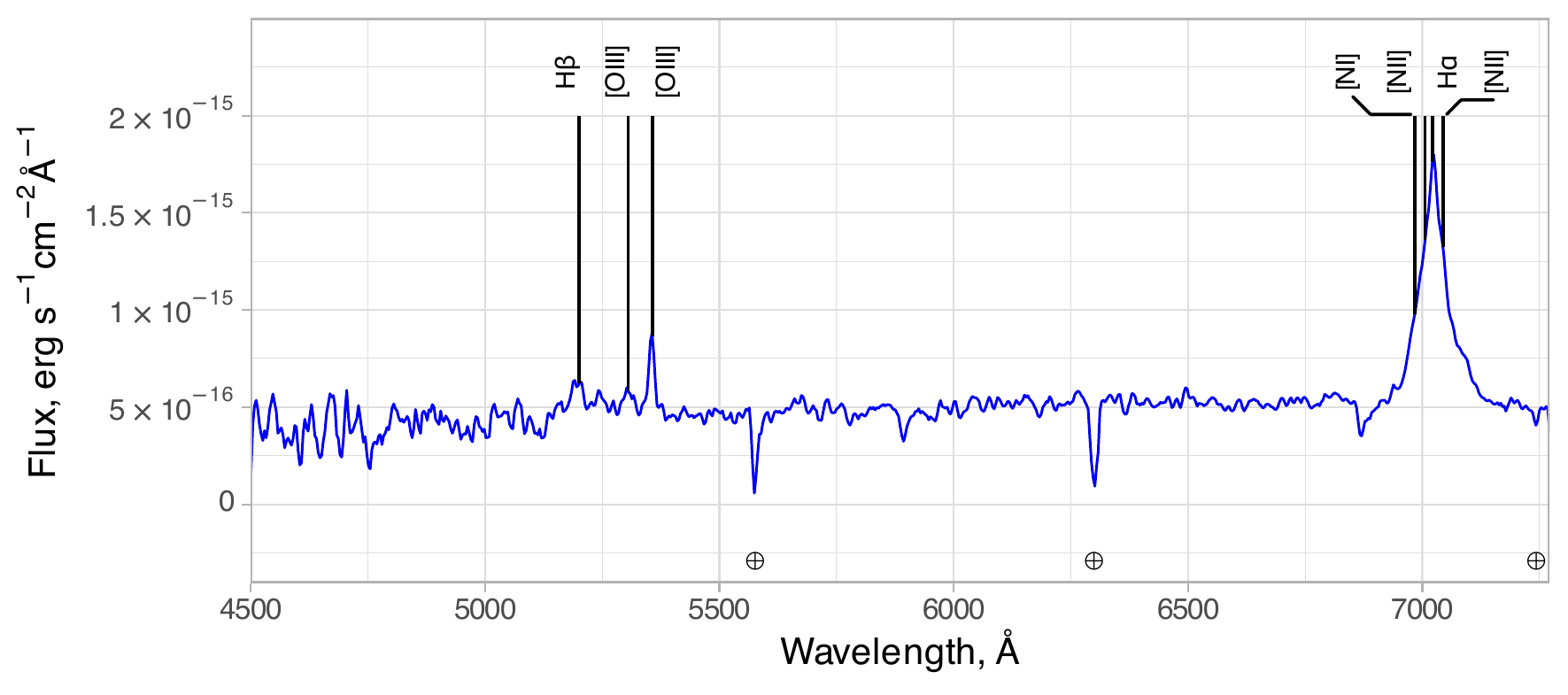}
  \caption{
  Optical spectra with the marked main emission and absorption lines.
  }
  \label{fig:spec0014_0107}
\end{figure*}
\addtocounter{figure}{-1}
\begin{figure*}
  \centering
  \vfill
  SRGA\,J021227.3+520953
  \includegraphics[width=.8\columnwidth]{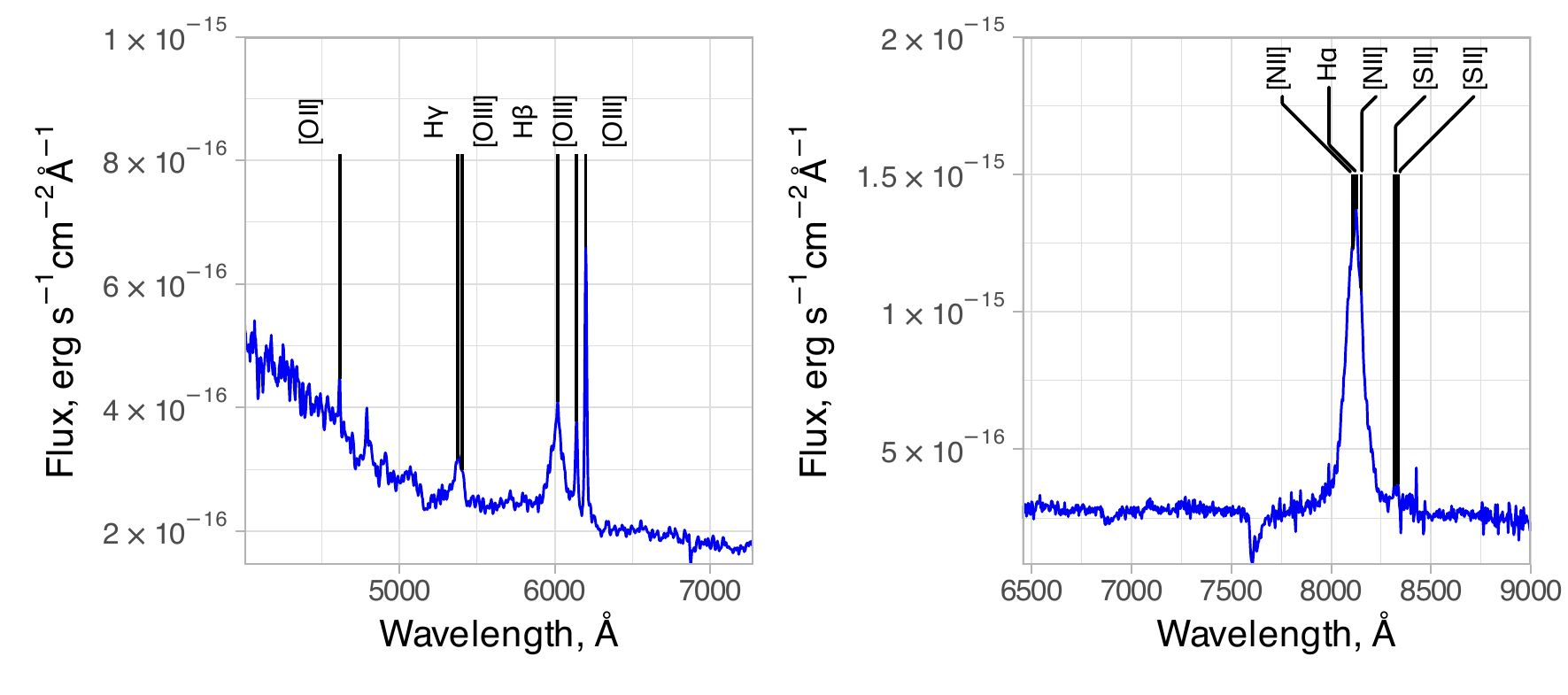}
  SRGA\,J025208.4+482955
  \includegraphics[width=.8\columnwidth]{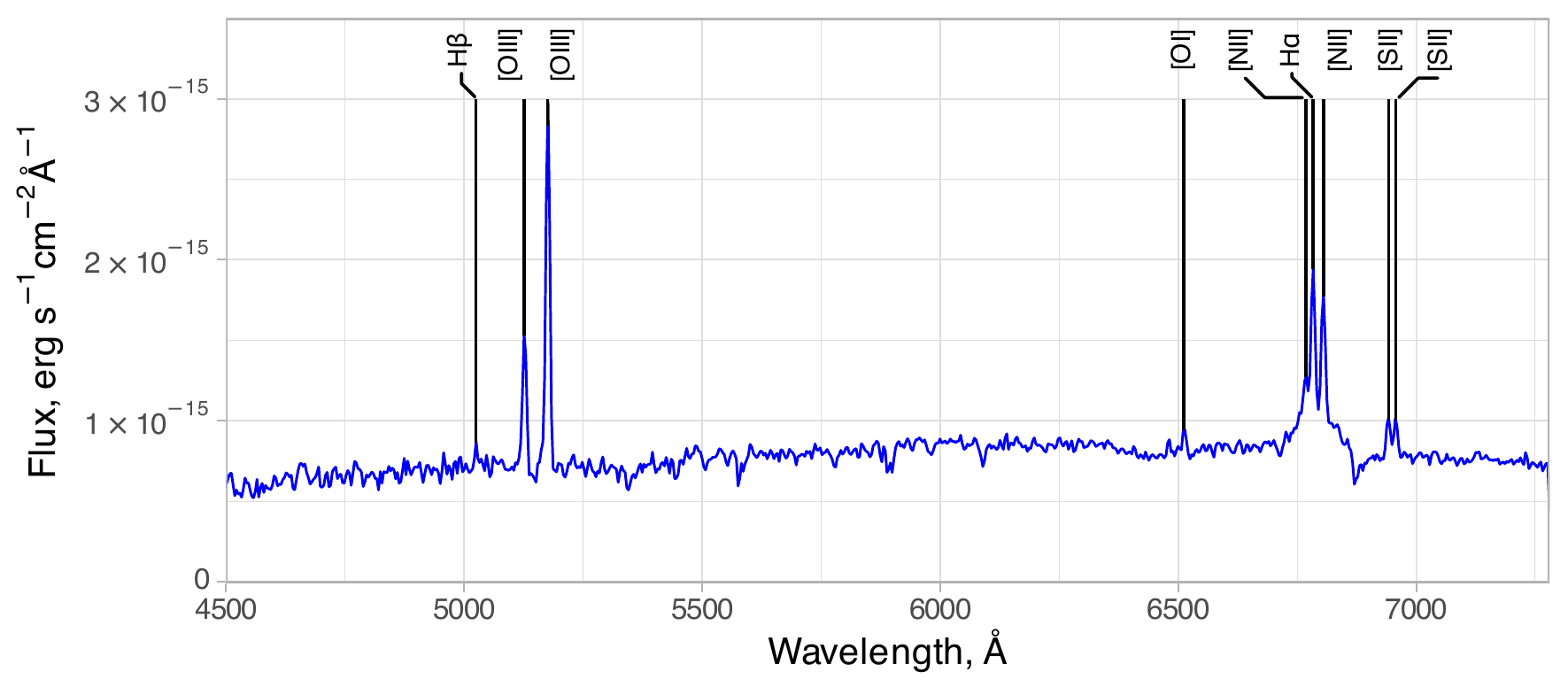}
   SRGA\,J045432.1+524003
  \includegraphics[width=.8\columnwidth]{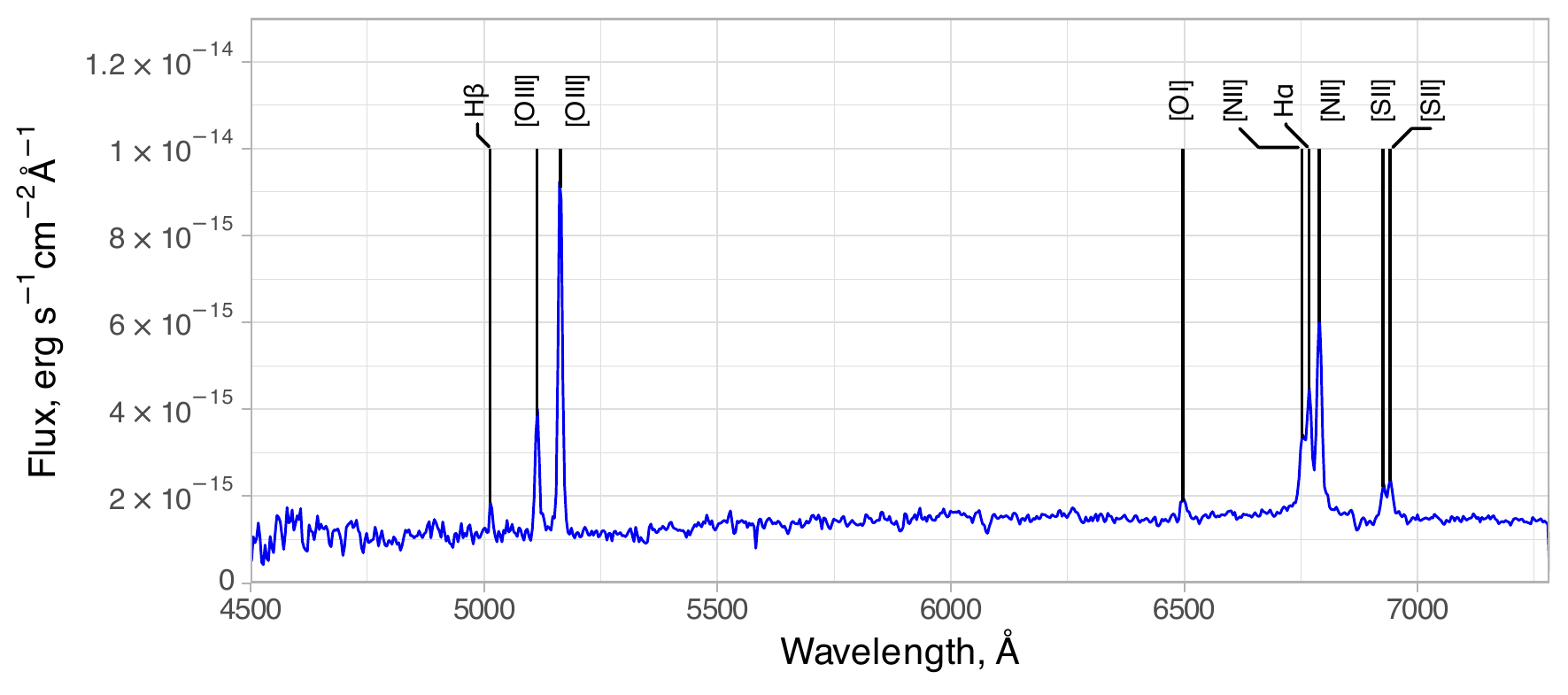}
  \caption{
  (Contd.)
  }
  \label{fig:spec0212_0252}
\end{figure*}
\addtocounter{figure}{-1}
\begin{figure*}
  \centering
  SRGA\,J051313.5+662747
  \includegraphics[width=.8\columnwidth]{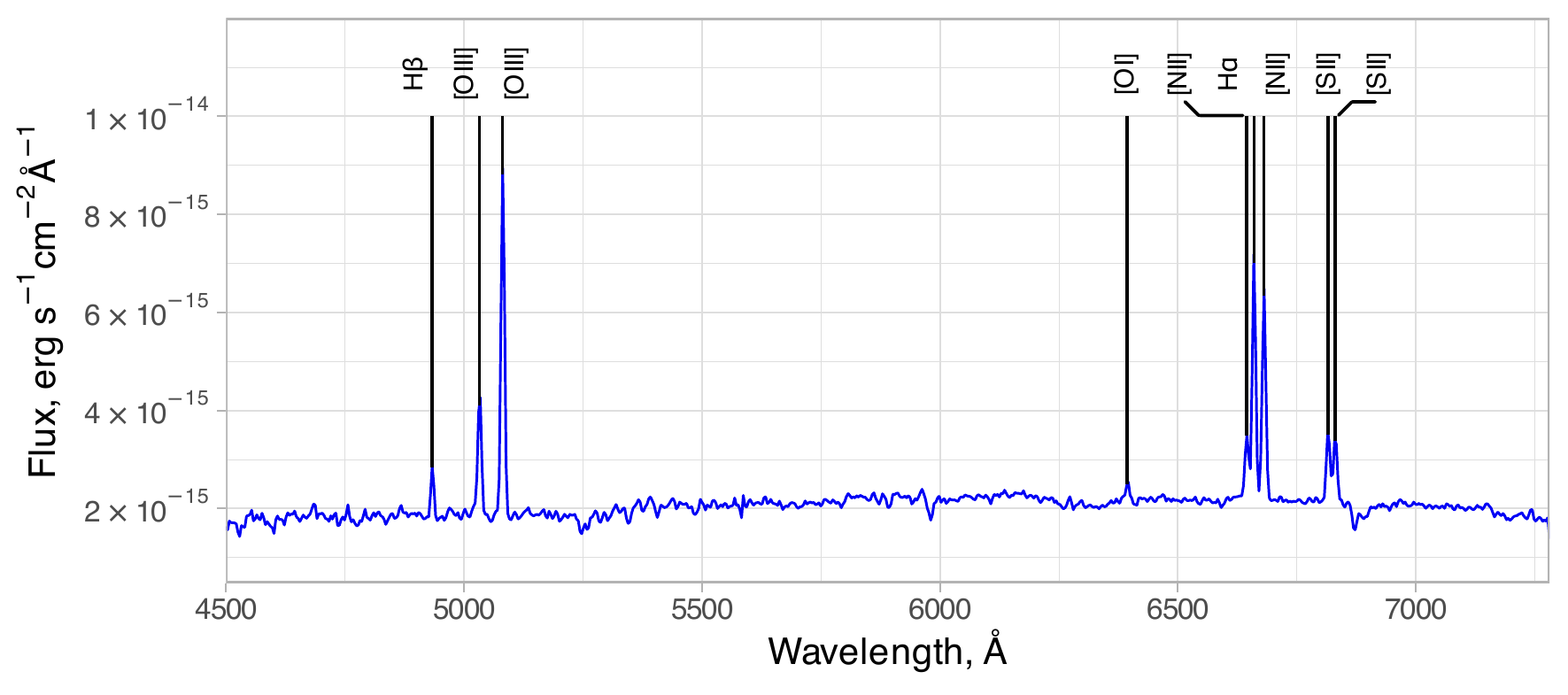}
  SRGA\,J110945.8+800815
  \includegraphics[width=.8\columnwidth]{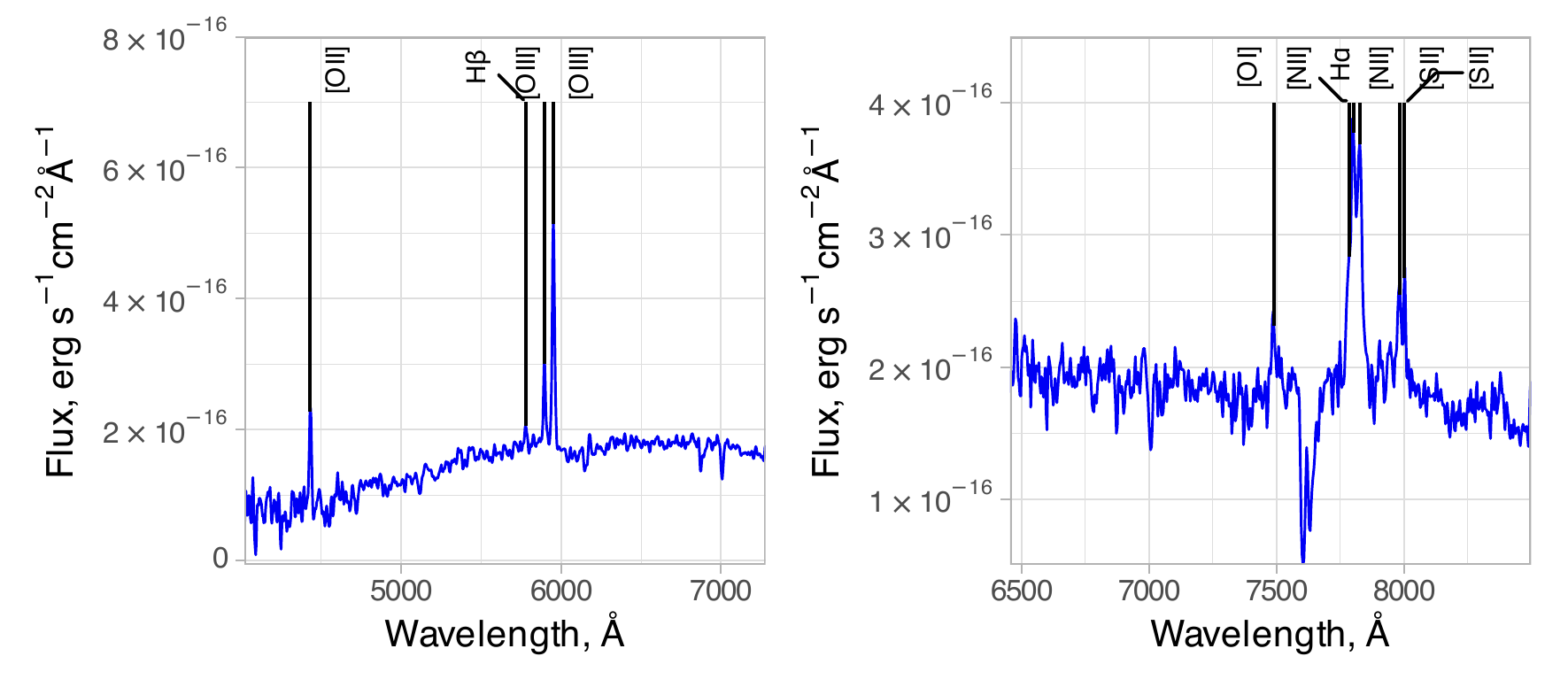}
   SRGA\,J161251.4$-$052100
  \includegraphics[width=.8\columnwidth]{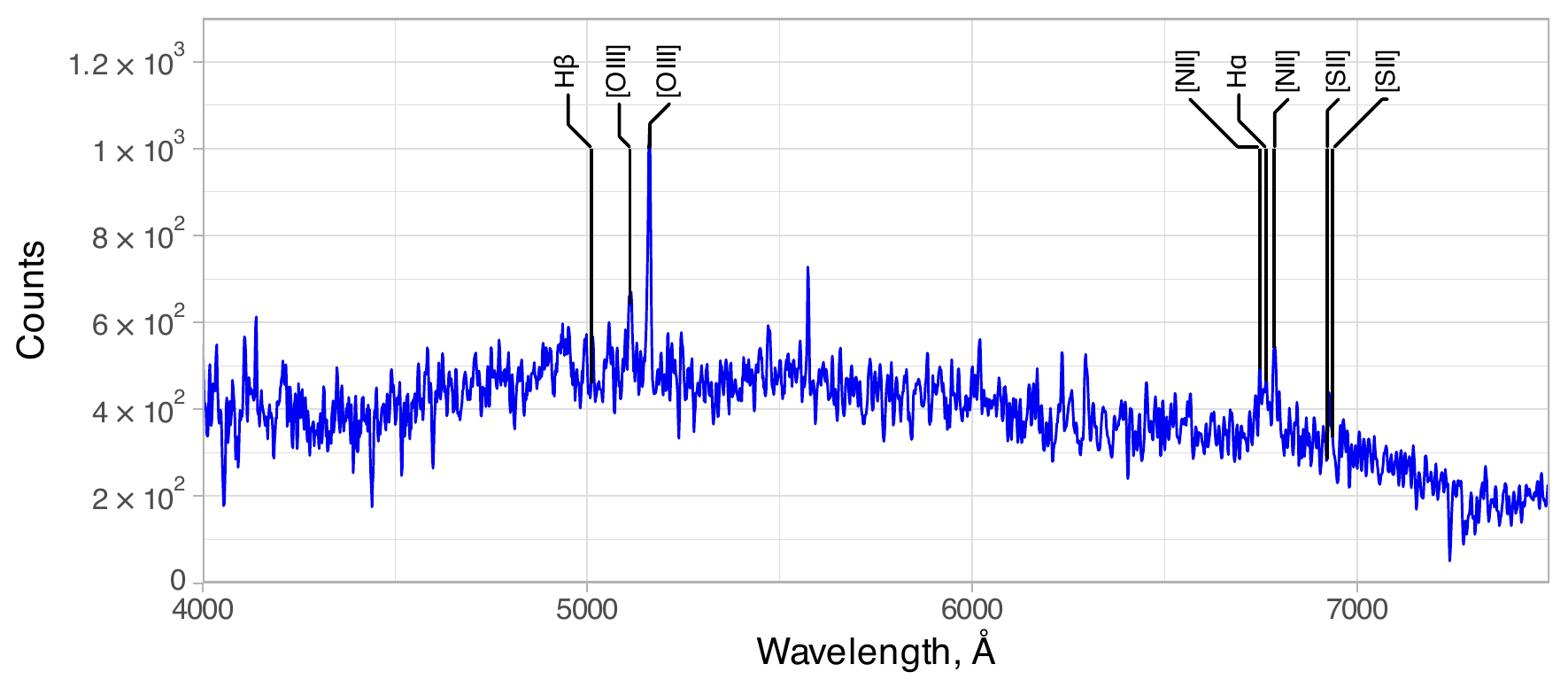}
  \caption{
  (Contd.)
  }
  \label{fig:spec0454_1109}
\end{figure*}
\addtocounter{figure}{-1}
\begin{figure*}
  \centering
  \vfill
  SRGA\,J161943.7$-$132609
  \includegraphics[width=.8\columnwidth]{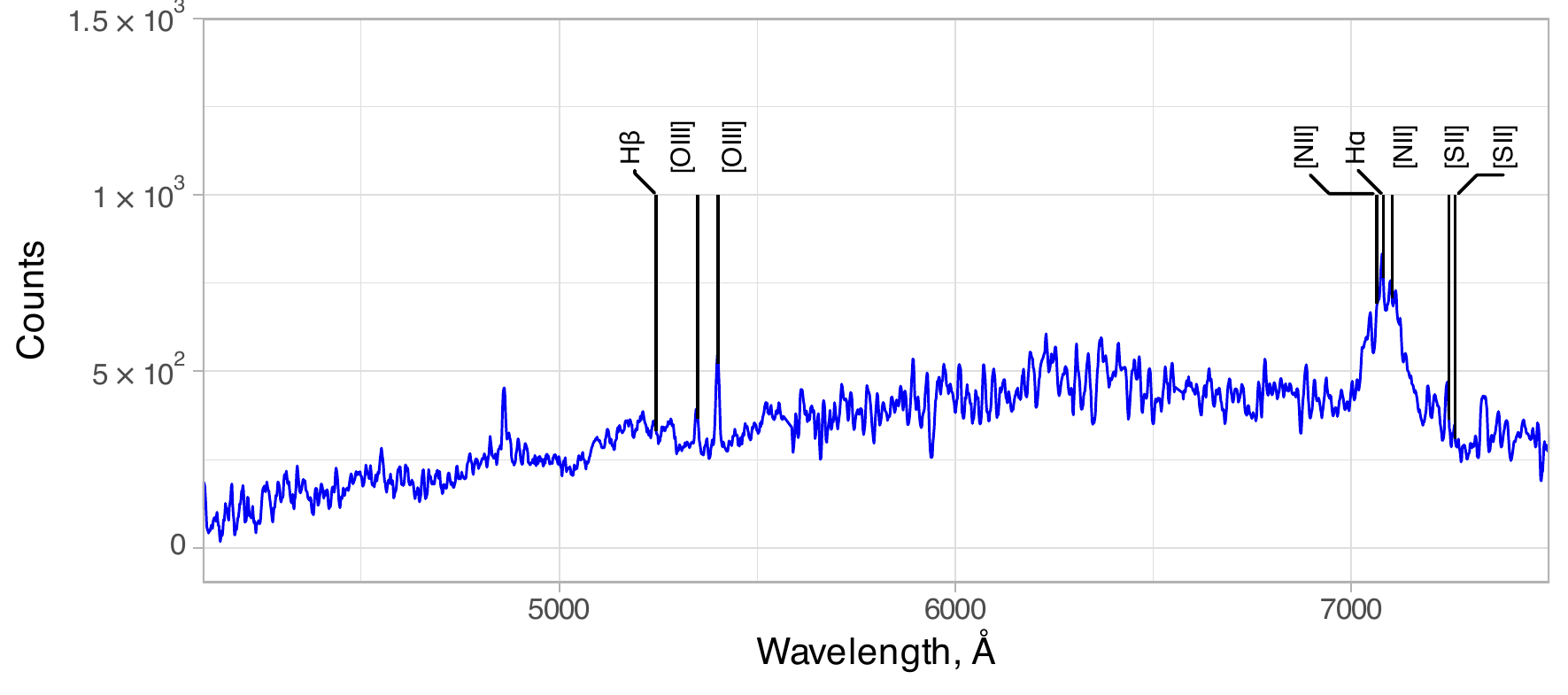}
  SRGA\,J182109.8+765819
  \includegraphics[width=.8\columnwidth]{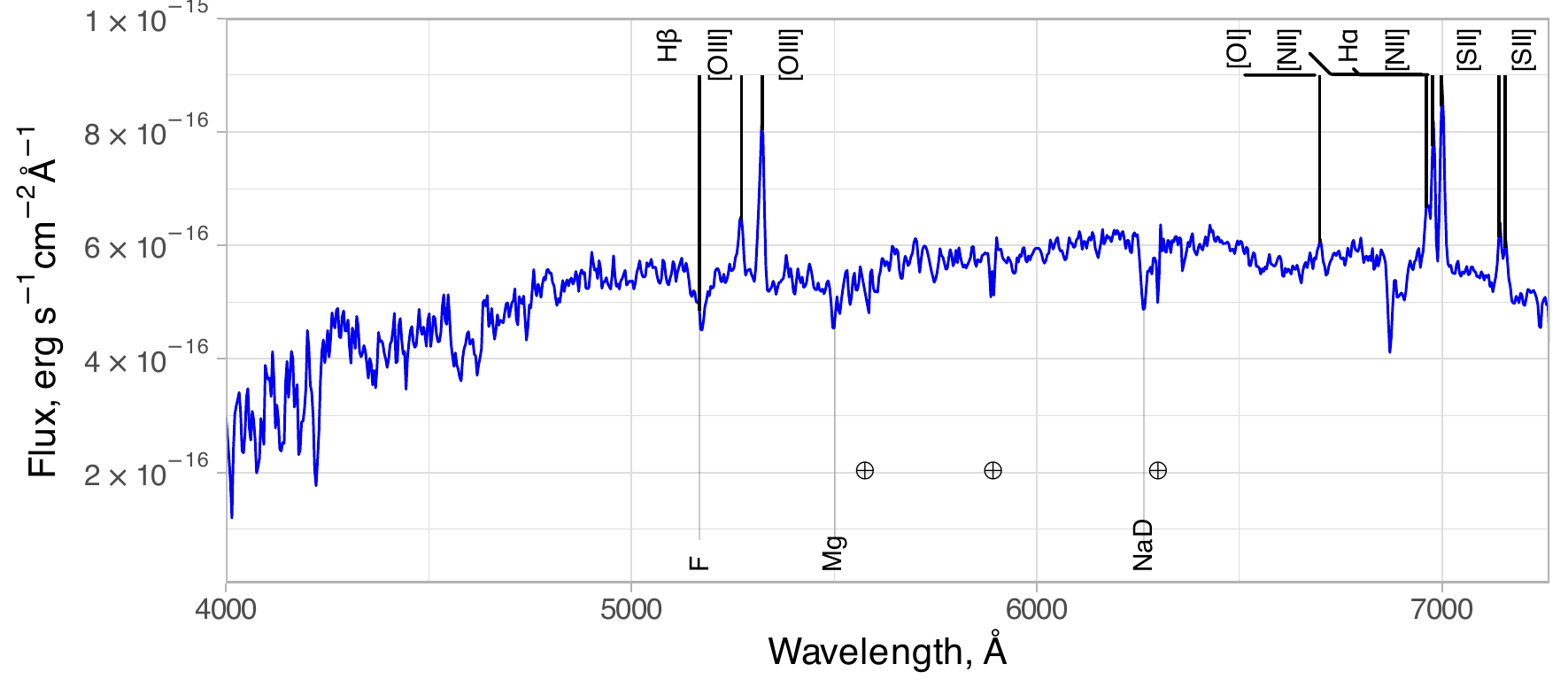}
  SRGA\,J193707.6+660816
  \includegraphics[width=.8\columnwidth]{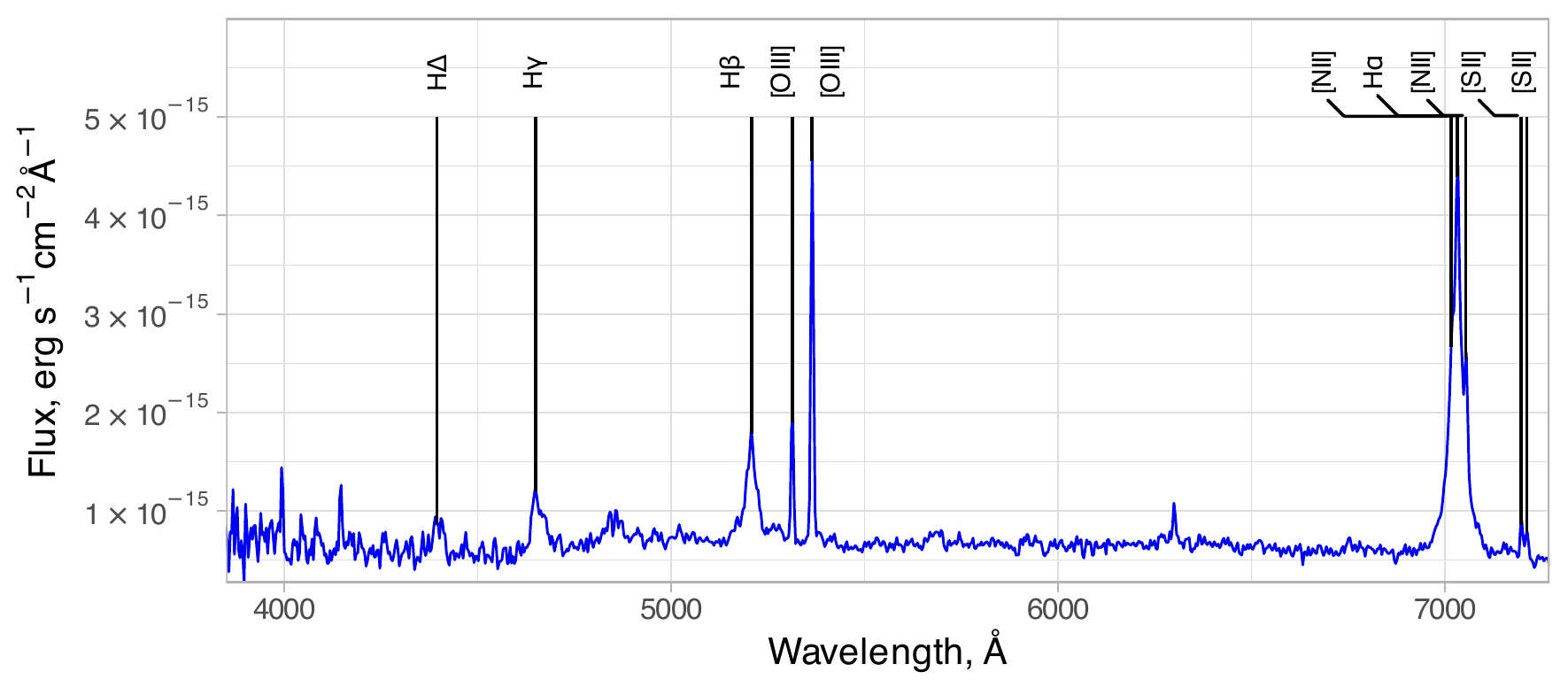}
  \caption{
  (Contd.)
  }
  \label{fig:spec1612_1821}
\end{figure*}
\addtocounter{figure}{-1}
\begin{figure*}
  \centering
  SRGA\,J200331.2+701332
  \includegraphics[width=.8\columnwidth]{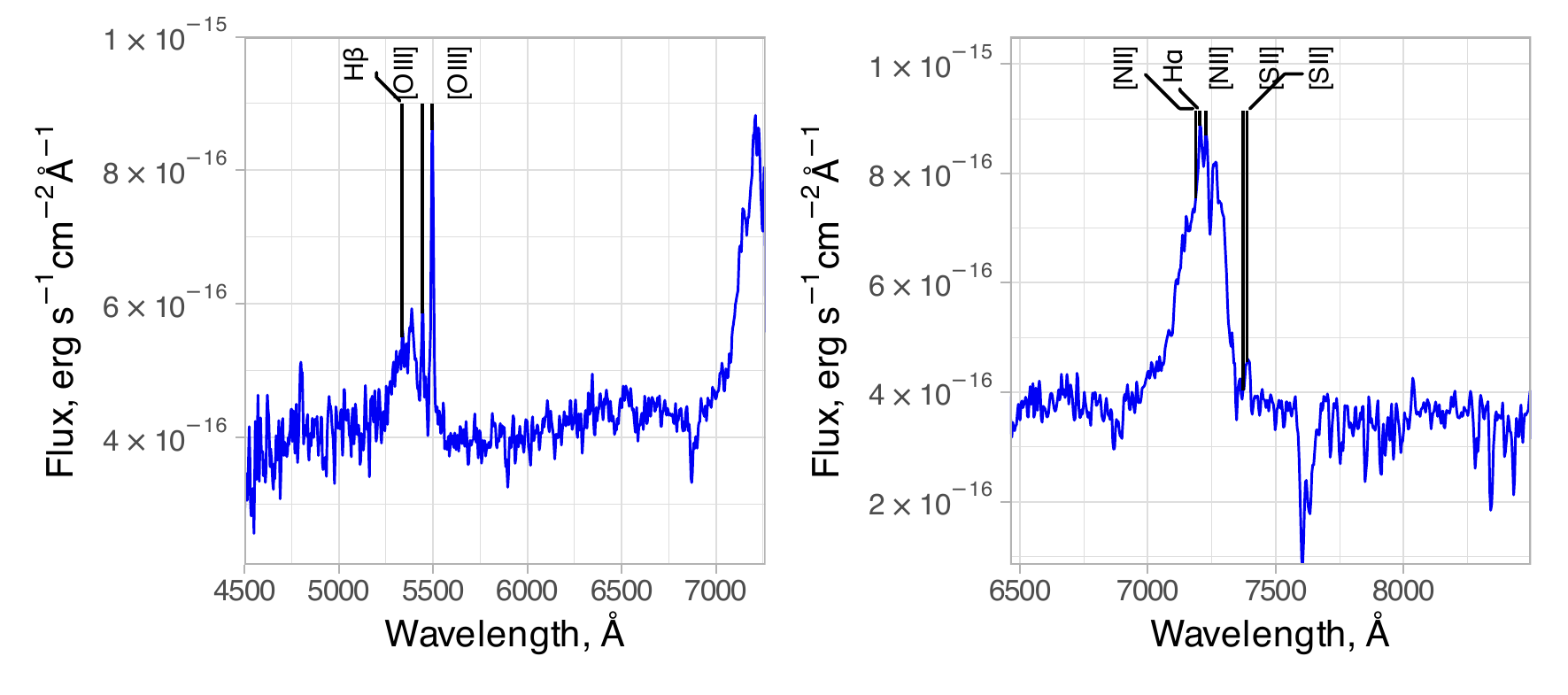}
  SRGA\,J211149.5+722815
  \includegraphics[width=.8\columnwidth]{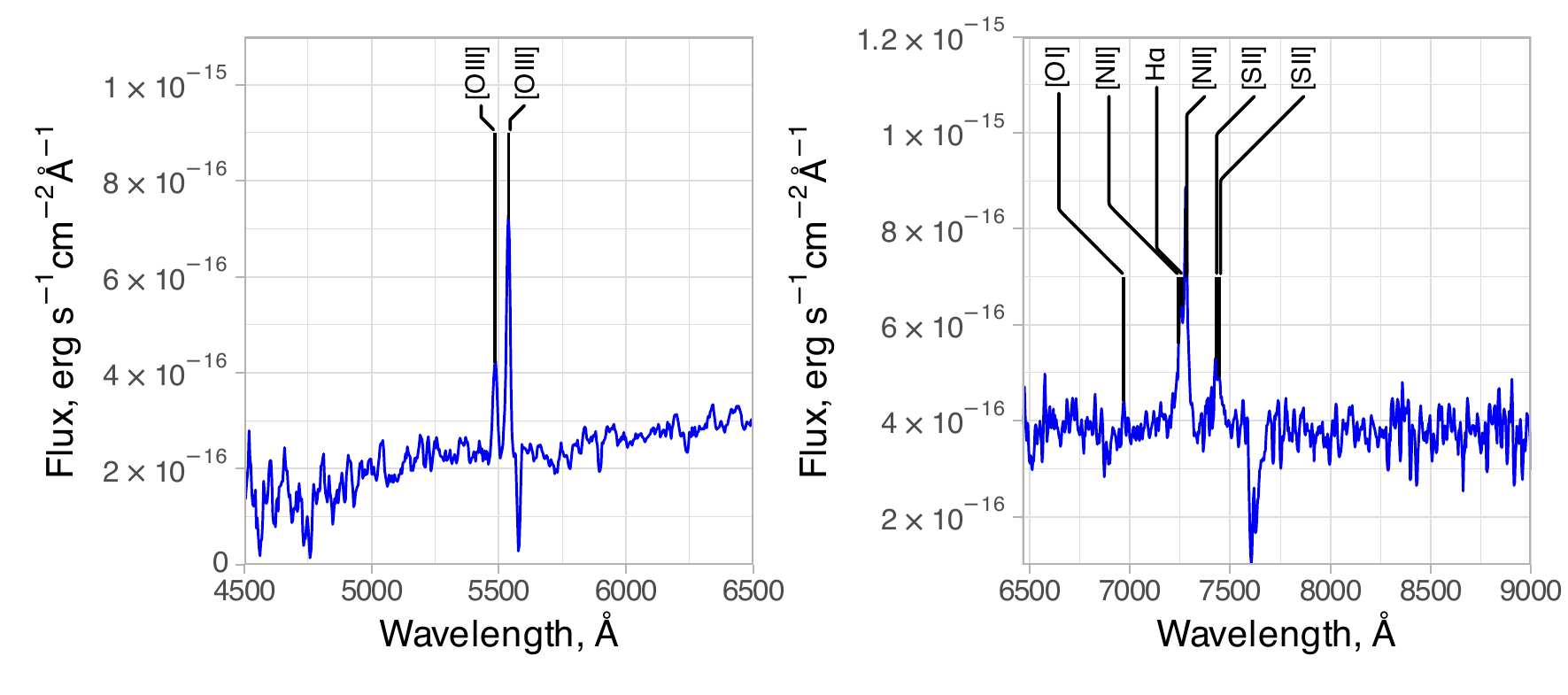}
  \caption{
  (Contd.)
  }
  \label{fig:spec1937_2111}
\end{figure*}

\begin{table*}
  \caption{\label{tab:spec_lines}
  Spectral features of the sources. The wavelengths are in the observer’s frame. The fluxes, equivalent widths, and FWHMs were obtained for the reference frames of the sources. The confidence intervals and the upper limits are given at the 1$\sigma$ and 2$\sigma$ confidence levels, respectively.
  }
  \centering
  \begin{tabular}[t]{lcccc}
  \toprule
  Line & Wavelength, \AA & Flux, $10^{-15}$~erg~s$^{-1}$~cm$^{-2}$ & Equiv. width, \AA & $FWHM$, $10^2$ km/s\\
  \midrule
  \multicolumn{5}{c}{SRGA\,J001439.6+183503}\\
  \addlinespace[0.3em]
  H$\alpha$ & $6681$ & $1.4\pm0.3$ & $-0.81\pm0.20$ & $-$\\
  {}[NII]$\lambda$6583 & $6701$ & $3.8\pm0.4$ & $-2.24\pm0.25$ & $-$\\
  \addlinespace[0.3em]
\multicolumn{5}{c}{SRGA\,J002240.8+804348}\\
\addlinespace[0.3em]
H$\gamma$, broad & $4847$ & $9.3\pm2.0$ & $-16\pm5$ & $46\pm9$\\
H$\beta$, broad & $5440$ & $35.6\pm1.7$ & $-48.1\pm2.5$ & $89\pm5$\\
H$\beta$ & $5440$ & $<0.7$ & $ >-0.9$ & $-$\\
{}[OIII]$\lambda$4959 & $5524$ & $1.0\pm0.4$ & $-1.4\pm0.5$ & $-$\\
{}[OIII]$\lambda$5007 & $5577$ & $2.6\pm0.5$ & $-3.5\pm0.7$ & $-$\\
H$\alpha$, broad & $7316$ & $128.8\pm2.1$ & $-221\pm4$ & $72.1\pm1.2$\\
\addlinespace[0.3em]

\multicolumn{5}{c}{SRGA\,J010742.9+574419}\\
\addlinespace[0.3em]
H$\beta$ & $5200$ & $1.7\pm0.4$ & $-3.4\pm1.1$ & $-$\\
{}[OIII]$\lambda$4959 & $5304$ & $1.4\pm0.4$ & $-2.9\pm0.8$ & $-$\\
{}[OIII]$\lambda$5007 & $5356$ & $5.4\pm0.5$ & $-11.2\pm1.2$ & $-$\\
{}[NII]$\lambda$6548 & $7007$ & $0.56\pm0.24$ & $-0.8\pm0.7$ & $-$\\
H$\alpha$, broad & $7023$ & $70.9\pm2.0$ & $-140\pm6$ & $37.5\pm1.2$\\
H$\alpha$ & $7023$ & $8.4\pm1.0$ & $-17\pm4$ & $-$\\
{}[NII]$\lambda$6583 & $7045$ & $1.67\pm0.24$ & $-2.3\pm2.0$ & $-$\\
\addlinespace[0.3em]
\multicolumn{5}{c}{SRGA\,J021227.3+520953}\\
\addlinespace[0.3em]
H$\beta$, broad & $6018$ & $11.8\pm0.7$ & $-51\pm3$ & $58\pm4$\\
H$\beta$ & $6018$ & $0.80\pm0.20$ & $-3.5\pm0.9$ & $-$\\
{}[OIII]$\lambda$4959 & $6140$ & $2.14\pm0.16$ & $-9.5\pm0.8$ & $-$\\
{}[OIII]$\lambda$5007 & $6200$ & $5.96\pm0.20$ & $-27.0\pm1.0$ & $-$\\
{}[NII]$\lambda$6548 & $8105$ & $0.37\pm0.08$ & $-1.14\pm0.24$ & $-$\\
H$\alpha$, broad & $8116$ & $79.8\pm0.8$ & $-244\pm4$ & $38.7\pm0.4$\\
H$\alpha$ & $8124$ & $2.75\pm0.26$ & $-8.4\pm0.8$ & $-$\\
{}[NII]$\lambda$6583 & $8149$ & $1.11\pm0.08$ & $-3.4\pm0.7$ & $-$\\
{}[SII]$\lambda$6716 & $8316$ & $0.55\pm0.14$ & $-1.8\pm0.5$ & $-$\\
{}[SII]$\lambda$6730 & $8333$ & $0.58\pm0.14$ & $-1.9\pm0.5$ & $-$\\
\addlinespace[0.3em]
\multicolumn{5}{c}{SRGA\,J025208.4+482955}\\
\addlinespace[0.3em]
H$\beta$ & $5026$ & $1.3\pm0.3$ & $-1.9\pm0.5$ & $-$\\
{}[OIII]$\lambda$4959 & $5127$ & $8.4\pm0.4$ & $-11.9\pm0.5$ & $-$\\
{}[OIII]$\lambda$5007 & $5176$ & $22.6\pm0.5$ & $-31.9\pm0.7$ & $-$\\
{}[OI]$\lambda$6300 & $6513$ & $1.27\pm0.22$ & $-1.6\pm0.3$ & $-$\\
{}[NII]$\lambda$6548 & $6769$ & $2.3\pm0.3$ & $-2.9\pm0.4$ & $-$\\
H$\alpha$, broad & $6784$ & $26.8\pm1.2$ & $-33.9\pm1.6$ & $41.7\pm2.4$\\
H$\alpha$ & $6784$ & $8.4\pm0.4$ & $-10.6\pm0.6$ & $-$\\
{}[NII]$\lambda$6583 & $6804$ & $7.4\pm0.4$ & $-9.3\pm0.5$ & $-$\\
{}[SII]$\lambda$6716 & $6942$ & $2.34\pm0.23$ & $-2.99\pm0.29$ & $-$\\
{}[SII]$\lambda$6730 & $6957$ & $2.20\pm0.22$ & $-2.81\pm0.29$ & $-$\\

  \bottomrule
  \end{tabular}
  \end{table*}

\addtocounter{table}{-1}

\begin{table*}
\caption{ (Contd.)}
  \centering
  \begin{tabular}[t]{lcccc}
  \toprule
  Line & Wavelength, \AA & Flux, $10^{-15}$~erg~s$^{-1}$~cm$^{-2}$ & Equiv. width, \AA & $FWHM$, $10^2$ km/s\\
  \midrule
\addlinespace[0.3em]
\multicolumn{5}{c}{SRGA\,J045432.1+524003}\\
\addlinespace[0.3em]
H$\beta$ & $5015$ & $5.8\pm1.7$ & $-5.1\pm1.5$ & $-$\\
{}[OIII]$\lambda$4959 & $5114$ & $29.5\pm1.7$ & $-25.5\pm1.5$ & $-$\\
{}[OIII]$\lambda$5007 & $5163$ & $88.3\pm2.2$ & $-76.1\pm2.5$ & $-$\\
{}[OI]$\lambda$6300 & $6498$ & $5.7\pm0.8$ & $-3.8\pm0.5$ & $-$\\
{}[NII]$\lambda$6548 & $6754$ & $17.0\pm0.5$ & $-11.3\pm0.4$ & $-$\\
H$\alpha$ & $6769$ & $28.3\pm1.3$ & $-19.0\pm0.9$ & $-$\\
H$\alpha$, broad & $6769$ & $52\pm3$ & $-35\pm2$ & $41\pm3$\\
{}[NII]$\lambda$6583 & $6789$ & $50.9\pm0.5$ & $-34.1\pm1.0$ & $-$\\
{}[SII]$\lambda$6716 & $6927$ & $8.9\pm0.7$ & $-6.0\pm0.5$ & $-$\\
{}[SII]$\lambda$6730 & $6942$ & $10.7\pm0.7$ & $-7.2\pm0.5$ & $-$\\
\addlinespace[0.3em]
\multicolumn{5}{c}{SRGA\,J051313.5+662747}\\
\addlinespace[0.3em]
H$\beta$ & $4933$ & $8.0\pm0.8$ & $-4.3\pm0.4$ & $-$\\
{}[OIII]$\lambda$4959 & $5033$ & $23.0\pm0.8$    & $-12.3\pm0.5$ & $-$\\
{}[OIII]$\lambda$5007 & $5081$ & $64\pm1$        & $-34.1\pm0.6$ & $-$\\
{}[OI]$\lambda$6300   & $6394$ & $3.8\pm0.5$     & $-1.78\pm0.23$ & $-$\\
{}[NII]$\lambda$6548  & $6645$ & $12.6\pm0.6$    & $-5.80\pm0.24$ & $-$\\
H$\alpha$             & $6660$ & $43.8\pm0.7$    & $-20.1\pm0.3$ & $-$\\
{}[NII]$\lambda$6583  & $6681$ & $38.8\pm0.7$    & $-17.8\pm0.4$ & $-$\\
{}[SII]$\lambda$6716  & $6816$ & $13.0\pm0.6$    & $-5.90\pm0.26$ & $-$\\
{}[SII]$\lambda$6730  & $6831$ & $11.8\pm0.6$    & $-5.38\pm0.29$ & $-$\\
\addlinespace[0.3em]
\multicolumn{5}{c}{SRGA\,J110945.8+800815}\\
\addlinespace[0.3em]
{}[OII]$\lambda$3727  & $4432$ & $2.3\pm0.4$   & $-27\pm6$ & $-$\\
H$\beta$              & $5779$ & $0.52\pm0.17$ & $-3.2\pm0.9$ & $-$\\
{}[OIII]$\lambda$4959 & $5896$ & $1.91\pm0.26$ & $-11.2\pm1.5$ & $-$\\
{}[OIII]$\lambda$5007 & $5952$ & $5.37\pm0.24$ & $-31.4\pm1.7$ & $-$\\
{}[OI]$\lambda$6300   & $7488$ & $0.91\pm0.19$ & $-5.1\pm1.1$ & $-$\\
{}[NII]$\lambda$6548  & $7781$ & $1.35\pm0.24$ & $-7.4\pm1.4$ & $-$\\
H$\alpha$             & $7802$ & $3.30\pm0.27$ & $-17.6\pm1.5$ & $-$\\
{}[NII]$\lambda$6583  & $7826$ & $3.28\pm0.25$ & $-17.5\pm1.4$ & $-$\\
{}[SII]$\lambda$6716  & $7980$ & $1.04\pm0.24$ & $-5.6\pm1.1$ & $-$\\
{}[SII]$\lambda$6730  & $8001$ & $0.99\pm0.24$ & $-5.2\pm1.1$ & $-$\\
  \addlinespace[0.3em]
\multicolumn{5}{c}{SRGA\,J161251.4$-$052100}\\
H$\alpha$ &  & & $-3.8\pm1.7$ & $-$\\
{}[NII]$\lambda$6548 & & & $-2.4\pm0.6$ & $-$\\
{}[NII]$\lambda$6583 & & &  $-7.2\pm1.9$ & $-$\\
H$\beta$ & & &  $ >-1.4$ & $-$\\
{}[OIII]$\lambda$4959 & & &  $-3.8\pm0.7$ & $-$\\
{}[OIII]$\lambda$5007 & & &  $-10.3\pm0.9$ & $-$\\
\addlinespace[0.3em]
\multicolumn{5}{c}{SRGA\,J161943.7$-$132609}\\
H$\alpha$ & &  & $ >-2.7$& $-$\\
H$\alpha$, broad & &  & $-87\pm7$ & $40.5\pm2.7$\\
{}[NII]$\lambda$6548 & &  & $ >-0.8$& $-$\\
{}[NII]$\lambda$6583 & &  & $-1.1\pm1.1$& $-$\\
H$\beta$ & & &  $ >-1.3$& $-$\\
{}[OIII]$\lambda$4959 & & &  $-3.7\pm0.7$& $-$\\
{}[OIII]$\lambda$5007 & & &  $-9.5\pm0.9$& $-$\\

  \bottomrule
  \end{tabular}
  \end{table*}
  
\addtocounter{table}{-1}

\begin{table*}
\caption{ (Contd.)}
  \centering
  \begin{tabular}[t]{lcccc}
  \toprule
  Line & Wavelength, \AA & Flux, $10^{-15}$~erg~s$^{-1}$~cm$^{-2}$ & Equiv. width, \AA & $FWHM$, $10^2$ km/s\\
  \midrule

  \addlinespace[0.3em]
\multicolumn{5}{c}{SRGA\,J182109.8+765819}\\
\addlinespace[0.3em]
H$\beta$              & $5168$ & $<0.4 $        & $ >-0.8$      & $-$\\
{}[OIII]$\lambda$4959 & $5269$ & $1.62\pm0.22$  & $-3.1\pm0.4$  & $-$\\
{}[OIII]$\lambda$5007 & $5321$ & $3.73\pm0.25$  & $-7.0\pm0.4$ & $-$\\
{}[OI]$\lambda$6300   & $6698$ & $0.38\pm0.14$  & $-0.69\pm0.22$ & $-$\\
{}[NII]$\lambda$6548  & $6963$ & $1.18\pm0.15$  & $-2.08\pm0.24$ & $-$\\
H$\alpha$             & $6979$ & $2.48\pm0.17$  & $-4.37\pm0.29$ & $-$\\
{}[NII]$\lambda$6583  & $7001$ & $3.46\pm0.17$  & $-6.1\pm0.3$ & $-$\\
{}[SII]$\lambda$6716  & $7143$ & $1.10\pm0.14$  & $-2.07\pm0.27$ & $-$\\
{}[SII]$\lambda$6730  & $7157$ & $0.83\pm0.14$  & $-1.59\pm0.25$ & $-$\\
\addlinespace[0.3em]
\multicolumn{5}{c}{SRGA\,J193707.6+660816}\\
\addlinespace[0.3em]
H$\delta$, broad    & $4397$ & $12.2\pm1.8$  & $-23\pm3$      & $23.5\pm2.0$\\
H$\gamma$, broad    & $4655$ & $17.5\pm1.5$  & $-28.0\pm2.6$  & $22.2\pm1.9$\\
H$\beta$, broad     & $5207$ & $29.4\pm1.3$  & $-40.9\pm2.1$  & $23.3\pm1.2$\\
H$\beta$              & $5207$ & $3.1\pm0.6$   & $-4.3\pm1.0$   & $-$\\
{}[OIII]$\lambda$4959 & $5313$ & $10.4\pm0.4$  & $-14.7\pm0.7$  & $-$\\
{}[OIII]$\lambda$5007 & $5364$ & $31.5\pm0.6$  & $-44.9\pm1.1$  & $-$\\
{}[NII]$\lambda$6548  & $7016$ & $1.64\pm0.25$ & $-2.6\pm0.4$   & $-$\\
H$\alpha$, broad    & $7032$ & $115.1\pm2.5$ & $-185\pm6$     & $21.5\pm0.4$\\
H$\alpha$             & $7032$ & $15.0\pm1.3$  & $-24.1\pm2.7$  & $-$\\
{}[NII]$\lambda$6583  & $7053$ & $4.93\pm0.25$ & $-8.1\pm1.1$   & $-$\\
{}[SII]$\lambda$6716  & $7197$ & $3.0\pm0.5$   & $-5.4\pm0.8$   & $-$\\
{}[SII]$\lambda$6730  & $7212$ & $2.1\pm0.4$   & $-3.8\pm0.9$   & $-$\\
\addlinespace[0.3em]
\multicolumn{5}{c}{SRGA\,J200331.2+701332}\\
\addlinespace[0.3em]
H$\beta$, broad     & $5362$ & $24.1\pm2.3$   & $-60\pm7$ & $103\pm11$\\
H$\beta$              & $5362$ & $<0.8$         & $ >-2.1$ & $-$\\
{}[OIII]$\lambda$4959 & $5445$ & $1.4\pm0.4$    & $-3.6\pm1.0$ & $-$\\
{}[OIII]$\lambda$5007 & $5496$ & $6.2\pm0.5$    & $-15.5\pm1.4$ & $-$\\
{}[NII]$\lambda$6548  & $7200$ & $0.7\pm0.4$    & $-1.9\pm1.5$ & $-$\\
H$\alpha$, broad    & $7216$ & $85\pm3$       & $-230\pm13$ & $82\pm3$\\
H$\alpha$             & $7216$ & $<1.1$         & $ >-2.9$ & $-$\\
{}[NII]$\lambda$6583  & $7239$ & $2.2\pm0.4$    & $-6\pm4$ & $-$\\
    \addlinespace[0.3em]
\multicolumn{5}{c}{SRGA\,J211149.5+722815}\\
\addlinespace[0.3em]
H$\beta$              & $5377$ & $<0.9$      & $ >-4$ & $-$\\
{}[OIII]$\lambda$4959 & $5486$ & $3.0\pm0.5$ & $-13.2\pm2.3$ & $-$\\
{}[OIII]$\lambda$5007 & $5537$ & $7.8\pm0.6$ & $-33\pm3$ & $-$\\
{}[NII]$\lambda$6548  & $7225$ & $1.1\pm0.6$ & $-5.1\pm2.0$ & $-$\\
H$\alpha$             & $7252$ & $5.5\pm0.8$ & $-12\pm3$ & $-$\\
{}[NII]$\lambda$6583  & $7279$ & $8.7\pm0.9$ & $-21\pm3$ & $-$\\
{}[SII]$\lambda$6716  & $7430$ & $2.4\pm0.7$ & $-5.6\pm1.6$ & $-$\\
{}[SII]$\lambda$6730  & $7450$ & $1.4\pm0.6$ & $-3.9\pm1.3$ & $-$\\
  \bottomrule
  \end{tabular}
  \end{table*}

\end{document}